\documentclass[preprintnumbers,floats,floatfix,amssymb,prd,twocolumn,nofootinbib]{revtex4}
\usepackage{amssymb,amsmath}
\usepackage{epsfig,hyperref}
\usepackage[svgnames]{xcolor}
\usepackage{pgf,tikz}
\usepackage{epstopdf}
\usepackage[british]{babel}
\usepackage{enumerate}

\makeatletter
\DeclareRobustCommand*{\bfseries}{%
  \not@math@alphabet\bfseries\mathbf
  \fontseries\bfdefault\selectfont
  \boldmath
}
\makeatother

\def\be{\begin{equation}}
\def\ee{\end{equation}}
\def\beq{\begin{eqnarray}}
\def\eeq{\end{eqnarray}}

\makeatletter
\renewcommand{\vec}[1]{\mbox{\boldmath$#1$}}
\newcommand{\arXiv}[2][]{\href{http://arxiv.org/abs/#2}{\texttt{arXiv:#2\@ifempty{#1}{}{ [#1]}}}}
\makeatother

\begin{document}
\title{Interior dynamics of neutral and charged black holes in $f(R)$ gravity}

\author{Jun-Qi Guo}%
\email{junqi.guo@tifr.res.in}
\author{Pankaj S. Joshi}%
\email{psj@tifr.res.in}
\affiliation{Department of Astronomy and Astrophysics, Tata Institute of Fundamental Research, Homi Bhabha Road, Mumbai 400005, India}

\date{\today}

\begin{abstract}
  In this paper, we explore the interior dynamics of neutral and charged black holes in $f(R)$ gravity. We transform $f(R)$ gravity from the Jordan frame into the Einstein frame and simulate scalar collapses in flat, Schwarzschild, and Reissner-Nordstr\"{o}m geometries. In simulating scalar collapses in Schwarzschild and Reissner-Nordstr\"{o}m geometries, Kruskal and Kruskal-like coordinates are used, respectively, with the presence of $f'$ and a physical scalar field being taken into account. The dynamics in the vicinities of the central singularity of a Schwarzschild black hole and of the inner horizon of a Reissner-Nordstr\"{o}m black hole is examined. Approximate analytic solutions for different types of collapses are partially obtained. The scalar degree of freedom $\phi$, transformed from $f'$, plays a similar role as a physical scalar field in general relativity. Regarding the physical scalar field in $f(R)$ case, when $d\phi/dt$ is negative (positive), the physical scalar field is suppressed (magnified) by $\phi$, where $t$ is the coordinate time. For dark energy $f(R)$ gravity, inside black holes, gravity can easily push $f'$ to $1$. Consequently, the Ricci scalar $R$ becomes singular, and the numerical simulation breaks down. This singularity problem can be avoided by adding an $R^2$ term to the original $f(R)$ function, in which case an infinite Ricci scalar is pushed to regions where $f'$ is also infinite. On the other hand, in collapse for this combined model, a black hole, including a central singularity, can be formed. Moreover, under certain initial conditions, $f'$ and $R$ can be pushed to infinity as the central singularity is approached. Therefore, the classical singularity problem, which is present in general relativity, remains in collapse for this combined model.
\end{abstract}

\maketitle

\section{Introduction\label{sec:introduction}}
The internal structure of black holes and spacetime singularities are key topics in gravitation and cosmology~\cite{Burko_1997_book,Brady_1999,Berger_2002,Joshi_2007,Henneaux_2008}, and are great platforms to explore the connection between classical and quantum physics. It is widely believed that rotating black holes exist in reality. According to Price's theorem, for a collapsing star, the gravitational radiation carries away all the initial features of the star's gravitational field, except the mass, charge, and angular momentum parameters~\cite{Price}. As a further step, it is natural to ask what the final state of the internal collapses might be.

Ever since the foundation of general relativity, people have been trying to go beyond it. This endeavor arises from unifying gravitation and quantum mechanics, and addressing some cosmological problems, including the singularity problem in the early Universe and the dark energy problem in the late Universe. Various modified gravity theories have been explored, including scalar-tensor theory, high-dimensional theory, and $f(R)$ gravity, etc. For a review of modified gravity theories, see Ref.~\cite{Clifton_1106}. For reviews of $f(R)$ theory, see Refs.~\cite{Sotiriou_0805,Felice_1002,Nojiri_1002,Capozziello_1108}.

Static and spherically symmetric black hole solutions in $f(R)$ gravity were explored in Ref.~\cite{Cruz_0907}. Charged Born-Infeld black holes for $f(R)$ theories were studied in Ref.~\cite{Olmo_1110}. Instabilities and (anti)-evaporation of Schwarzschild-de Sitter and Reissner-Nordstr\"{o}m black holes in modified gravity were discussed in Refs.~\cite{Nojiri_1301,Sebastiani_1305,Nojiri_1405}.

Gravitational collapses in some modified gravity theories have been studied numerically. Spherical collapse of a neutral scalar field in a given spherical, charged black hole in Brans-Dicke theory was investigated in Ref.~\cite{Avelino_2009}. Spherical collapses of a charged scalar field in dilaton gravity and $f(R)$ gravity were explored in Refs.~\cite{Borkowska_2011} and~\cite{Hwang_1110}, respectively. Spherical scalar collapse in $f(R)$ gravity was simulated in Ref.~\cite{Guo_1312}. Asymptotic analysis was implemented in the vicinity of the singularity of a formed black hole.

\subsection{Mass inflation}
In the vicinity of the central singularity inside a Schwarzschild black hole, the tidal force diverges, and maximal globally hyperbolic region defined by initial data is inextendible. However, inside charged (Reissner-Nordstr\"{o}m) and rotating (Kerr) black holes, the central singularity is timelike. The globally hyperbolic region is up to the Cauchy horizon, and the spacetime is extendible beyond this horizon to a larger manifold. The Reissner-Nordstr\"{o}m inner (Cauchy) horizon is a surface of infinite blueshift, which in turn may cause the inner horizon unstable~\cite{Simpson_1973}. Furthermore, the strong cosmic censorship conjecture was proposed, which states that for generic asymptotically flat initial data, the maximal Cauchy development is future inextendible. For mathematical explorations of the internal structures of charge black holes, see Refs.~\cite{Dafermos_2003,Dafermos_2014}. For reviews on the Cauchy problem in general relativity and strong cosmic censorship, see Refs.~\cite{Ringstrom_2015,Isenberg_2015}, respectively.

The backreaction of the radiative tail from a gravitational collapse on the inner horizon of a Reissner-Nordstr\"{o}m black hole was investigated by Poisson and Israel~\cite{Poisson_1989,Poisson_1990}. It was shown that due to the divergence of the tail's energy density occurring on the inner horizon, the effective internal gravitational-mass parameter becomes unbounded. This phenomenon is usually called mass inflation. These arguments were extended to the rotating black hole case in Ref.~\cite{Barrabes_1990}.

In Refs.~\cite{Poisson_1989,Poisson_1990}, approximate analytic expressions were obtained by considering a simplified model in which the perturbations were modeled by cross-flowing radial streams of infalling and outgoing lightlike particles. To get more information, some numerical simulations in more realistic models have been performed. The dynamics of a spherical, charged black hole perturbed nonlinearly by a self-gravitating massless scalar field was numerically studied in Refs.~\cite{Gnedin_1991,Gnedin_1993,Brady_1995,Burko_1997,Burko_1997b,Hansen_2005}. Under the influence of the scalar field, the inner horizon of a charged black hole contracts to zero volume, and the center becomes a spacelike singularity. The mass inflation phenomenon was observed. In Refs.~\cite{Hod_1997,Oren_2003}, with regular initial data, spherical collapse of a charged scalar field was simulated. An apparent horizon was formed. A null, weak mass-inflation singularity along the Cauchy horizon and a final, spacelike, central singularity were obtained. Spherical collapses in Brans-Dicke theory, dilaton gravity, and $f(R)$ gravity were investigated in Refs.~\cite{Avelino_2009,Borkowska_2011,Hwang_1110}. Mass inflation phenomena were also reported.

It is important to connect approximate analytic candidate expressions with numerical results. In Refs.~\cite{Burko_1998,Burko_1999}, the features of the Cauchy horizon singularity in charge scattering were studied. Analytic and numerical results were compared at some steps.

In this paper, we use the following notations:
\begin{enumerate}[(i)]
  \item Neutral collapse: neutral scalar collapse toward a black hole formation.
  \item Neutral scattering: neutral scalar collapse in a (neutral) Schwarzschild geometry.
  \item Charge scattering: neutral scalar collapse in a (charged) Reissner-Nordstr\"{o}m geometry. In this process, the scalar field is scattered by the inner horizon of a Reissner-Nordstr\"{o}m black hole.
\end{enumerate}

\subsection{New results}
In this paper, we explore neutral scalar collapses in flat, Schwarzschild, and Reissner-Nordstr\"{o}m geometries in $f(R)$ gravity, taking the well-known Hu-Sawicki model as an example~\cite{Hu_0705}. We seek approximate analytic solutions. A generalized Misner-Sharp energy in $f(R)$ gravity in the Jordan frame was defined in Ref.~\cite{Cai_0910}. In this paper, for ease of operation, we mainly work in the Einstein frame and compute the Misner-Sharp mass function of the general relativity version, instead. Moreover, we will investigate a dark energy $f(R)$ singularity problem.

We explore scalar collapses in both general relativity and $f(R)$ gravity. For convenience, we transform the dynamical system in $f(R)$ gravity from the Jordan frame into the Einstein frame. In the new system, we equivalently work in Einstein gravity to which a scalar degree of freedom $\phi(\equiv\sqrt{3/2}\ln{f'}/\sqrt{8{\pi}G})$ and a physical scalar field $\psi$ are coupled. Basically, $\phi$ plays a similar role as what a physical scalar field $\psi$ does in Einstein gravity. While in $f(R)$ gravity, the physical scalar field $\psi$ is suppressed (magnified) when $d\phi/dt$ is negative (positive), where $t$ is the coordinate time. For simplicity, the results in general relativity are presented in a separate paper~\cite{Guo_1507}, and we focus on collapse in $f(R)$ gravity in this paper.

According to the strength of the scalar field, charge scattering can be classified into five types as follows:
\begin{enumerate}[(i)]
  \item Type I: spacelike scattering. When the scalar field is very strong, the inner horizon can contract to zero volume rapidly, and the central singularity becomes spacelike. The dynamics near the spacelike singularity is similar to that in neutral collapse.
  \item Type II: null scattering. When the scalar field is intermediate, the inner horizon can contract to a place close to the center or reach the center. For each variable (the metric elements and physical scalar field), the spatial and temporal derivatives are almost equal. In the case of the center being reached, the central singularity is null. This type has two stages: early/slow and late/fast. In the early stage, the inner horizon contracts slowly, and the scalar field also varies slowly. In the late stage, the inner horizon contracts quickly, and the dynamics is similar to that in the spacelike scattering case.
  \item Type III: critical scattering. This case is on the edge between the above two cases. The central singularity becomes null.
  \item Type IV: weak scattering. When the scalar field is very weak, the inner horizon contracts but not much. Then the central singularity remains timelike.
  \item Type V: tiny scattering. When the scalar field is very tiny, the influence of the scalar field on the internal geometry is negligible.
\end{enumerate}
In this paper, we will explore the dynamics of Types I, II, and IV, and obtain approximate analytic solutions for the first two.

By comparing the dynamics in a Reissner-Nordstr\"{o}m geometry and charge scattering, we investigate the causes of mass inflation and seek further approximate analytic solutions with the following improvements. Usually, double-null coordinates are used in studies of mass inflation in spherical symmetry. In the line element of double-null coordinates, the two null coordinates $u$ and $v$ are present in the form of product $dudv$. In the equations of motion, mixed derivatives of $u$ and $v$ are present quite often. In this paper, we use a slightly modified line element, in which one coordinate is timelike and the rest are spacelike. In this case, in the equations of motion, spatial and temporal derivatives are usually separated. This simplifies the numerical formalism and helps to obtain approximate analytic solutions. In addition, we compare numerical results and approximate analytic solutions closely at each step. We compare the dynamics for Schwarzschild black holes, Reissner-Nordstr\"{o}m black holes, neutral collapse, and charge scattering. We treat the system as a mathematical dynamical system rather than a physical one, examining the contributions from all the terms in the equations of motion.

In Ref.~\cite{Hwang_1110} where spherical charged scalar collapse in $f(R)$ gravity was simulated, a singularity problem was reported. When a dark energy $f(R)$ model is used, $f'$ can be pushed to $1$ easily. Correspondingly, the Ricci scalar $R$ becomes singular. This singularity problem can be avoided when an $R^2$ model is used instead. However, the causes of this singularity problem were not explained. In this paper, we will consider a simpler case. Instead of simulating the collapse of a charged scalar field, we study neutral scalar collapse in a Reissner-Nordstr\"{o}m geometry. The same singularity problem is found. By analyzing the contributions from all the terms in the equations of motion for $\phi({\equiv}\sqrt{3/2}{\ln}f'/\sqrt{{8\pi}G})$ with $f'{\equiv}df/dR$, we interpret the causes for the singularity problem. Basically, near the inner horizon, in the equation of motion for $\phi$, the scalar field $\phi$ and the geometry construct a positive feedback system. Depending on initial conditions, $\phi$ can be accelerated either in positive or in negative directions, until singularities are met. In the negative case, $\phi$ can be accelerated to negative infinity. Correspondingly, $f'$ goes to zero as the central singularity is approached. However, in the positive case, $\phi$ can be pushed to zero in a short time. Correspondingly, $f'$ and the Ricci scalar $R$ are pushed to $1$ and infinity, respectively. This is the cause of the singularity problem. Taking into account quantum-gravitational effects at high curvature scale, one may obtain an additional $R^2$ term to the Lagrangian for gravity~\cite{Starobinsky_1980,Vilenkin_1985}. When this $R^2$ term is added to the $f(R)$ function, a singular $R$ is pushed to regions where $f'$ is also singular. Therefore, the singularity problem can be avoided~\cite{Starobinsky_1980,Vilenkin_1985,Nojiri_0804,Bamba_0807,Capozziello_0903,Appleby_0909,Bamba_1012,Bamba_1101}.

Although the dark energy $f(R)$ singularity problem is avoided in the combined model (a combination of a dark energy $f(R)$ model and the $R^2$ model), the classical singularity problem, which is present in general relativity, remains in collapse for this model. Under certain initial conditions, near the central singularity, $d\phi/dt$ can be positive. Then the positive feedback system in the equation of motion for $\phi$ can push $\phi$ and $R$ to positive infinity.

This paper is organized as follows. In Sec.~\ref{sec:framework}, we build the framework for charge scattering, including action for charge scattering, the coordinate system, and the $f(R)$ model. In Sec.~\ref{sec:set_up}, we set up the numerical formalism for charge scattering. In Secs.~\ref{sec:neutral_collapse}, \ref{sec:neutral_scattering}, and \ref{sec:results}, scalar collapses in flat, Schwarzschild, and Reissner-Nordstr\"{o}m geometries will be explored, respectively. In Sec.~\ref{sec:weak_scattering}, we consider weak charge scattering. In Sec.~\ref{sec:singularity}, we discuss the causes and avoidance of the singularity problem. In Sec.~\ref{sec:summary}, the results will be summarized.

In this paper, we set $G=c=4\pi\epsilon_0=1$.

\section{Framework\label{sec:framework}}
In this section, we build the framework for charge scattering in $f(R)$ gravity, in which a self-gravitating massless scalar field collapses in a Reissner-Nordstr\"{o}m geometry in $f(R)$ gravity. Compared to general relativity, in this process, there is one extra scalar degree of freedom $f'{\equiv}df/dR$. For convenience, $f(R)$ gravity is transformed from the Jordan frame into the Einstein frame. For comparison and verification considerations, we use Kruskal-like coordinates, and set up the initial conditions by modifying those in a Reissner-Nordstr\"{o}m geometry with a physical scalar field, a scalar degree of freedom $f'$, and the potential for $f'$. The Hu-Sawicki model is used as an example.

\subsection{Action}
The action for charge scattering in $f(R)$ gravity can be written as follows:
\be S=\int d^{4}x \sqrt{-g}\left[\frac{f(R)}{16\pi G}+\mathcal{L}_{\psi}+\mathcal{L}_{F}\right], \label{f_R_action} \ee
with
\begin{align}
\mathcal{L}_{\psi}&=-\frac{1}{2}g^{\alpha\beta}\psi_{,\alpha}\psi_{,\beta},\\
\nonumber\\
\mathcal{L}_{F}&=-\frac{F_{\mu\nu}F^{\mu\nu}}{4}.
\end{align}
$f(R)/(16{\pi}G)$, $\mathcal{L}_{\psi}$, and $\mathcal{L}_{F}$ are the Lagrange densities for $f(R)$ gravity, a physical scalar field $\psi$, and the electric field for a Reissner-Nordstr\"{o}m black hole, respectively. $f(R)$ is a certain function of the Ricci scalar $R$, and $G$ is the Newtonian gravitational constant. $F_{\mu\nu}$ is the electromagnetic-field tensor for the electric field of a Reissner-Nordstr\"{o}m black hole.

The energy-momentum tensor for the massless scalar field $\psi$ is
\be T^{(\psi)}_{\mu\nu}
\equiv-\frac{2}{\sqrt{|g|}}\frac{\delta(\sqrt{|g|}\mathcal{L}_{\psi})}{\delta g^{\mu\nu}}
=\psi_{,\mu}\psi_{,\nu}-\frac{1}{2}g_{\mu\nu}g^{\alpha\beta}\psi_{,\alpha}\psi_{,\beta}.\label{energy_tensor_psi_JF}\ee
The electric field of a Reissner-Nordstr\"{o}m black hole can be treated as a static electric field of a point charge of strength $q$ sitting at the origin $r=0$. In the Reissner-Nordstr\"{o}m metric, the only nonvanishing components of $F_{\mu\nu}$ are $F_{tr}=-F_{tr}=-q/r^2$. The corresponding energy-momentum tensor for the electric field is~\cite{Poisson_2004}
\begin{eqnarray}
{T^{(F)}}^{\mu}_{\nu}
&\equiv&-\frac{2}{\sqrt{|g|}}\frac{\delta(\sqrt{|g|}\mathcal{L}_{F})}{\delta{g_{\mu}^{\nu}}}\nonumber\\
&=&\frac{1}{4\pi}\left(F^{\mu\rho}F_{\nu\rho}-\frac{1}{4}\delta^{\mu}_{\nu}F^{\alpha\beta}F_{\alpha\beta}\right)\nonumber\\
&=&\frac{q^2}{8{\pi}r^4}\cdot\mbox{diag}(-1,-1,1,1).
\label{em_tensor}
\end{eqnarray}
Although Eq.~(\ref{em_tensor}) is obtained in the Reissner-Nordstr\"{o}m metric, it is valid in any coordinate system, since as seen by static observers, the electromagnetic field should be purely electric and radial~\cite{Poisson_1990,Poisson_2004}.

\subsection{$f(R)$ theory}
The equivalent of the Einstein equation in $f(R)$ gravity reads
\be f'R_{\mu\nu}-\frac{1}{2}f g_{\mu\nu} -\left(\nabla_{\mu}\nabla_{\nu}-g_{\mu\nu} \Box\right) f'
= 8\pi T_{\mu\nu},\label{gravi_eq_fR} \ee
where $f'{\equiv}df/dR$ and $\Box\equiv\nabla_{\alpha}\nabla^{\alpha}$. The trace of Eq.~(\ref{gravi_eq_fR}) describes the dynamics of $f'$,
\be \Box f'-\frac{2f-f'R}{3}-\frac{8\pi}{3}T=0, \label{trace_eq1}\ee
where $T$ is the trace of the stress-energy tensor $T_{\mu\nu}$. Defining a new variable $\chi$ by
\be \chi\equiv\frac{df}{dR}, \label{f_prime}\ee
and a potential $U(\chi)$ by
\be U'(\chi)\equiv\frac{dU}{d\chi}=\frac{2f-f'R}{3},\label{v_prime}\ee
one can rewrite Eq.~(\ref{trace_eq1}) as
\be \Box\chi-U'(\chi)-\frac{8\pi}{3}T=0.\label{trace_eq2}\ee

The field equations for $f(R)$ gravity (\ref{gravi_eq_fR}) are somewhat different from the more familiar corresponding ones in general relativity. Therefore, for convenience, we transform $f(R)$ gravity from the current frame, which is usually called the Jordan frame, into the Einstein frame, in which the formalism can be formally treated as Einstein gravity coupled to a scalar field.

Rescaling $\chi$ by
\be \kappa\phi\equiv\sqrt{\frac{3}{2}}\ln\chi,\label{rescale_f_prime}\ee
one obtains the corresponding action of $f(R)$ gravity in the Einstein frame~\cite{Felice_1002}
\begin{eqnarray}
S_{E} & = & \int d^4 x\sqrt{-\tilde g}\left[\frac{1}{2\kappa^2} \tilde R -\frac{1}{2} \tilde g^{\mu\nu}\partial_\mu\phi\partial_\nu\phi-V(\phi)\right]\nonumber\\
 && + \int d^4 x\mathcal{L}_M\left(\frac{\tilde g_{\mu\nu}}{{\chi(\phi)}},\psi,q\right),
\end{eqnarray}
where $\kappa=\sqrt{8\pi G}$, $\tilde g_{\mu\nu}=\chi\cdot g_{\mu\nu}$, $V(\phi)\equiv(\chi R-f)/(2\kappa^{2}\chi^2)$, and a tilde denotes that the quantities are in the Einstein frame. The Einstein field equations are
\be
\tilde G_{\mu\nu} = \kappa^2 \left[\tilde T_{\mu\nu}^{(\phi)}+\tilde T_{\mu\nu}^{(M)}\right],
\ee
where
\begin{align}
\tilde T_{\mu\nu}^{(\phi)} &= \partial_\mu\phi\partial_\nu\phi-\tilde g_{\mu\nu}\left[\frac{1}{2}\tilde g^{\alpha\beta}\partial_\alpha\phi\partial_\beta\phi+V(\phi)\right], \\
\tilde T_{\mu\nu}^{(M)} &= \frac{T_{\mu\nu}^{(M)}}{\chi}.
\end{align}
$T_{\mu\nu}^{(M)}$ is the ordinary energy-momentum tensor for the physical matter fields in terms of $g_{\mu\nu}$ in the Jordan frame. With the expression for the energy-momentum tensor for the scalar field $\psi$ in the Jordan frame, shown in Eq.~(\ref{energy_tensor_psi_JF}), the corresponding expression in the Einstein frame can be written as
\begin{eqnarray}
\tilde T_{\mu\nu}^{(\psi)} & = & \frac{1}{\chi} \left(\partial_\mu \psi \partial_\nu \psi - \frac{1}{2}g_{\mu\nu}g^{\alpha\beta}\partial_\alpha\psi\partial_\beta \psi\right)\nonumber\\
 & = & \frac{1}{\chi} \left(\partial_\mu \psi \partial_\nu \psi - \frac{1}{2}\tilde g_{\mu\nu} \tilde g^{\alpha\beta}\partial_\alpha\psi\partial_\beta \psi\right),
\end{eqnarray}
which gives
\be
\tilde T^{(\psi)}=\tilde g^{\mu\nu} \tilde T^{(\psi)}_{\mu\nu}
=-\frac{\tilde g^{\mu\nu}\tilde \partial_{\mu}\psi\tilde\partial_{\nu}\psi}{\chi}=\frac{T^{(\psi)}}{\chi^2}.
\label{T_JF_EF}\ee

In the Jordan frame, in any coordinate system, the energy-momentum tensor for the static electric field of a point charge of strength $q$ sitting at the origin $r=0$ can be expressed as [see Eq.~(\ref{em_tensor})]
\be T^{(q)\mu}_{\hphantom{ddd}\nu}=\frac{q^2}{8\pi{r_{\mbox{\tiny JF}}}^4}\cdot\mbox{diag}(-1,-1,1,1).\ee
Then we have in the Einstein frame,
\begin{eqnarray}
{{\tilde T}^{(q)\mu}}_{\hphantom{ddd}\nu} & \equiv & \tilde g^{\mu\alpha}\tilde T^{(q)}_{\alpha\nu}\nonumber\\
 & = & \frac{g^{\mu\alpha}}{\chi}\cdot\frac{T^{(q)}_{\alpha\nu}}{\chi}\nonumber\\
 & = & \frac{q^2}{8\pi {\chi^2 r_{\mbox{\tiny JF}}}^4}\cdot\mbox{diag}(-1,-1,1,1)\nonumber\\
 & = & \frac{q^2}{8\pi {r_{\mbox{\tiny EF}}}^4}\cdot\mbox{diag}(-1,-1,1,1),
\end{eqnarray}
where $r_{\mbox{\tiny JF}}$ and $r_{\mbox{\tiny EF}}$ are the quantity $r$ in the Jordan and Einstein frames, respectively. Since we mainly work in the Einstein frame in this paper, we simply use $r$ for $r_{\mbox{\tiny EF}}$. We denote the total energy-momentum tensor for the source fields as
\be {\tilde T^{(\mbox{total})\mu}}_{\hphantom{ddddddd}\nu}
={\tilde T^{(\phi)\mu}}_{\hphantom{ddd}\nu}
+{\tilde T^{(\psi)\mu}}_{\hphantom{ddd}\nu}
+{\tilde T^{(q)\mu}}_{\hphantom{ddd}\nu}.\ee

The equations of motion for $\phi$ and $\psi$ can be derived from the Lagrange equations as
\be
\tilde\Box\phi - V'(\phi)-\frac{1}{\sqrt{6}}\kappa\tilde T^{(\psi)} = 0,
\label{fR:Box_phi}
\ee
\be
\tilde\Box\psi -\sqrt{\frac{2}{3}} ~\kappa \tilde g^{\mu\nu} \partial_\mu \phi\partial_\nu \psi = 0.
\label{fR:Box_psi}
\ee
Alternatively, Eqs.~(\ref{fR:Box_phi}) and (\ref{fR:Box_psi}) can be obtained from the corresponding ones in the Jordan frame. Some details are given in the Appendix.

In the Einstein frame, the potential for $\phi$ can be written as
\be V(\phi)=\frac{\chi R-f}{2\kappa^2\chi^2}. \label{potential_EF}\ee
Then we have
\be V'(\phi)=\frac{dV}{d\chi}\cdot \frac{d\chi}{d\phi}
=\frac{2f-\chi R}{{\sqrt{6}}\kappa\chi^2}.\label{V_prime_EF}\ee

\begin{figure}
  \epsfig{file=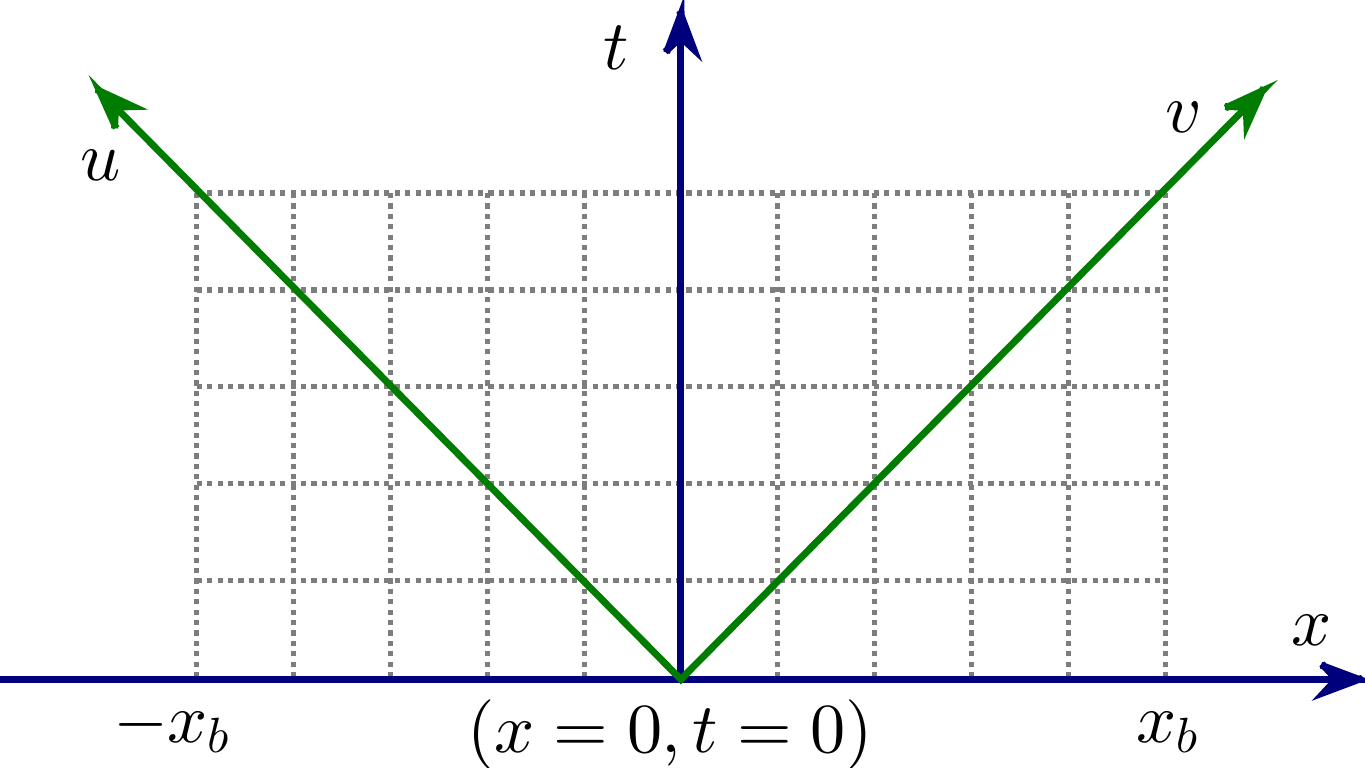, height=4cm}
  \caption{Initial and boundary conditions for charge scattering. Initial slice is at $t=0$. Definition domain for $x$ is $[-x_b~x_b]$.
  $u=(t-x)/2$ and $v=(t+x)/2$.}
  \label{fig:ic_and_bc}
\end{figure}

\begin{figure}
  \epsfig{file=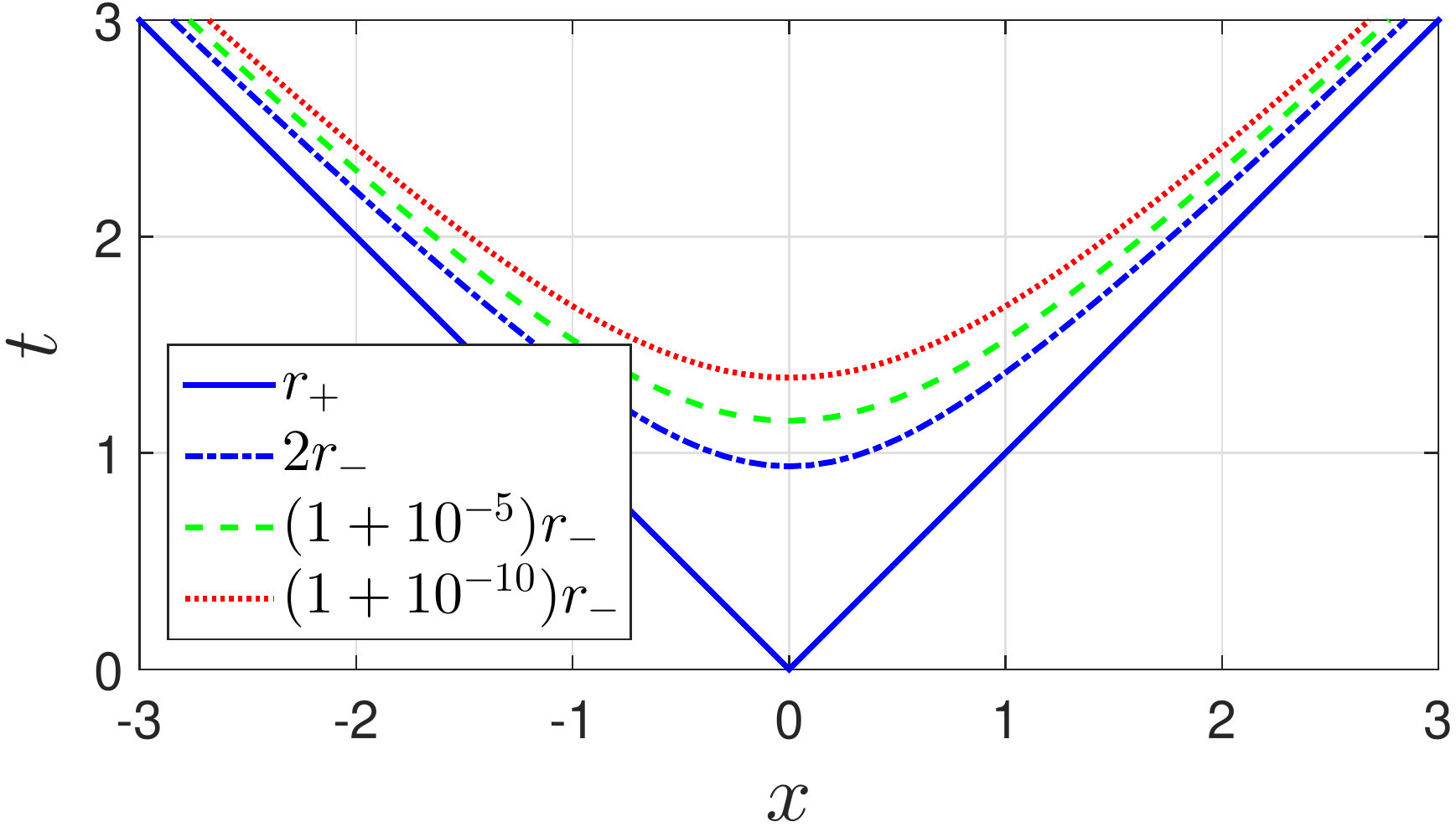, height=4cm}
  \caption{Contour lines for $r$ defined by Eq.~(\ref{r_RN_metric}) in a Reissner-Nordstr\"{o}m geometry with $m=1$ and $q=0.7$. Although the exact inner horizon is at regions where $uv$ and $(t^2-x^2)$ are infinite, $r$ can be very close to the inner horizon $r=r_{-}$ even when $uv$ and $(t^2-x^2)$ take moderate values.}
  \label{fig:contour_lines_r}
\end{figure}

\subsection{Coordinate system}
In the studies of mass inflation, the double-null coordinates described by Eq.~(\ref{double_null_metric_dudv}) are usually used,
\be
ds^{2} = -4e^{-2\sigma}dudv+r^2d\Omega^2,
\label{double_null_metric_dudv}
\ee
where $\sigma$ and $r$ are functions of the coordinates $u$ and $v$. $u$ and $v$ are outgoing and ingoing characteristics (trajectories of photons), respectively. For convenience, in this paper, we use a slightly modified form described by Eq.~(\ref{double_null_metric_dtdx}), obtained by defining $u=(t-x)/2$ and $v=(t+x)/2$~\cite{Frolov_2004},
\be
ds^{2} = e^{-2\sigma}(-dt^2+dx^2)+r^2d\Omega^2.
\label{double_null_metric_dtdx}
\ee
This set of coordinates is illustrated in Fig.~\ref{fig:ic_and_bc}. Similar to the Schwarzschild metric, the Reissner-Nordstr\"{o}m metric can be expressed in Kruskal-like coordinates~\cite{Graves_1960} (also see Refs.~\cite{Poisson_1989,Poisson_1990,Poisson_2004,Reall_2015}). So for ease and intuitiveness, we set the initial conditions close to those of the Reissner-Nordstr\"{o}m metric in Kruskal-like coordinates, taking into account the presence of a physical scalar field $\psi$, a scalar degree of freedom $f'$, and the potential $U(f')$.

In the form of Ref.~\cite{Reall_2015}, the Reissner-Nordstr\"{o}m metric in Kruskal-like coordinates in the region of $r>r_{-}$ can be written as
\be ds^2=\frac{r_{+}r_{-}}{k_{+}^2r^{2}}e^{-2k_{+}r}\left(\frac{r}{r_{-}}-1\right)^{1+\frac{k_{+}}{|k_{-}|}}(-dt^2+dx^2)+r^2d\Omega^2,
\label{RN_metric_text}\ee
where $r_{\pm}(=m\pm\sqrt{m^2-q^2})$ and $k_{\pm}[=(r_{\pm}-r_{\mp}]/(2r_{\pm}^2)]$ are the locations and surface gravities for the outer and inner horizons of a Reissner-Nordstr\"{o}m black hole, respectively.
$r(t,x)$ is defined implicitly below~\cite{Reall_2015},
\be 4uv=t^2-x^2=e^{2k_{+}r}\left(1-\frac{r}{r_{+}}\right)\left(\frac{r}{r_{-}}-1\right)^{-\frac{k_{+}}{|k_{-}|}}.\label{r_RN_metric}\ee
In this set of coordinates, as implied by Eq.~(\ref{r_RN_metric}), the exact inner horizon is at regions where $uv$ and $(t^2-x^2)$ are infinite. However, it is found that, even when $uv$ and $(t^2-x^2)$ take moderate values, $r$ still can be very close to the inner horizon, e.g., $r=(1+10^{-10})r_{-}$. (See Fig.~\ref{fig:contour_lines_r}.) Therefore, at such regions, the interaction between the scalar fields and the inner horizon still can be very strong, then we can investigate mass inflation numerically.

This formalism has several advantages as follows:
\begin{enumerate}[(i)]
  \item In the line element (\ref{double_null_metric_dtdx}), one coordinate is timelike and the rest are spacelike. This is a conventional setup. It is more convenient and more intuitive to use this set of coordinates. For the set of coordinates described by Eq.~(\ref{double_null_metric_dudv}), in the equations of motion, many terms are mixed derivatives of $u$ and $v$; while for the set of coordinates described by Eq.~(\ref{double_null_metric_dtdx}), in the equations of motion, spatial and temporal derivatives are usually separated.
  \item We set initial conditions close to those in the Reissner-Nordstr\"{o}m metric. Consequently, with the terms related to the scalar fields being removed, we can test our code by comparing the numerical results to the analytic ones in the Reissner-Nordstr\"{o}m case conveniently. Moreover, by comparing dynamics for scalar collapse to that in the Reissner-Nordstr\"{o}m case, we can obtain intuitions on how the scalar fields affect the geometry.
  \item The interactions between scalar fields and the geometry are local effects. In Refs.~\cite{Gnedin_1991,Gnedin_1993}, the space between the inner and outer horizons are compactified into finite space. This overcompactification, at least to us, makes it a bit hard to understand the dynamics. In the configuration that we choose, the space is partially compactified, and the picture of charge scattering turns out to be simpler.
\end{enumerate}

\subsection{$f(R)$ model}
For a viable dark energy $f(R)$ model, $f'$ has to be positive to avoid ghosts~\cite{Nunez}, and $f''$ has to be positive to avoid the Dolgov-Kawasaki instability~\cite{Dolgov}. The model should also be able to generate a cosmological evolution compatible with the observations~\cite{Amendola,Guo_1305} and to pass the Solar System tests~\cite{Hu_0705,Justin1,Justin2,Chiba,Tamaki_0808,Tsujikawa_0901,Guo_1306}. Equivalently, general relativity should be restored at high curvature scale, and the $f(R)$ model mainly deviates from general relativity at low curvature scale comparable to the cosmological constant. In this paper, we take a typical dark energy $f(R)$ model, the Hu-Sawicki model, as an example. This model reads~\cite{Hu_0705}
\be f(R)=R-R_{0}\frac{D_{1}R^n}{D_{2}R^n+R_{0}^n},\label{f_R_Hu_Sawicki_general}\ee
where $n$ is a positive parameter, $D_1$ and $D_2$ are dimensionless parameters, $R_{0}=8\pi\bar{\rho}_0/3$, and $\bar{\rho}_0$ is the average matter density of the current Universe. We consider one of the simplest versions of this model, i.e., $n=1$,
\be f(R)=R-\frac{DR_{0} R}{R+R_{0}}, \label{f_R_Hu_Sawicki}\ee
where $D$ is a dimensionless parameter. In this model,
\be f'=1-\frac{DR_{0} ^2}{(R+R_{0})^2},\label{f_prime_Hu_Sawicki}\ee
\be V'(\phi)=\frac{R^{3}}{\sqrt{6}\kappa f'^{2}(R+R_{0})^{2}}\left[1+(1-D)\frac{R_{0}}{R}\left(2+\frac{R_{0}}{R}\right) \right].
\label{V_prime_Hu_Sawicki}\ee

\begin{figure*}[t!]
  \epsfig{file=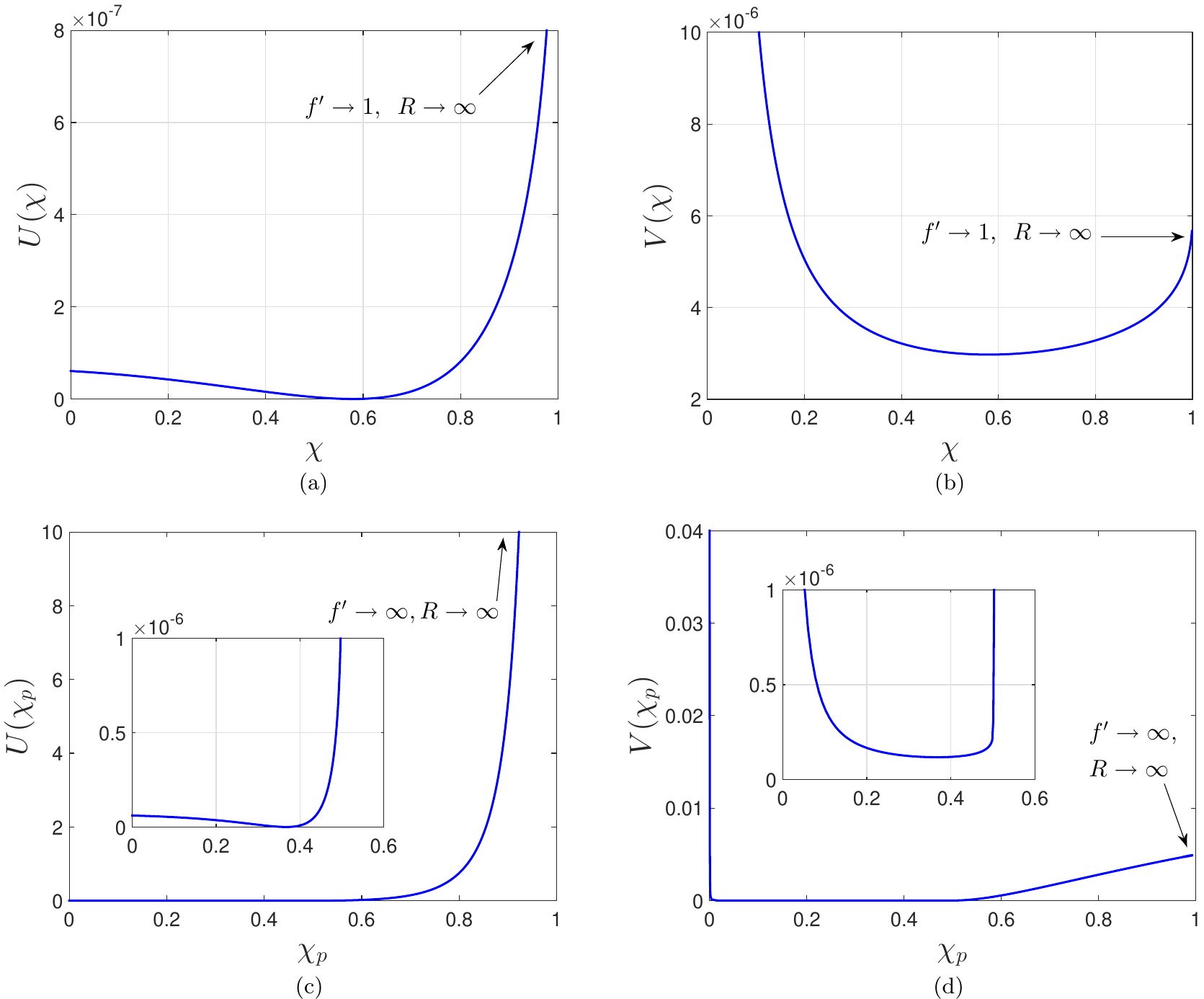, width=0.8\textwidth}
  \caption{Potentials for $f(R)$ models. $\chi{\equiv}f'$ and $\chi_p{\equiv}\chi/(1+\chi)$. $U(\chi)$ and $V(\chi)$ are the potentials in the Jordan and Einstein frames and can be obtained from Eqs.~(\ref{v_prime}) and (\ref{potential_EF}), respectively. (a) and (b) are for the Hu-Sawicki model~(\ref{f_R_Hu_Sawicki}), $f(R)=R-DR_{0}R/(R+R_{0})$, while (c) and (d) are for the combined model~(\ref{f_R_combined}), $f(R)=R-DR_{0}R/(R+R_{0})+{\alpha}R^2$. $D=1.2$, $R_{0}=10^{-5}$, and $\alpha=1$.}
  \label{fig:potential}
\end{figure*}

As implied in Eq.~(\ref{V_prime_Hu_Sawicki}), to make sure that the de Sitter curvature, for which $V'(\phi)=0$, has a positive value, the parameter $D$ needs to be greater than $1$. In this paper, we set $D$ to $1.2$ and set $R_{0}$ to $10^{-5}$ or $10^{-6}$. Then, together with Eqs.~(\ref{potential_EF}) and (\ref{V_prime_Hu_Sawicki}), these values imply that the radius of the de Sitter horizon is about $\sqrt{1/R_0}\sim 10^{3}$. Moreover, in the configuration of the initial conditions described in Secs.~\ref{sec:set_up_ic} and \ref{sec:results}, the radii of the outer apparent horizons of the formed black holes are about $2.1$ and $3.7$, respectively. [See Figs.~\ref{fig:neutral_collapse_evolutions}(f) and \ref{fig:evolutions}(f).] The potentials in the Jordan and Einstein frames are plotted in Figs.~\ref{fig:potential}(a) and \ref{fig:potential}(b), respectively.

\section{Numerical setup for charge scattering\label{sec:set_up}}
In this section, we set up the numerical formalisms for charge scattering in $f(R)$ gravity, including field equations, initial conditions, boundary conditions, discretization scheme, and tests of numerical codes.

\subsection{Field equations\label{sec:field_eqs}}
In double-null coordinates (\ref{double_null_metric_dtdx}), using
\be \tilde G^{t}_{t}+\tilde G^{x}_{x}
=8\pi\left[{\tilde T^{(\mbox{total})t}}_{\hphantom{ddddddd}t}+{\tilde T^{(\mbox{total})x}}_{\hphantom{ddddddd}x}\right],\nonumber\ee
one obtains the equation of motion for $r$,
\be r(-r_{,tt}+r_{,xx}) -r_{,t}^2+r_{,x}^2 = e^{-2\sigma}\left(1-8\pi r^2V-\frac{q^2}{r^2}\right), \label{equation_r}\ee
where $r_{,t}\equiv dr/dt$ and other quantities are defined analogously. For simplicity, we define $\eta\equiv r^2$ and integrate the equation of motion for $\eta$, instead~\cite{Frolov_2004}. The equation of motion for $\eta$ can be obtained by rewriting Eq.~(\ref{equation_r}) as
\be -\eta_{,tt}+\eta_{,xx}=2e^{-2\sigma}\left(1-8\pi r^2V-\frac{q^2}{r^2}\right).\label{equation_r_2}\ee
$G^{\theta}_{\theta}=8\pi\tilde T^{(\mbox{total})\theta}_{\hphantom{ddddddd}\theta}$ provides the equation of motion for $\sigma$,
\be
\begin{split}
&-\sigma_{,tt}+\sigma_{,xx}+\frac{r_{,tt}-r_{,xx}}{r}\\
&+4\pi\left(\phi_{,t}^2-\phi_{,x}^2 + \frac{\psi_{,t}^2-\psi_{,x}^2}{\chi}-2e^{-2\sigma}V\right)\\
&+e^{-2\sigma}\frac{q^2}{r^4}=0.
\end{split}
\label{equation_sigma}
\ee

In double-null coordinates, the equations of motion for $\phi$ (\ref{fR:Box_phi}) and $\psi$ (\ref{fR:Box_psi}) become, respectively,
\be
\begin{split}
&-\phi_{,tt}+\phi_{,xx}+\frac{2}{r}(-r_{,t}\phi_{,t}+r_{,x}\phi_{,x})\\
&=e^{-2\sigma}\left[V'(\phi)+\frac{1}{\sqrt{6}}\kappa\tilde{T}^{(\psi)}\right],
\end{split}
\label{equation_phi}
\ee
\be
\begin{split}
&-\psi_{,tt}+\psi_{,xx}+\frac{2}{r}(-r_{,t}\psi_{,t}+r_{,x}\psi_{,x})\\
&=\sqrt{\frac{2}{3}}\kappa(-\phi_{,t}\psi_{,t}+\phi_{,x}\psi_{,x}),
\end{split}
\label{equation_psi}
\ee
where
\be \tilde{T}^{(\psi)}
=\tilde g^{\mu\nu} \tilde T^{(\psi)}_{\mu\nu}
=-\frac{\tilde g^{\mu\nu}\tilde \partial_{\mu}\psi\tilde\partial_{\nu}\psi}{\chi}
=\frac{1}{\chi}e^{2\sigma}(\psi_{,t}^2-\psi_{,x}^2).\label{T_psi}\ee

The $\{uu\}$ and $\{vv\}$ components of the Einstein equations yield the constraint equations
\be r_{,uu}+2\sigma_{,u}r_{,u}+4\pi r\left(\phi_{,u}^2+\frac{\psi_{,u}^2}{\chi}\right)=0, \label{constraint_eq_uu}\ee
\be r_{,vv}+2\sigma_{,v}r_{,v}+4\pi r\left(\phi_{,v}^2+\frac{\psi_{,v}^2}{\chi}\right)=0. \label{constraint_eq_vv}\ee
Via the definitions of $u=(t-x)/2$ and $v=(t+x)/2$, the constraint equations can be expressed in $(t,x)$ coordinates.
Equations $(\ref{constraint_eq_vv})-(\ref{constraint_eq_uu})$ and $(\ref{constraint_eq_vv})+(\ref{constraint_eq_uu})$ generate the constraint equations
for the $\{tx\}$ and $\{tt\}+\{xx\}$ components, respectively,
\be r_{,tx}+r_{,t}\sigma_{,x}+r_{,x}\sigma_{,t}+4\pi r\left(\phi_{,t}\phi_{,x}+\frac{\psi_{,t}\psi_{,x}}{\chi}\right)=0,
\label{constraint_eq_xt}
\ee
\be
\begin{split}
&r_{,tt}+r_{,xx}+2(r_{,t}\sigma_{,t}+r_{,x}\sigma_{,x})\\
&+4\pi r\left(\phi_{,t}^2+\phi_{,x}^2+\frac{\psi_{,t}^2+\psi_{,x}^2}{\chi}\right)=0.
\end{split}
\label{constraint_eq_xx_tt}
\ee

\subsection{Initial conditions\label{sec:set_up_ic}}
We set the initial data to be time symmetric:
\be r_{,t}=\sigma_{,t}=\phi_{,t}=\psi_{,t}=0 \hphantom{ddd} \mbox{at} \hphantom{d} t=0. \label{ic_text}\ee
Therefore, in this configuration, the constraint equation~(\ref{constraint_eq_xt}) is satisfied identically. Note that, in this configuration, the values of $r_{,t}$ and $\sigma_{,t}$ at $t=0$ are the same as those in the Reissner-Nordstr\"{o}m metric case.

We set the initial value for $\psi$ as
\be \psi(x,t)|_{t=0}=a\cdot\exp\left[-\frac{(x-x_{0})^2}{b}\right].\ee
In this paper, we give $\phi$ two sets of initial conditions:
\begin{align}
\mbox{set 1}:\hphantom{d}&\phi(x,t)|_{t=0}=\phi_{0},  \\
\nonumber\\
\mbox{set 2}:\hphantom{d}&\phi(x,t)|_{t=0}=\phi_{0}+ae^{-(x-x_{0})^2},
\end{align}
where $\phi_{0}$ is de Sitter value, defined by $V'(\phi_0)=0$.
The initial value for $\sigma$ is defined to be the same as the corresponding one in the Reissner-Nordstr\"{o}m case~(\ref{RN_metric_text}),
\be e^{-2\sigma}\big|_{t=0}
=e^{-2\sigma}\big|^{\scriptsize{\mbox{RN}}}_{t=0}
=\frac{r_{+}r_{-}}{k_{+}^2r^{2}}e^{-2k_{+}r}\left(\frac{r}{r_{-}}-1\right)^{1+\frac{k_{+}}{|k_{-}|}},
\label{sigma_RN_metric}
\ee
where $r$ is defined by Eq.~(\ref{r_RN_metric}) with $t=0$. We obtain the initial value for $r$ in charge scattering by combining Eqs.~(\ref{equation_r}) and (\ref{constraint_eq_xx_tt}),
\be
\begin{split}
r_{,xx}=
&-r_{,t}\sigma_{,t}-r_{,x}\sigma_{,x}+\frac{r_{,t}^2-r_{,x}^2}{2r}\\
&-2{\pi}r\left(\phi_{,t}^2+\phi_{,x}^2+\frac{\psi_{,t}^2+\psi_{,x}^2}{\chi}\right)\\
&+\frac{1}{2r}e^{-2\sigma}\left(1-8{\pi}r^{2}V-\frac{q^2}{r^2}\right).
\end{split}
\label{r_xx_ic}\ee

We set $r_{,x}=\sigma_{,x}=0$ at the origin $(x=0,t=0)$ as in the Reissner-Nordstr\"{o}m metric case. The definition domain for the spatial coordinate $x$ is $[-x_{b}~x_{b}]$. Then $r(x,t)|_{t=0}$ can be obtained by integrating Eq.~(\ref{r_xx_ic}) via the fourth-order Runge-Kutta method from $x=0$ to $x={\pm}x_{b}$, respectively. The initial values of $r$, $\sigma$, $f'$, and $\psi$ are shown in Fig.~\ref{fig:evolutions}.

In this paper, we employ the finite difference method. The leapfrog integration scheme is implemented, which is a three-level scheme and requires initial data on two different time levels. With the initial data at $t=0$, we compute the data at $t=\Delta t$ with a second-order Taylor series expansion as done in Ref.~\cite{Pretorius}. Take the variable $\psi$ as an example,
\be \psi|_{t={\Delta}t}=\psi|_{t=0}+\psi_{,t}|_{t=0}\Delta t+\frac{1}{2}\psi_{,tt}|_{t=0}(\Delta t)^2.\label{ic_taylor_expansion}\ee
The values of $\psi|_{t=0}$ and $\psi_{,t}|_{t=0}$ are set up as discussed above, and the value of $\psi_{,tt}|_{t=0}$ can be obtained from the equation of motion for $\psi$ (\ref{equation_psi}).

Up to this point, the initial conditions are fixed, with all the field equations being taken into account. The first-order time derivatives of $r$, $\sigma$, $\phi$, and $\psi$ at $t=0$ described by Eq.~(\ref{ic_text}) ensure that the constraint equation (\ref{constraint_eq_xt}) is satisfied. The equation for $r_{,xx}$ at $t=0$ expressed by (\ref{r_xx_ic}) implies that the constraint equation (\ref{constraint_eq_xx_tt}) is satisfied. Computations of $r$, $\sigma$, $\phi$, and $\psi$ at $t={\Delta}t$ via a second-order Taylor series expansion, as expressed by Eq.~(\ref{ic_taylor_expansion}) for the case of $\psi$, satisfy all the equations of motion.

Note that the region of $x<0$ is included in the initial conditions. This may not be physical. However, we are mainly interested in the interior dynamics of black holes, and then it is not important where the scalar field originally comes from. This setup makes it convenient for us to compare the results of charge scattering to the known solutions of the Reissner-Nordstr\"{o}m geometry. Therefore, we use this setup as a toy model.

\subsection{Boundary conditions}
The values of $r$, $\sigma$, $\phi$, and $\psi$ at the boundaries of $x={\pm}x_{b}$ are obtained via extrapolations. In fact, since we are mainly concerned with the dynamics around $x=0$, the boundary conditions will not affect the dynamics in this region, as long as $x_b$ is large enough.

\begin{figure}
\hspace{-15.0pt}
\includegraphics[width=9cm,height=7.2cm]{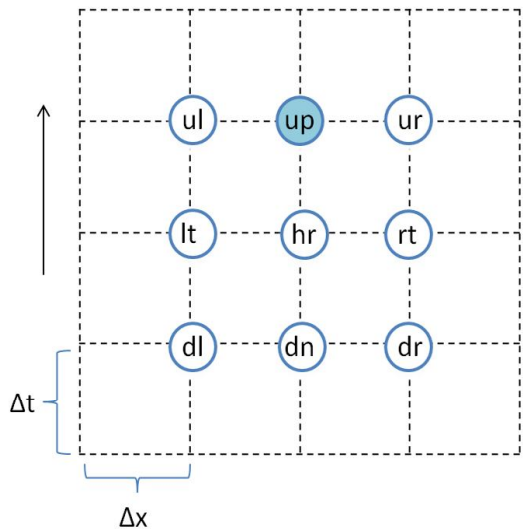}
\caption{Numerical evolution scheme.}
\label{fig:grid_scheme}
\end{figure}

\begin{figure*}[t!]
  \epsfig{file=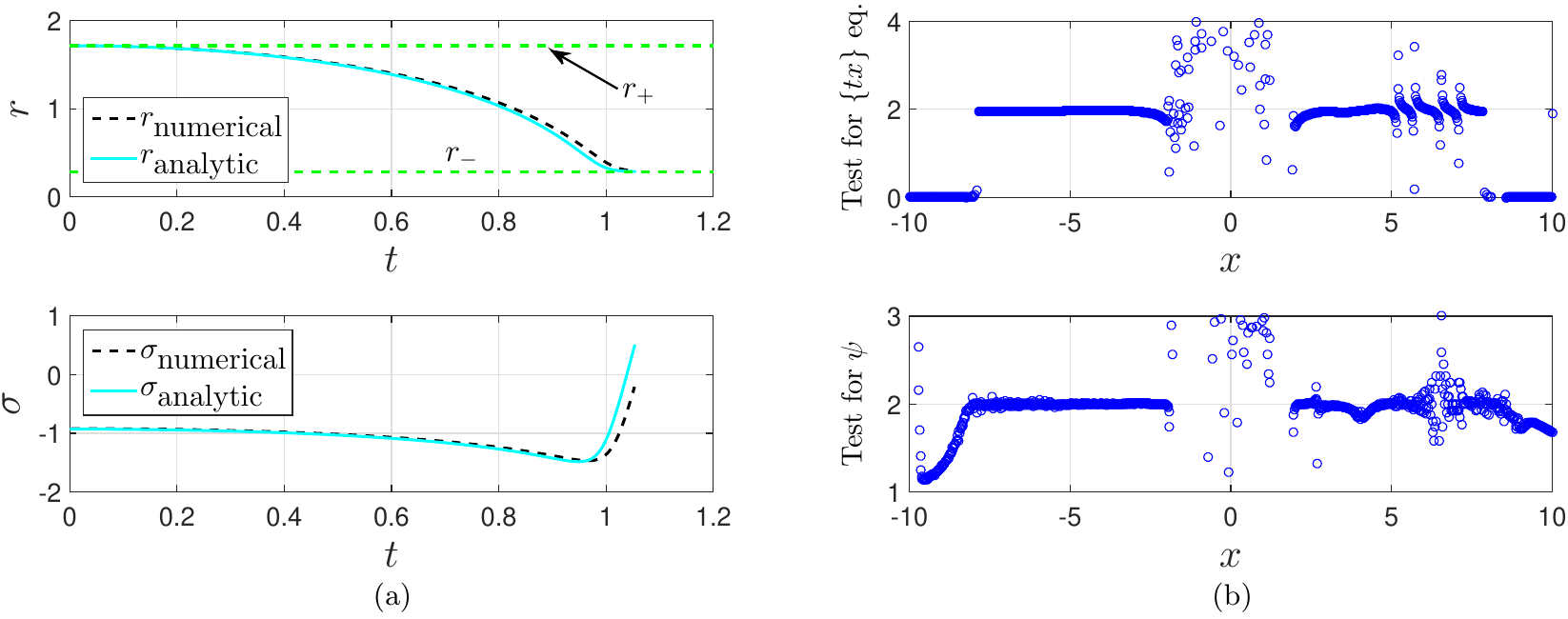, width=0.9\textwidth}
  \caption{Tests of numerical code for charge scattering. (a) Numerical vs analytic results for a Reissner-Nordstr\"{o}m black hole. $m=1$, $q=0.7$, and ${\Delta}x={\Delta}t=10^{-4}$. The slice is for $(x=3{\Delta}x,t=t)$. This is a special case of charge scattering with contributions of scalar fields being set to zero. Numerical and analytic results match well at an early stage, while at a later stage gravity and electric field become stronger, the numerical evolutions have a time delay, compared to analytic solutions. (b) Numerical tests for the $\{tx\}$ constraint equation (\ref{constraint_eq_xt}) and the evolution of $\psi$ on the slice $(x=x,t=0.65)$. They are both second-order convergent.}
  \label{fig:numerical_tests}
\end{figure*}

\subsection{Discretization scheme}
In this paper, we implement the leapfrog integration scheme, which is second-order accurate and nondissipative. We let the temporal and spatial grid spacings be equal, ${\Delta}t={\Delta}x$.

The equations of motion for $\phi$ (\ref{equation_phi}) and $\psi$ (\ref{equation_psi}) are coupled. Newton's iteration method is employed to solve this problem~\cite{Pretorius}. With the illustration of Fig.~\ref{fig:grid_scheme}, the initial conditions provide the data at the levels of {\lq\lq}down\rq\rq~and {\lq\lq}here\rq\rq.~We need to obtain the data on the level of {\lq\lq}up\rq\rq.~We take the values at the level of {\lq\lq}here\rq\rq~to be the initial guess for the level of {\lq\lq}up\rq\rq.~Then, taking $\psi$ as an example, we update the values at the level of {\lq\lq}up\rq\rq~using the following iteration:
\be \psi^{\text{new}}_{\text{up}}=\psi_{\text{up}}-\frac{G(\psi_{\text{up}})}{J(\psi_{\text{up}})},\nonumber\ee
where ${G}(\psi_{\text{up}})$ is the residual of the differential equation for the function $\psi_{\text{up}}$, and $J(\psi_{\text{up}})$ is the Jacobian defined by
\be J(\psi_{\text{up}})=\frac{\partial G(\psi_{\text{up}})}{\partial\psi_{\text{up}}}.\nonumber\ee
We do the iterations for the coupled equations one by one, and run the iteration loops until the desired accuracies are achieved.

\subsection{Tests of numerical code}
To make sure that the numerical results are trustworthy, one needs to test the numerical code. We compare the numerical results obtained by the code with the analytic ones for the dynamics in a Reissner-Nordstr\"{o}m geometry, and examine the convergence of the constraint equations and dynamical equations in charge scattering.

The dynamics in a Reissner-Nordstr\"{o}m geometry is a special case for charge scattering, in which the contributions from the scalar fields are set to zero. This special case has analytic solutions expressed by Eqs.~(\ref{RN_metric_text}) and (\ref{r_RN_metric}). Therefore, we can test our code by comparing the numerical and analytic results in the Reissner-Nordstr\"{o}m geometry. Set $m=1$, $q=0.7$, and ${\Delta}x={\Delta}t=10^{-4}$. We plot the evolutions of $r$ and $\sigma$ along the slice $(x=3{\times}10^{-4},t=t)$ in Fig.~\ref{fig:numerical_tests}(a). As shown in Fig.~\ref{fig:numerical_tests}(a), numerical and analytic results match well at an early stage; while at a late stage where gravity and electric field become strong, the numerical evolutions have a time delay compared to the analytic solutions.

When the numerical results are obtained, we substitute the numerical results into the discretized equations of motion and the constraint equations, and find that they are well satisfied. See Figs.~\ref{fig:eom_spacelike} and \ref{fig:eom_null}, for example. Moreover, the convergence of the constraint equations (\ref{constraint_eq_xt}) and (\ref{constraint_eq_xx_tt}) is examined. We assume one constraint equation is $n$th-order convergent: residual=$\mathcal{O}(h^{n})$, where $h$ is the grid size. Therefore, the convergence rate of the discretized constraint equations can be obtained from the ratio between residuals with two different step sizes,
\be n=\log_{2}\left[\frac{\mathcal{O}(h^{n})}{\mathcal{O}\left(\left(\frac{h}{2}\right)^{n}\right)}\right].\label{accuracy_test}\ee
Our numerical results show that both of the constraint equations are about second-order convergent. As a representative, we plot the results for the $\{tx\}$ constraint equation (\ref{constraint_eq_xt}) in Fig.~\ref{fig:numerical_tests}(b) for the slice $(x=x,t=0.65)$.

Convergence tests via simulations with different grid sizes are also implemented~\cite{Sorkin,Golod}. If the numerical solution converges, the relation between the numerical solution and the real one can be expressed by
\be F_{\mbox{real}}=F^{h}+\mathcal{O}(h^{n}),\nonumber\ee
where $F^{h}$ is the numerical solution. Then, for step sizes equal to $h/2$ and $h/4$, we have
\be F_{\mbox{real}}=F^{\frac{h}{2}}+\mathcal{O}\left[\left(\frac{h}{2}\right)^{n}\right],\nonumber\ee
\be F_{\mbox{real}}=F^{\frac{h}{4}}+\mathcal{O}\left[\left(\frac{h}{4}\right)^{n}\right].\nonumber\ee
Defining $c_1\equiv F^{h}-F^{\frac{h}{2}}$ and $c_2\equiv F^{\frac{h}{2}}-F^{\frac{h}{4}}$, one obtains the convergence rate
\be n=\log_{2}\left(\frac{c_1}{c_2}\right). \label{convergence_test}\ee
The convergence tests for $\eta\equiv r^{2}$, $\sigma$, $\phi$, and $\psi$ are investigated. They are all second-order convergent. As a representative, the results for $\psi$ are plotted in Fig.~\ref{fig:numerical_tests}(b) for the slice $(x=x,t=0.65)$. The values of the parameters in charge scattering in this section are described at the beginning of Sec.~\ref{sec:results}. We use the spatial range of $x\in[-10~10]$ and the grid spacings of $h={\Delta}x={\Delta}t=0.02$.

\section{Neutral scalar collapse\label{sec:neutral_collapse}}
In this section, we consider neutral collapse in flat geometry in $f(R)$ gravity and discuss the mass inflation which happens in the vicinity of the central singularity of the formed black hole.

\subsection{Numerical setup}
The numerical setup in neutral scalar collapse in $f(R)$ gravity is discussed in Ref.~\cite{Guo_1312}. The dynamical equations for $r$, $\eta$, $\sigma$, $\phi$, and $\psi$ can be obtained by setting the terms related to the electric field in the corresponding equations in Sec.~\ref{sec:field_eqs} to zero:
\be r(-r_{,tt}+r_{,xx})-r_{,t}^2+r_{,x}^2 = e^{-2\sigma}(1-8{\pi}r^{2}V),\label{equation_r_collapse}\ee
\be -\eta_{,tt}+\eta_{,xx}=2e^{-2\sigma}(1-8{\pi}r^{2}V),\label{equation_eta_collapse}\ee
\be
\begin{split}
&-\sigma_{,tt}+\sigma_{,xx}+\frac{r_{,tt}-r_{,xx}}{r}\\
&+4\pi\left(\phi_{,t}^2-\phi_{,x}^2 + \frac{\psi_{,t}^2-\psi_{,x}^2}{\chi}-2e^{-2\sigma}V\right)=0,
\end{split}
\label{equation_sigma_collapse}
\ee
\be
\begin{split}
&-\phi_{,tt}+\phi_{,xx}+\frac{2}{r}(-r_{,t}\phi_{,t}+r_{,x}\phi_{,x})\\
&=e^{-2\sigma}\left[V'(\phi)+\frac{1}{\sqrt{6}}\kappa\tilde{T}^{(\psi)}\right],
\end{split}
\label{equation_phi_collapse}
\ee
\be
\begin{split}
&-\psi_{,tt}+\psi_{,xx}+\frac{2}{r}(-r_{,t}\psi_{,t}+r_{,x}\psi_{,x})\\
&=\sqrt{\frac{2}{3}}\kappa(-\phi_{,t}\psi_{,t}+\phi_{,x}\psi_{,x}),
\end{split}
\label{equation_psi_collapse}
\ee
where $\tilde{T}^{(\psi)}=e^{2\sigma}(\psi_{,t}^2-\psi_{,x}^2)/\chi$.

In the equation of motion for $\sigma$ (\ref{equation_sigma_collapse}), the term $(r_{,tt}-r_{,xx})/r$ can create big errors near the center $x=r=0$. To avoid such a problem, we use the constraint equation (\ref{constraint_eq_uu}) alternatively~\cite{Frolov_2004}. Defining a new variable $g$
\be g\equiv-2\sigma-\ln(-r_{,u}), \label{g_definition}\ee
one can rewrite Eq.~(\ref{constraint_eq_uu}) as the equation of motion for $g$,
\be g_{,u}=4{\pi}\cdot\frac{r}{r_{,u}}\cdot\left(\phi_{,u}^2+\frac{\psi_{,u}^2}{\chi}\right). \label{equation_g}\ee
In the numerical integration, once the value of $r$ at the advanced level is obtained, the value of $\sigma$ at the current level will be computed using Eq.~(\ref{g_definition}).

We set the initial data as
\be r_{,tt}=r_{,t}=\sigma_{,t}=\phi_{,t}=\psi_{,t}=0 \hphantom{ddd} \mbox{at} \hphantom{d} t=0.\ee
The initial values for $\chi[\equiv\exp(\sqrt{2/3}\kappa\phi)]$ and $\psi(r)$ are defined as
\begin{align}
\chi(r)|_{t=0}&=a\cdot[1-\tanh(r-r_1)^2]+\chi_{0},\\
\nonumber\\
\psi(r)|_{t=0}&=b\cdot\tanh(r-r_2)^2,
\end{align}
with $a=0.2$, $b=0.1$, $r_{1}=r_{2}=4$, and $U'(\chi_{0})=0$. The parameters for the Hu-Sawicki model~(\ref{f_R_Hu_Sawicki}) are set as $D=1.2$ and $R_{0}=10^{-6}$.

The local Misner-Sharp mass $m$ is defined as~\cite{Misner}
\be g^{\mu\nu}r_{,\mu}r_{,\nu}=e^{2\sigma}(-r_{,t}^2+r_{,x}^2){\equiv}1-\frac{2m}{r}.\label{mass_Misner}\ee
(See Ref.~\cite{Hayward} for details on various properties of the Misner-Sharp mass/energy in spherical symmetry.) Then on the initial slice $(x=x,t=0)$, the equations for $r$, $m$, and $g$ are~\cite{Frolov_2004,Guo_1312}
\begin{align}
r_{,x}&=\left(1-\frac{2m}{r}\right)e^g,\label{r_ic_collapse}\\
\nonumber\\
m_{,r}&=4{\pi}r^2\left[V+\frac{1}{2}\left(1-\frac{2m}{r}\right)\left(\phi_{,r}^2+\frac{\psi_{,r}^2}{\chi}\right)\right],\label{m_ic_collapse}\\
\nonumber\\
g_{,r}&=4{\pi}r\left(\phi_{,r}^2 + \frac{\psi_{,r}^2}{\chi}\right).\label{g_ic_collapse}
\end{align}
Set $r=m=g=0$ at the origin $(x=0,t=0)$. Then the values of $r$, $m$, and $g$ on the initial slice $(x=x,t=0)$ can be obtained by integrating Eqs.~(\ref{r_ic_collapse})-(\ref{g_ic_collapse}) from the center $x=0$ to the outer boundary $x=x_{b}$ via the fourth-order Runge-Kutta method. The values of $r$, $\sigma$, $\phi$, and $\psi$ at $t={\Delta}t$ can be obtained with a second-order Taylor series expansion, as discussed in Sec.~\ref{sec:set_up_ic}. The value of $g$ at $t={\Delta}t$ can be obtained using Eq.~(\ref{g_definition}).

The range for the spatial coordinate is defined to be $x\in[0 \mbox{ } 20]$. At the inner boundary $x=0$, $r$ is always set to zero. The terms $2(-r_{,t}\phi_{,t}+r_{,x}\phi_{,x})/r$ in Eq.~(\ref{equation_phi_collapse}) and $2(-r_{,t}\psi_{,t}+r_{,x}\psi_{,x})/r$ in Eq.~(\ref{equation_psi_collapse}) need to be regular at $x=r=0$. Since $r$ is always set to zero at the center, so is $r_{,t}$. Then we enforce $\phi$ and $\psi$ to
satisfy $\phi_{,x}=\psi_{,x}=0$ at $x=0$. The value of $g$ at $x=0$ is obtained via extrapolation. We set up the outer boundary conditions at $x=20$ via extrapolation. The temporal and the spatial grid spacings are ${\Delta}t={\Delta}x=0.005$.

The numerical code is second-order convergent and is the one developed in Ref.~\cite{Guo_1312}.

\begin{figure*}[t!]
  \epsfig{file=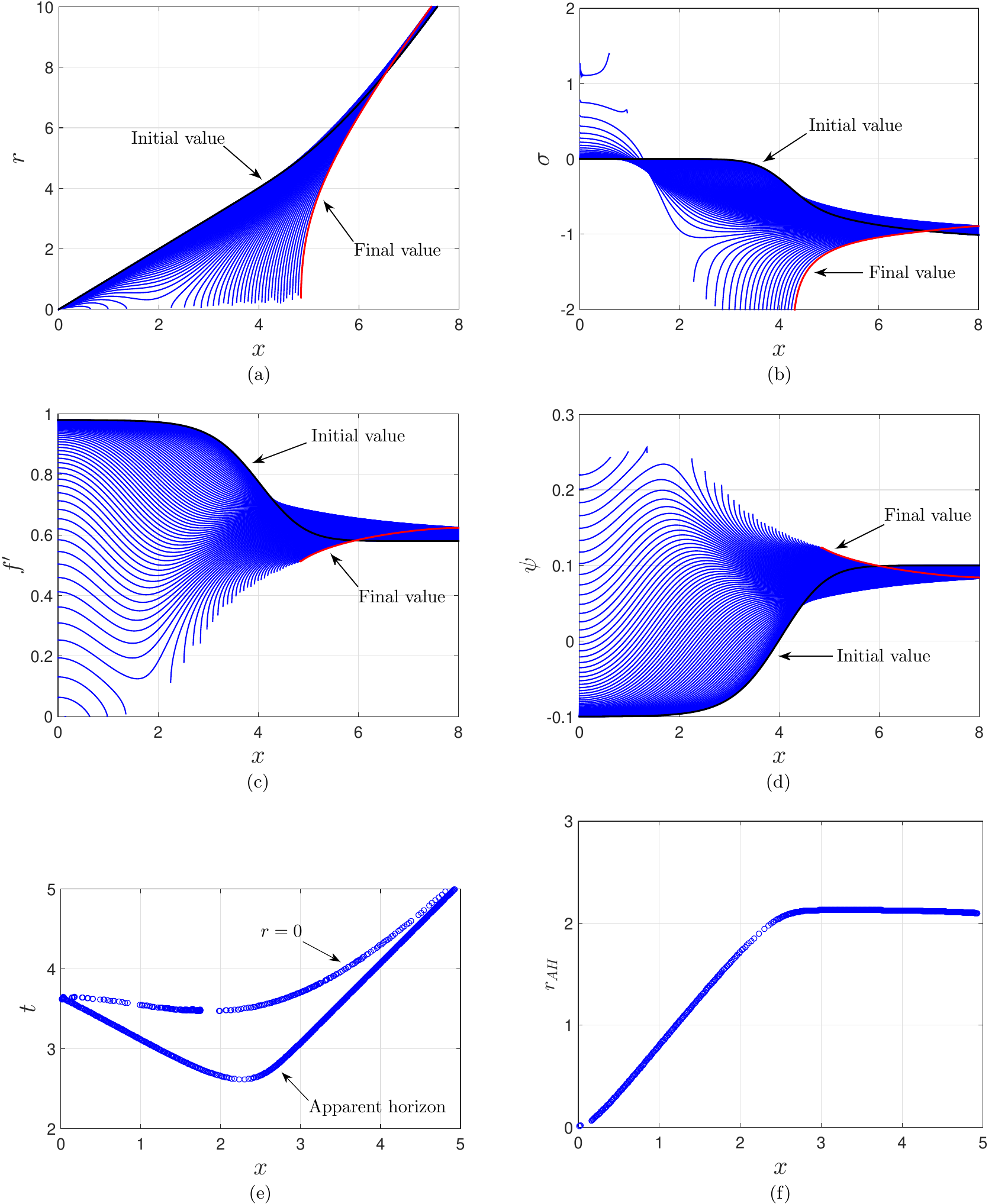, width=0.9\textwidth}
  \caption{Evolutions in neutral collapse for the Hu-Sawicki model~(\ref{f_R_Hu_Sawicki}), $f(R)=R-DR_{0}R/(R+R_{0})$. In (a)-(d), the time interval between two consecutive slices is $10{\Delta}t=0.05$. (e) and (f) are for the apparent horizon and the singularity curve of the formed black hole.}
  \label{fig:neutral_collapse_evolutions}
\end{figure*}

\begin{figure*}[t!]
  \epsfig{file=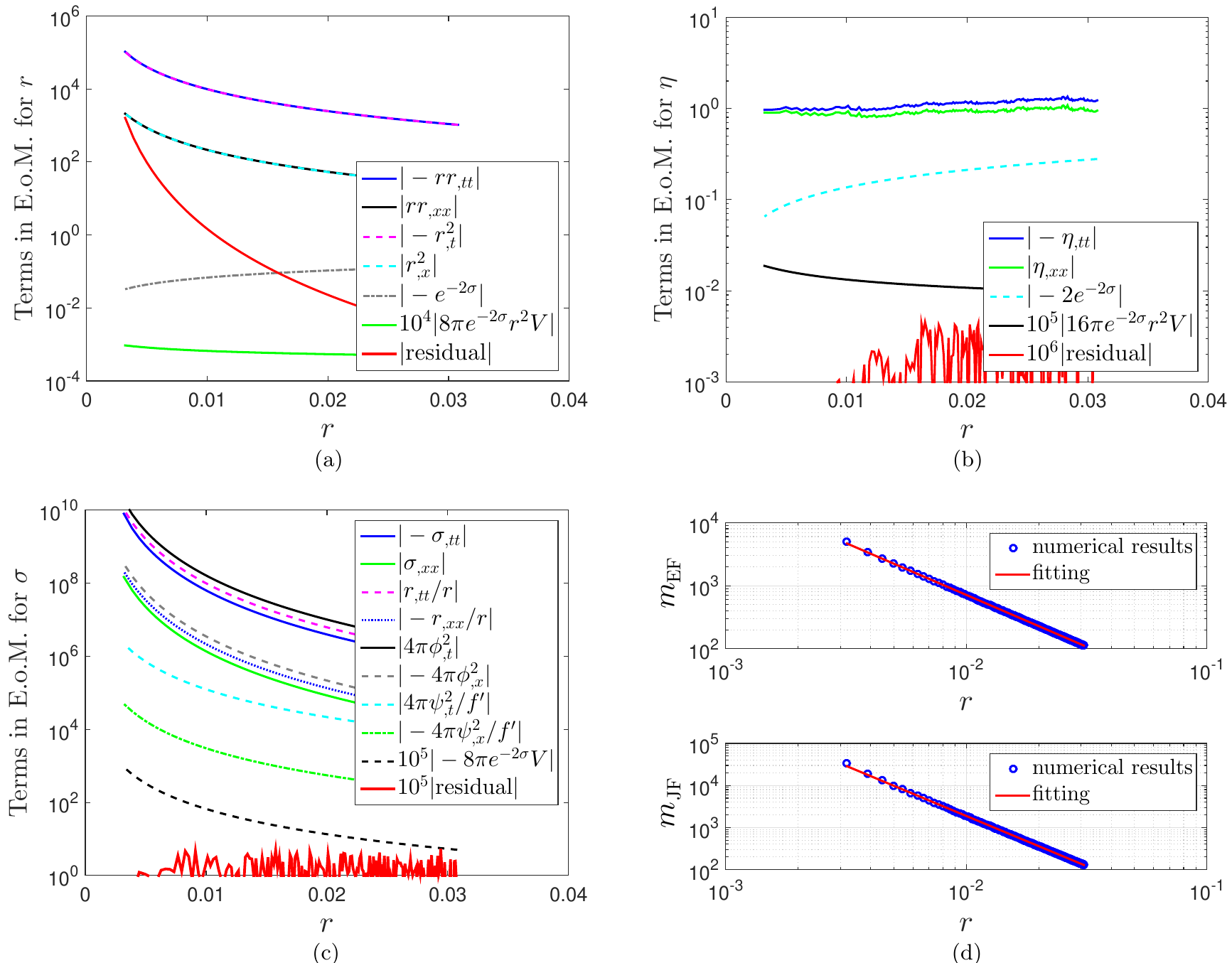, width=0.9\textwidth}
  \caption{(color online). Dynamics on the slice $(x=1,t=t)$ in neutral collapse for the Hu-Sawicki model~(\ref{f_R_Hu_Sawicki}).
  (a)-(c): dynamical equations for $r$, $\eta$, and $\sigma$.
  (b) Near the central region, due to accumulation, the scalar field $\phi$ is strong. $\sigma$ is positive. As a result, in the equation of motion
  for $\eta$~(\ref{equation_eta_collapse}), the terms $2e^{-2\sigma}$ and $16{\pi}e^{-2\sigma}r^{2}V$ are negligible. The equation is reduced to $\eta_{,tt}\approx\eta_{,xx}$.
  (d) $\ln(m_{\scriptsize{\mbox{EF}}})=a{\ln}r+b$, $a=-1.6438\pm0.0008$, $b=-0.998\pm0.003$.
  $\ln(m_{\scriptsize{\mbox{JF}}})=a{\ln}r+b$, $a=-2.385\pm0.002$, $b=-3.443\pm0.009$.}
  \label{fig:neutral_collapse_x_equal_1}
\end{figure*}

\begin{figure*}[t!]
  \epsfig{file=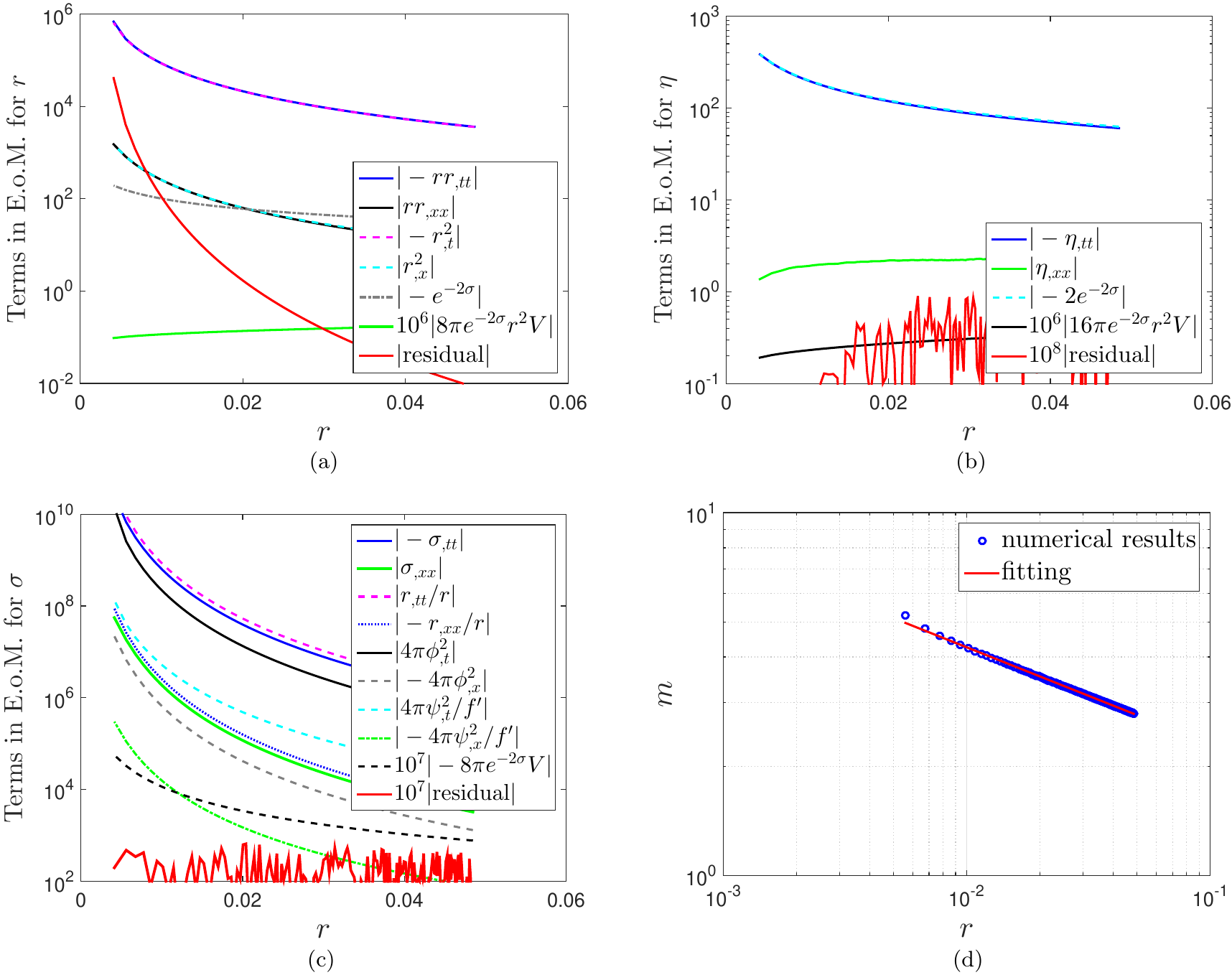, width=0.9\textwidth}
  \caption{(color online). Dynamics on the slice $(x=2,t=t)$ in neutral collapse for the Hu-Sawicki model.
  (a)-(c): dynamical equations for $r$, $\eta$, and $\sigma$.
  (b) At large-$x$ regions, the scalar field $\phi$ is weak, and $\sigma$ is negative. As a result, in the equation of motion for $\eta$~(\ref{equation_eta_collapse}), the term $2e^{-2\sigma}$ is important. The equation is reduced to $\eta_{,tt}\approx-2e^{-2\sigma}$.
  (d) $\ln m=a{\ln}r+b$, $a=-0.2673\pm0.0008$, $b=0.217\pm0.003$.}
  \label{fig:neutral_collapse_x_equal_2}
\end{figure*}

\subsection{Black hole formation}
The evolutions of $r$, $\sigma$, $\phi$, and $\psi$ are plotted in Figs.~\ref{fig:neutral_collapse_evolutions}(a)-\ref{fig:neutral_collapse_evolutions}(d), respectively. On the apparent horizon of a black hole, the expansion of the outgoing null geodesics orthogonal to the apparent horizon is zero~\cite{Baumgarte}. Then in double-null coordinates, on the apparent horizon, there is~\cite{Csizmadia}
\be g^{\mu\nu}r_{,\mu}r_{,\nu}=e^{2\sigma}(-r_{,t}^2+r_{,x}^2)=1-\frac{2m}{r}=0.\ee
Using this property, we locate the apparent horizon and plot it in Figs.~\ref{fig:neutral_collapse_evolutions}(e) and \ref{fig:neutral_collapse_evolutions}(f). As shown in Fig.~\ref{fig:neutral_collapse_evolutions}(a), the central singularity is also approached in the collapse. Therefore, a black hole is formed.

\subsection{Asymptotic dynamics in the vicinity of the central singularity of the formed black hole\label{sec:asymptotics_collapse}}
We focus on the dynamics in the vicinity of the central singularity of the formed black hole. We will discuss that, in the vicinity of the singularity, due to the backreaction of the scalar fields on the geometry, the Misner-Sharp mass diverges. In other words, in addition to the inner horizons of Reissner-Nordstr\"{o}m and Kerr black holes, mass inflation also happens in the vicinity of the central singularity of a Schwarzschild black hole.

In the vicinity of the central singularity, the field equation can be reduced to the following forms~\cite{Guo_1312}:
\begin{align}
rr_{,tt}&\approx-r_{,t}^2,\label{equation_r_asymptotic}\\
\nonumber\\
\sigma_{,tt}&\approx\frac{r_{,tt}}{r}+4\pi\phi_{,t}^2,\label{equation_sigma_asymptotic}\\
\nonumber\\
\phi_{,tt}&\approx-\frac{2}{r}r_{,t}\phi_{,t},\label{equation_phi_asymptotic}\\
\nonumber\\
\psi_{,tt}&\approx-\frac{2}{r}r_{,t}\psi_{,t}+\sqrt{\frac{2}{3}}\kappa\phi_{,t}\psi_{,t}.\label{equation_psi_asymptotic}
\end{align}
The asymptotic solutions to Eqs.~(\ref{equation_r_asymptotic})-(\ref{equation_phi_asymptotic}) are~\cite{Guo_1312}
\begin{align}
r&\approx A\xi^{\beta},\label{r_asymptotic_collapse}\\
\nonumber\\
\sigma&\approx B\ln\xi+\sigma_{0}\approx[\beta(1-\beta)-4{\pi}C^2]\ln\xi+\sigma_{0},\label{sigma_asymptotic_collapse}\\
\nonumber\\
\phi&\approx C\ln\xi.\label{phi_asymptotic_collapse}
\end{align}
The variable $\xi$ is defined as $\xi{\equiv}t_0-t$, where $t_0$ is the coordinate time on the singularity curve.

As implied in Eq.~(\ref{equation_psi_asymptotic}), $\psi$ is suppressed (magnified) by $\phi$ when $\phi_{,t}$ is negative (positive). Due to the complex competition between gravity and dark energy, an approximate analytic expression for $\psi$ is not obtained. Substituting Eq.~(\ref{r_asymptotic_collapse}) into Eq.~(\ref{equation_r_asymptotic}) yields
\be (1-\beta)\xi^{2(\beta-1)}\approx\beta\xi^{2(\beta-1)}.\nonumber\ee
Then we have
\be \beta\approx\frac{1}{2}.\label{beta_asymptotic_collapse}\ee

Using Eqs.~(\ref{rescale_f_prime}) and (\ref{phi_asymptotic_collapse}), there is
\be \chi\approx\xi^{\sqrt{\frac{2}{3}}{\kappa}C}.\ee
So as shown in Fig.~\ref{fig:neutral_collapse_evolutions}(c), when $C$ is positive, $\chi$ also approaches zero as $\xi$ and $r$ approach zero. Then the transformation between the Jordan and Einstein frames, $g^{\scriptsize{\mbox{(EF)}}}_{\mu\nu}=\chi{\cdot}g^{\scriptsize{\mbox{(JF)}}}_{\mu\nu}$, breaks down. Actually this is not a serious problem, because numerical simulation stops anyway when the central singularity is approached. Moreover, in this paper, we discuss the dynamics in the vicinity of the central singularity rather than on the central singularity.

\subsection{Mass inflation}
In the vicinity of the singularity, in the equation of motion for $\sigma$~(\ref{equation_sigma_asymptotic}), because of the contribution from $\phi$, $\sigma(x,t)$ is greater than the corresponding value in the Schwarzschild black hole case. This makes the mass function divergent near the singularity as will be discussed below.

Near the singularity, using Eqs.~(\ref{mass_Misner}), (\ref{r_asymptotic_collapse})-(\ref{beta_asymptotic_collapse}), the mass function can be written as
\be
\begin{split}
m&=\frac{r}{2}[1+e^{2\sigma}(r_{,t}^2-r_{,x}^2)]\\
&\approx\left[\frac{1}{8}(1-K^{2})A^{3}e^{2\sigma_0}\right]\xi^{-8{\pi}C^2}\\
&\approx\left[\frac{1}{8}(1-K^{2})A^{3+16{\pi}C^2}e^{2\sigma_0}\right]r^{-16{\pi}C^2}.
\end{split}
\label{mass_analytic_collapse}
\ee
where $K\equiv|r_{,x}/r_{,t}|$. In the Schwarzschild black hole case, $C=0$. The mass function is always constant and is equal to the black hole mass. In neutral collapse, the parameter $\beta$ does not change much and remains about $1/2$. However, the parameter $C$ is not zero. Then the metric quantity $\sigma$ is modified. [See Eq.~(\ref{sigma_asymptotic_collapse}).] As a result, the delicate balance between $r$ and $e^{2\sigma}(-r_{,t}^2+r_{,x}^2)$ is broken. Consequently, as implied in Eq.~(\ref{mass_analytic_collapse}), near the singularity, the mass function diverges: mass inflation occurs.

During the collapse, before the black hole is formed, the energy of the scalar fields accumulates in the central region. As a result, the scalar fields near $x=0$ are stronger than those at large-$x$ regions. Next we discuss three consequences. As a support, we examine the dynamics in the vicinity of the singularity via mesh refinement that was implemented in Refs.~\cite{Guo_1312,Garfinkel}, and plot two sample sets of results on the slices $(x=1,t=t)$ and
$(x=2,t=t)$ in Figs.~\ref{fig:neutral_collapse_x_equal_1} and \ref{fig:neutral_collapse_x_equal_2}, respectively.
\begin{enumerate}[(i)]
  \item Values of $\sigma$. Due to the backreaction of the scalar fields on the geometry, $\sigma$ in small-$x$ regions is greater than in large-$x$ regions. In fact, $\sigma$ is positive in small-$x$ regions, while negative in large-$x$ regions. [See Fig.~\ref{fig:neutral_collapse_evolutions}(b).] Our numerical results of the parameter $C$ in $\phi{\approx}C\ln\xi$ are $C\approx0.18$ at $x=1$ and $C\approx0.07$ at $x=2$.
      Then we have $4{\pi}C^2\approx0.41>1/4$ and $\sigma>0$ at $x=1$, while $4{\pi}C^2\approx0.06<1/4$ and $\sigma<0$ at $x=2$.
      [See Eq.~(\ref{sigma_asymptotic_collapse}).]
  \item Equation of motion for $\eta$~(\ref{equation_eta_collapse}). For positive $\sigma$, the terms $2e^{-2\sigma}$ and $16{\pi}e^{-2\sigma}r^{2}V$ in Eq.~(\ref{equation_eta_collapse}) are negligible, compared to the other two. Then Eq.~(\ref{equation_eta_collapse}) is reduced to $\eta_{,tt}\approx\eta_{,xx}$. [See Fig.~\ref{fig:neutral_collapse_x_equal_1}(b).] However, for negative $\sigma$, the term $2e^{-2\sigma}$ is important, and Eq.~(\ref{equation_eta_collapse}) is reduced to $\eta_{,tt}\approx-2e^{-2\sigma}$. [See Fig.~\ref{fig:neutral_collapse_x_equal_2}(b).]
  \item Growth of the mass function. As implied in Eq.~(\ref{mass_analytic_collapse}), the mass function grows faster in the strong scalar field case than in the weak one. [See Figs.~\ref{fig:neutral_collapse_x_equal_1}(d) and \ref{fig:neutral_collapse_x_equal_2}(d).]
\end{enumerate}

Since $f(R)$ gravity is defined in the Jordan frame, it is interesting to examine the mass function in the Jordan frame. A generalized Misner-Sharp energy in $f(R)$ gravity in the Jordan frame was defined in Ref.~\cite{Cai_0910}. However, due to the complexity of some integrals, an explicit quasi-local form is usually not available with the exceptions of a Friedmann-–Robertson–-Walker universe and static spherically symmetric solutions with constant scalar curvature.
In this paper, for simplicity, we remain to use the conventional format of the Misner-Sharp function. Considering the transformation that we used, $g^{\scriptsize{\mbox{(EF)}}}_{\mu\nu}=\chi{\cdot}g^{\scriptsize{\mbox{(JF)}}}_{\mu\nu}$, there are $e^{2\sigma}|_{\scriptsize{\mbox{JF}}}=\chi{\cdot}e^{2\sigma}|_{\scriptsize{\mbox{EF}}}$
and $r_{\scriptsize{\mbox{JF}}}=\chi^{-1/2}{\cdot}r_{\scriptsize{\mbox{EF}}}$. Considering that near the central singularity~\cite{Guo_1312,Guo_1507}
\be K_{\scriptsize{\mbox{JF}}}\equiv\Big|\frac{r_{,x}}{r_{,t}}\Big|_{\scriptsize{\mbox{JF}}}\approx\Big|\frac{\phi_{,x}}{\phi_{,t}}\Big|,\ee
one can obtain that $K_{\scriptsize{\mbox{JF}}}{\approx}K_{\scriptsize{\mbox{EF}}}$. Then the mass function in the Jordan frame can be written as
\be
\begin{split}
m_{\scriptsize{\mbox{JF}}}&=\frac{r_{\scriptsize{\mbox{JF}}}}{2}\left[1+e^{2\sigma_{\scriptsize{\mbox{JF}}}}(r_{,t}^2-r_{,x}^2)\big|_{\scriptsize{\mbox{JF}}}\right]\\
&\approx\left[\frac{1}{8}\left(1-\sqrt{\frac{2}{3}}{\kappa}C\right)^{2}(1-K^{2})A^{3}e^{2\sigma_0}\right]\xi^{-8{\pi}C^2-\sqrt{\frac{1}{6}}{\kappa}C}\\
&\approx\left[\frac{1}{8}\left(1-\sqrt{\frac{2}{3}}{\kappa}C\right)^{2}(1-K^{2})A^{3+16{\pi}C^2+\sqrt{\frac{2}{3}}{\kappa}C}e^{2\sigma_0}\right]\\
& \hphantom{ddd} {\cdot}r^{-16{\pi}C^2-\sqrt{\frac{2}{3}}{\kappa}C},
\end{split}
\label{mass_analytic_collapse_JF}
\ee
In the case of $C>0$, due to the factor $r^{-\sqrt{2/3}{\kappa}C}$, $m_{\scriptsize{\mbox{JF}}}$ is greater than $m_{\scriptsize{\mbox{EF}}}$.
The $m_{\scriptsize{\mbox{JF}}}$ along the slice $(x=1,t=t)$ is plotted in Fig.~\ref{fig:neutral_collapse_x_equal_1}(d).

For stationary black holes (e.g., Schwarzschild and Reissner-Nordstr\"{o}m), the Misner-Sharp mass function is always equal to the black hole mass. In spherical symmetry, at spatial infinity, the mass function describes the total energy/mass of an asymptotically flat spacetime~\cite{Hayward}. In gravitational collapse case, it means the total mass of the collapsing system.

In the vicinities of the central singularity of a Schwarzschild black hole and the inner horizon of a Reissner-Nordstr\"{o}m or Kerr black hole, the dynamics and some quantities are \emph{local}. The mass function is just a parameter which varies at each point, not giving \emph{global} information on the black hole mass.

\section{Neutral scalar scattering\label{sec:neutral_scattering}}
In this section, we consider neutral scattering, in which a neutral scalar field collapses in a Schwarzschild geometry in $f(R)$ gravity. The numerical formalism is a simpler version of the one in charge scattering that has been constructed in Sec.~\ref{sec:set_up}, and it can be obtained by removing the electric terms in the field equations presented in Sec.~\ref{sec:field_eqs} and replacing the Reissner-Nordstr\"{o}m geometry with a Schwarzschild one.

\subsection{A dark energy $f(R)$ singularity problem}
In neutral scattering, for usual initial conditions of the scalar degree of freedom $f'$, $f'$ asymptotes to zero as the central singularity is approached, which is similar to what happens in the neutral scalar collapse discussed in Sec.~\ref{sec:neutral_collapse}. Details are skipped here. On the other hand, when the initial velocity or acceleration of $f'$ is large enough, $f'$ can become $1$ before the central singularity is approached. As implied in Eq.~(\ref{f_R_Hu_Sawicki_general}), for dark energy $f(R)$ models, this means that the Ricci scalar becomes infinite, and the simulation breaks down. In this section, we will focus on this singular circumstance.

The parameters are set as follows:
\begin{enumerate}[(i)]
  \item Schwarzschild geometry: $m=1$.
  \item Physical scalar field: $\psi(x,t)|_{\scriptsize{t=0}}=a\cdot\exp\left[-(x-x_{0})^2/b\right]$, $a=0.08$, $b=1$, and $x_{0}=4$.
  \item $f(R)$ model: $f(R)=R-DR_{0}R/(R+R_{0})$, $D=1.2$, and $R_{0}=10^{-5}$.
  \item Scalar degree of freedom: $\phi(x,t)|_{\scriptsize{t=0}}=a+b\cdot\exp\left[-(x-x_{0})^2/c\right]$, $a=-0.02$, $b=-0.1$, $c=1$, and $x_{0}=-2$.
  \item Grid. Spatial range: $x\in[-10~10]$. Grid spacings: ${\Delta}x={\Delta}t=0.005$.
\end{enumerate}

\begin{figure*}[t!]
  \epsfig{file=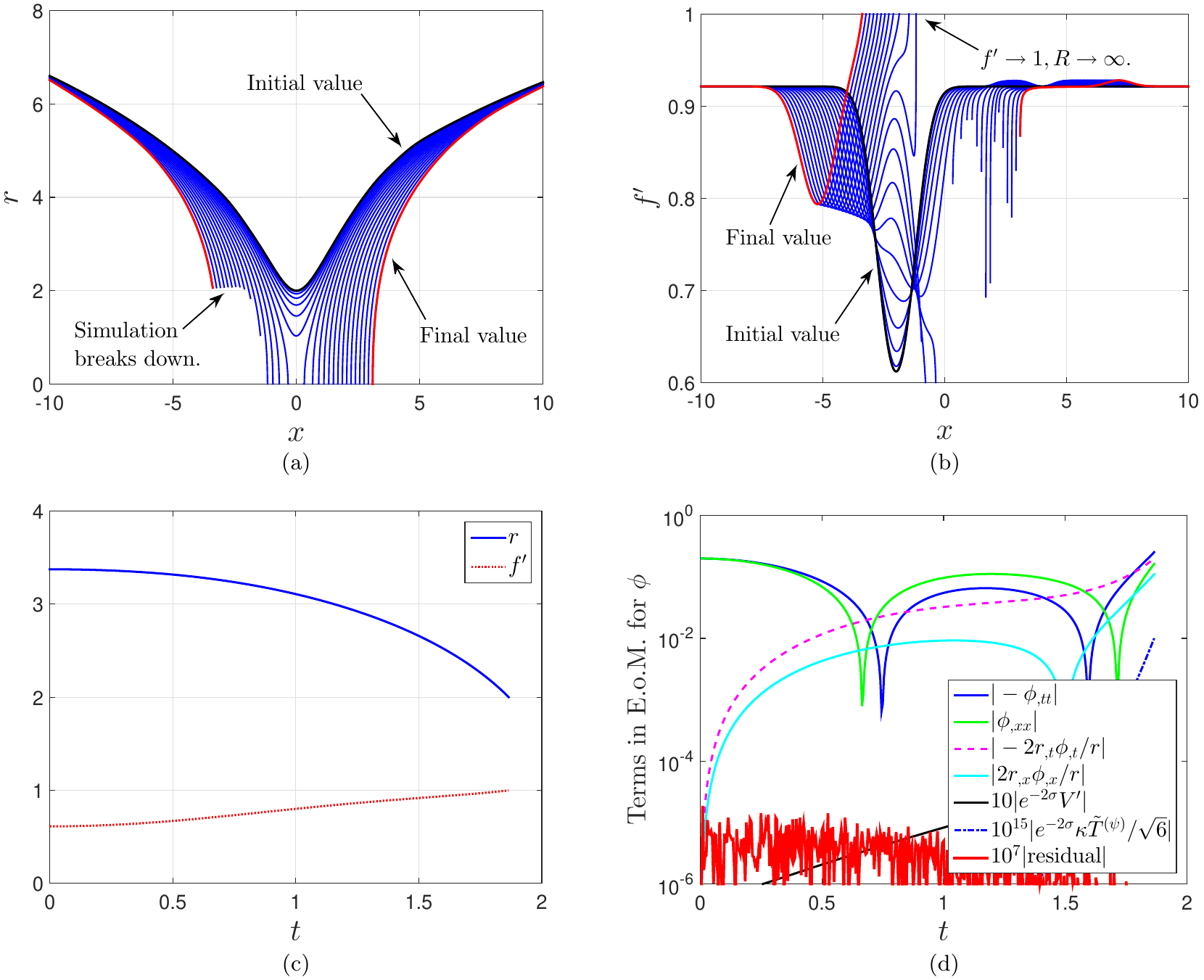, width=0.9\textwidth}
  \caption{(color online). Results for neutral scalar collapse in a Schwarzschild geometry for the Hu-Sawicki model.
  (a) and (b): evolutions of $r$ and $f'$. The time interval between two consecutive slices is $30{\Delta}t=0.15$.
  (c) evolutions of $r$ and $f'$ on the slice $(x=-2,t=t)$.
  (d) dynamical equation for $\phi$ on the slice $(x=-2,t=t)$. $\phi$ is mainly accelerated by the spatial derivative $\phi_{,xx}$ and the geometrical term $-2r_{,t}\phi_{,t}/r$. Eventually, $\phi$ and $f'$ approach $0$ and $1$, respectively. Then the Ricci scalar becomes singularity, and the simulation stops.}
  \label{fig:neutral_scattering_singularity}
\end{figure*}

The results in this circumstance are plotted in Fig.~\ref{fig:neutral_scattering_singularity}. We plot the dynamics of $\phi$ on the slice $(x=-2,t=t)$ in Fig.~\ref{fig:neutral_scattering_singularity}(d), from which one can see that $\phi$ is mainly accelerated by the spatial derivative $\phi_{,xx}$ and the geometrical term $-2r_{,t}\phi_{,t}/r$. Eventually, $\phi$ and $f'$ approach $0$ and $1$, respectively. Then the Ricci scalar becomes singularity, and the simulation stops, as shown in Fig.~\ref{fig:neutral_scattering_singularity}(a).

\subsection{Avoidance of the singularity problem}
In fact, it has been argued that such a singularity problem can also be caused in cosmology and compact stars~\cite{Frolov_2008}. This problem can be avoided by adding an $R^2$ term to the
dark energy $f(R)$ model~\cite{Starobinsky_1980,Vilenkin_1985,Nojiri_0804,Bamba_0807,Capozziello_0903,Appleby_0909,Bamba_1012,Bamba_1101}. In this paper, we add the $R^2$ term to the Hu-Sawicki model,
\be f(R)=R-\frac{DR_{0} R}{R+R_{0}}+{\alpha}R^2. \label{f_R_combined}\ee
In this combined model, at high curvature scale, $f'\approx2{\alpha}R$. So a singular $R$ is pushed to regions where $f'$ and $\phi$ are also singular. Then the singularity problem is avoided. The parameters take the same values as in the last subsection. In addition, the new parameter $\alpha$ in the $f(R)$ model~(\ref{f_R_combined}) is set to $1$.

\begin{figure*}[t!]
  \epsfig{file=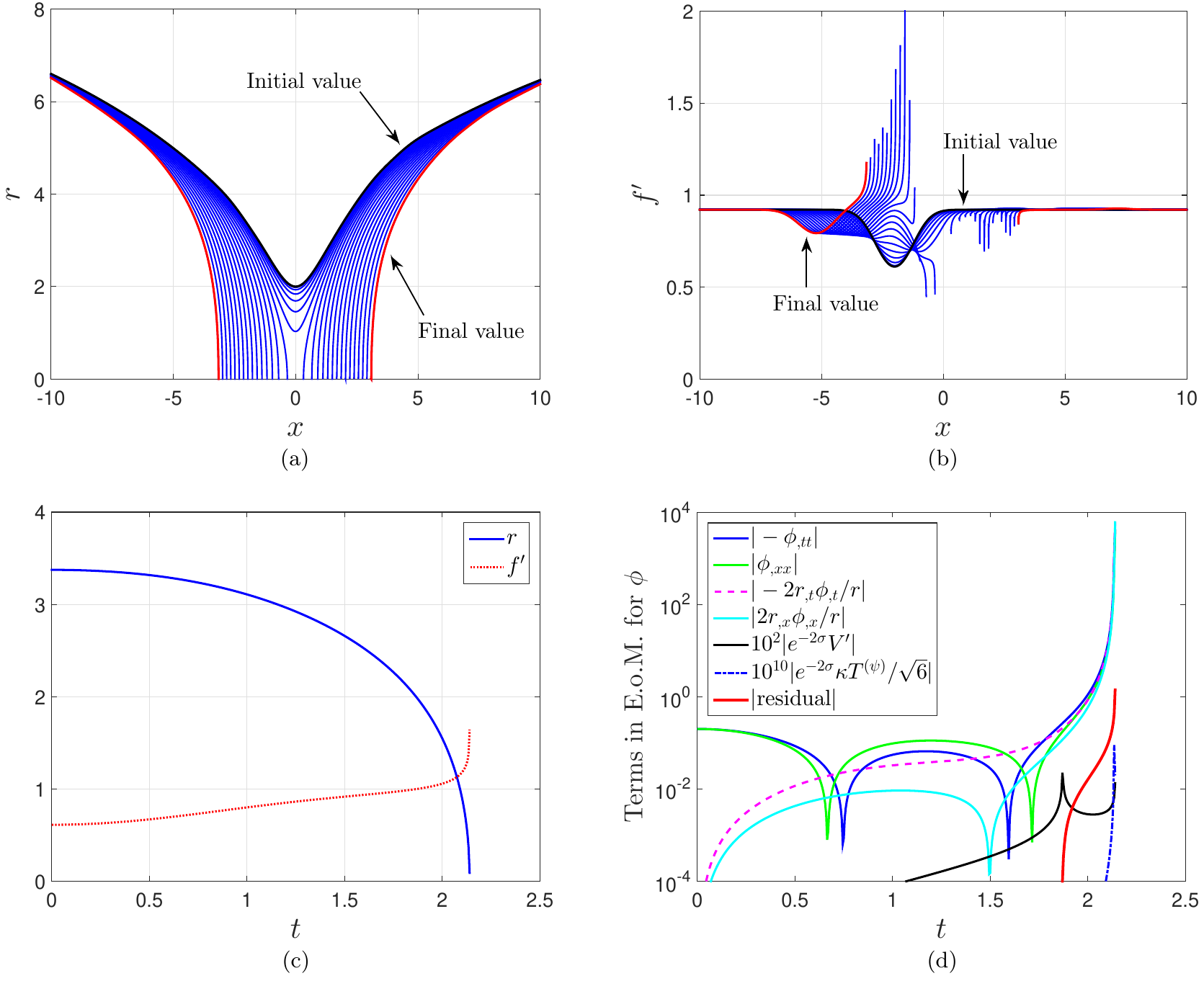, width=0.9\textwidth}
  \caption{(color online). Results for neutral scalar collapse in a Schwarzschild geometry in a combined $f(R)$ model~(\ref{f_R_combined}).
  (a) and (b): evolutions of $r$ and $f'$. The time interval between two consecutive slices is $30{\Delta}t=0.15$.
  (c) evolutions of $r$ and $f'$ on the slice $(x=-2,t=t)$.
  (d) dynamical equation for $\phi$ on the slice $(x=-2,t=t)$.}
  \label{fig:neutral_scattering_singularity_avoidance}
\end{figure*}

The numerical results for neutral scattering for this modified model are plotted in Fig.~\ref{fig:neutral_scattering_singularity_avoidance}. As shown in Fig.~\ref{fig:neutral_scattering_singularity_avoidance}(b), for this model, $f'$ can cross $1$ without difficulty. What are the limits that $f'$ and $R$ can reach? As shown in Fig.~\ref{fig:neutral_scattering_singularity_avoidance}(d), as the central singularity is approached, there is
\be \phi_{,tt}\approx-\frac{2}{r}r_{,t}\phi_{,t}.\ee
Since $r_{,t}$ is negative, the above equation describes a positive feedback system of $\phi_{,tt}$ and $\phi_{,t}$: $|\phi_{,t}|$ produces more of $|\phi_{,tt}|$, and in turn $|\phi_{,tt}|$ produces more of $|\phi_{,t}|$. Since $\phi_{,t}$ is positive, $\phi$ can be rapidly accelerated to positive infinity. Then $f'$ and $R$ will go to infinity as the central singularity is approached. This feature surely deserves some attention as discussed below.

The Starobinsky model, $f(R)=R+{\alpha}R^2$, was obtained by taking into account quantum-gravitational effects. It could cause inflation in the early Universe and is conventionally believed to be singularity-free~\cite{Starobinsky_1980}. The combined model is reduced to the Starobinsky model at high curvature scale. It is found that, in neutral scattering for the combined model, a new black hole, including a new central singularity, can be formed. Near the central singularity, gravity dominates other terms, including the potential related to the $R^2$ term, such that the Ricci scalar $R$ can be pushed to infinity by gravity. We also simulate scalar collapse for the Starobinsky model in flat geometry. Similar results are obtained~\cite{Guo_2015}. Therefore, the classical singularity problem, which is present in general relativity, remains in collapse for these models. Further details are skipped.

\section{Results for charge scattering\label{sec:results}}
In this section, we explore charge scattering: scalar collapse in a Reissner-Nordstr\"{o}m geometry. We study the evolutions of the metric components and scalar fields and obtain approximate analytic solutions. We closely compare the dynamics in Schwarzschild black holes, Reissner-Nordstr\"{o}m black holes, neutral scalar collapse, and charge scattering.

In this section, the parameters are set as follows:
\begin{enumerate}[(i)]
  \item Reissner-Nordstr\"{o}m geometry: $m=1$, and $q=0.7$.
  \item Physical scalar field:
  $\psi(x,t)|_{\scriptsize{t=0}}=a\cdot\exp\left[-(x-x_{0})^2/b\right]$, $a=0.08$, $b=1$, and $x_{0}=4$.
  \item $f(R)$ model:
  $f(R)=R-DR_{0} R/(R+R_{0})$, $D=1.2$, and $R_{0}=10^{-5}$.
  \item Scalar degree of freedom: $\phi(x,t)|_{t=0}=\phi_0$, with $V'(\phi_0)=0$.
  \item Grid. Spatial range: $x\in[-10~10]$. Grid spacings: ${\Delta}x={\Delta}t=0.005$ for Secs.~\ref{sec:evolutions} and \ref{sec:spacelike}, and ${\Delta}x={\Delta}t=0.0025$ for Secs.~\ref{sec:fast_stage} and \ref{sec:slow_stage}.
\end{enumerate}

\begin{figure*}[t!]
  \epsfig{file=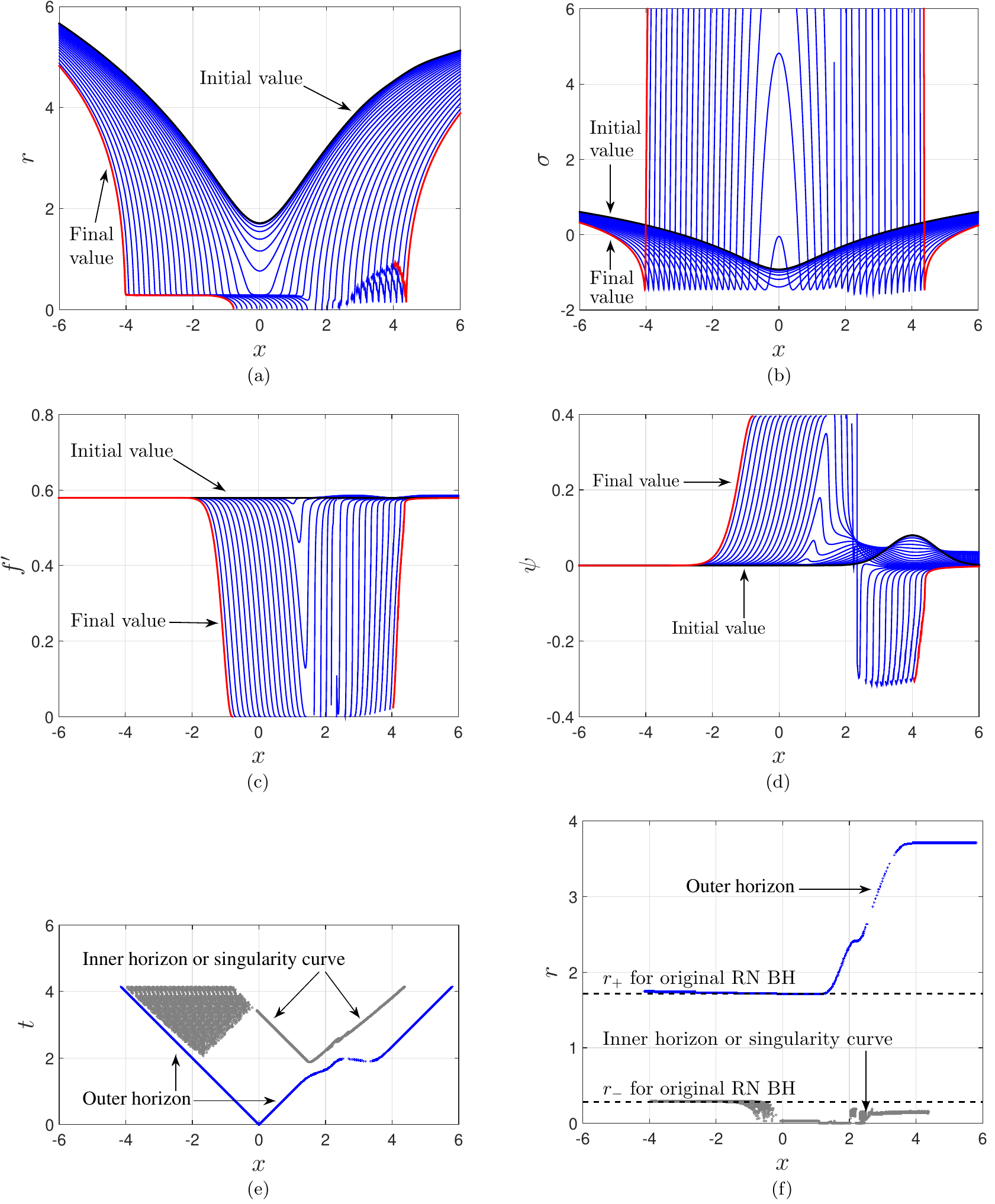, width=0.93\textwidth}
  \caption{Evolutions in charge scattering.
  (a)-(d): evolutions of $r$, $\sigma$, $f'$, and $\psi$. The time interval between two consecutive slices is $30{\Delta}t=0.15$.
  (e) and (f) are for the apparent horizon and the singularity curve of the black hole.
  When the scalar field is strong enough (around $x=2$), the inner horizon can be pushed to the center, and the central singularity becomes spacelike. When the scalar field is weak enough (e.g., $-4<x<-1$), the inner horizon does not change much. At the intermediate state (e.g., $0<x<1.5$), the inner horizon contracts to zero, and the central singularity becomes null. The results for the inner horizon, especially for $x>2$, are not that accurate. We are aware that the inner horizon is actually at infinity, while $r$ still can be very close to $r_{-}$ when $x$ and $t$ take moderate values.}
  \label{fig:evolutions}
\end{figure*}

\subsection{Evolutions\label{sec:evolutions}}
\subsubsection{Outline}
In this subsection, we describe the evolutions of $r$, $\sigma$, $f'$, and $\psi$ that are plotted in Fig.~\ref{fig:evolutions}. Examining the equations of motion (\ref{equation_r})-(\ref{equation_psi}) and the numerical results plotted in Figs.~\ref{fig:eom_spacelike} and \ref{fig:eom_null}, one can see that in the charge scattering dynamical system, there are three types of quantities as follows:
\begin{enumerate}[(i)]
  \item Metric components: $r$ and $\sigma$. They contribute as gravity.
  \item Scalar fields: $\phi$ and $\psi$. They contribute as self-gravitating fields.
  \item Electric field and $V(\phi)$. As implied in Eq.~(\ref{equation_r}), they are repulsive forces. However, the numerical results show that in charge scattering, compared to the contributions from other quantities, the contribution from $V(\phi)$ is negligible.
\end{enumerate}
Furthermore, these quantities can be separated into two sides: the gravitating side ($r$, $\sigma$, $\phi$, and $\psi$) and the repulsive side [electric field and $V(\phi)$]. The dynamics in charge scattering consists mainly of how these variables interact and how the gravitating and repulsive sides compete.

According to the strength of the scalar field, charge scattering can be classified into five types as follows:
\begin{enumerate}[(i)]
  \item Type I: spacelike scattering. When the scalar field is very strong, the inner horizon can contract to zero volume rapidly, and the central singularity becomes spacelike. Sample slice: $(x=1.5,t=t)$ in Fig.~\ref{fig:evolutions}. See Sec.~\ref{sec:spacelike}.
  \item Type II: null scattering. When the scalar field is intermediate, the inner horizon can contract to a place close to the center or reach the center. For each variable, the spatial and temporal derivatives are almost equal. In the case of the center being reached, the central singularity becomes null. This type has two stages: early/slow and late/fast. In the early stage, the inner horizon contracts slowly, and the scalar field also varies slowly. In the late stage, the inner horizon contracts quickly, and the dynamics is similar to that in the spacelike case. Sample slice: $(x=0.5,t=t)$ in Fig.~\ref{fig:evolutions}. See Secs.~\ref{sec:fast_stage} and \ref{sec:slow_stage}.
  \item Type III: critical scattering. This case is on the edge between the above two cases. When the central singularity is reached, it becomes null. Sample slice: $(x=1.4225,t=t)$ in Fig.~\ref{fig:evolutions}. Due to the similarity to that in general relativity discussed in Ref.~\cite{Guo_1507}, details on this type of scattering in $f(R)$ gravity are skipped in this paper.
  \item Type IV: weak scattering. When the scalar field is very weak, the inner horizon does not contract much.
        Sample case: Fig.~\ref{fig:evolutions_weak}. See Sec.~\ref{sec:weak_scattering}.
  \item Type V: tiny scattering. When the scalar field is very tiny, the influence of the scalar field on the geometry is negligible. Sample slice: $(x=-3,t=t)$ in Fig.~\ref{fig:evolutions}.
\end{enumerate}
In this paper, we will discuss Types I, II, and IV.

\subsubsection{Causes of mass inflation and evolutions}
The local Misner-Sharp mass for a charge black hole hole is
\be g^{\mu\nu}r_{,\mu}r_{,\nu}=e^{2\sigma}(-r_{,t}^2+r_{,x}^2)=1-\frac{2m}{r}+\frac{q^2}{r^2}. \label{mass_definition_RN}\ee
In a Reissner-Nordstr\"{o}m geometry, in Kruskal-like coordinates expressed by Eq.~(\ref{RN_metric_text}), near the inner horizon, although $\sigma$ asymptotes to positive infinity, $(r_{,t}^2-r_{,x}^2)$ is much less than $e^{-2\sigma}$. Consequently, $e^{2\sigma}(r_{,t}^2-r_{,x}^2)$ approaches zero. Then as implied in Eq.~(\ref{mass_definition_RN}), the mass function $m$ takes a finite value and is equal to the black hole mass. For more details, see Ref.~\cite{Guo_1507}.

In charge scattering, the equations of motion for $\eta{\equiv}r^2$
\\
\\
\\
\\
\\
and $\sigma$ are
\be -\eta_{,tt}+\eta_{,xx}=2e^{-2\sigma}\left(1-8\pi r^2V-\frac{q^2}{r^2}\right),\label{equation_r_scattering}\ee
\be
\begin{split}
&-\sigma_{,tt}+\sigma_{,xx}+\frac{r_{,tt}-r_{,xx}}{r}\\
&+4\pi\left[\phi_{,t}^2-\phi_{,x}^2 + \frac{\psi_{,t}^2-\psi_{,x}^2}{\chi}-2e^{-2\sigma}V\right]\\
&+e^{-2\sigma}\frac{q^2}{r^4}=0.
\end{split}
\label{equation_sigma_scattering}
\ee
At the beginning, $|\phi_{,t}|$ and $|\psi_{,t}|$ may be less than $|\phi_{,x}|$ and $|\psi_{,x}|$, respectively. However, as $r$ decreases toward the central singularity, gravity becomes stronger. Then $|\phi_{,t}|$ and $|\psi_{,t}|$ become greater than $|\phi_{,x}|$ and $|\psi_{,x}|$, respectively.
[See Fig.~\ref{fig:eom_null}.] As a result, in this case,
the repulsive \lq\lq{force}\rq\rq, $4\pi[\phi_{,t}^2-\phi_{,x}^2+(\psi_{,t}^2-\psi_{,x}^2)/\chi]+e^{-2\sigma}q^2/r^4$, is greater than the corresponding one, $e^{-2\sigma}q^2/r^4$, in a Reissner-Nordstr\"{o}m geometry. This makes $\sigma$ accelerate faster than in the Reissner-Nordstr\"{o}m geometry. Consequently, the repulsive force from $2e^{-2\sigma}(q^2/r^2-1)$ for $\eta{\equiv}r^2$ is much weaker than the corresponding value in the Reissner-Nordstr\"{o}m geometry. As a result, near the inner horizon, $|r_{,t}|$ is much greater than the corresponding one in the Reissner-Nordstr\"{o}m case. Then $(r_{,t}^2-r_{,x}^2)$ moves from extremely tiny values in the Reissner-Nordstr\"{o}m metric case to moderate values, and $r$ crosses the inner horizon $r=r_{-}$ for the given Reissner-Nordstr\"{o}m geometry. With Eq.~(\ref{mass_definition_RN}), the mass parameter grows dramatically: mass inflation takes place. In other words, regarding the causes of mass inflation in charge scattering, the scalar fields' backreaction on $r$ is more important than that on $\sigma$. The evolutions of $r$, $\sigma$, $f'$, and $\psi$ are plotted in Figs.~\ref{fig:evolutions}(a)-\ref{fig:evolutions}(d).

\begin{figure*}[t!]
  \epsfig{file=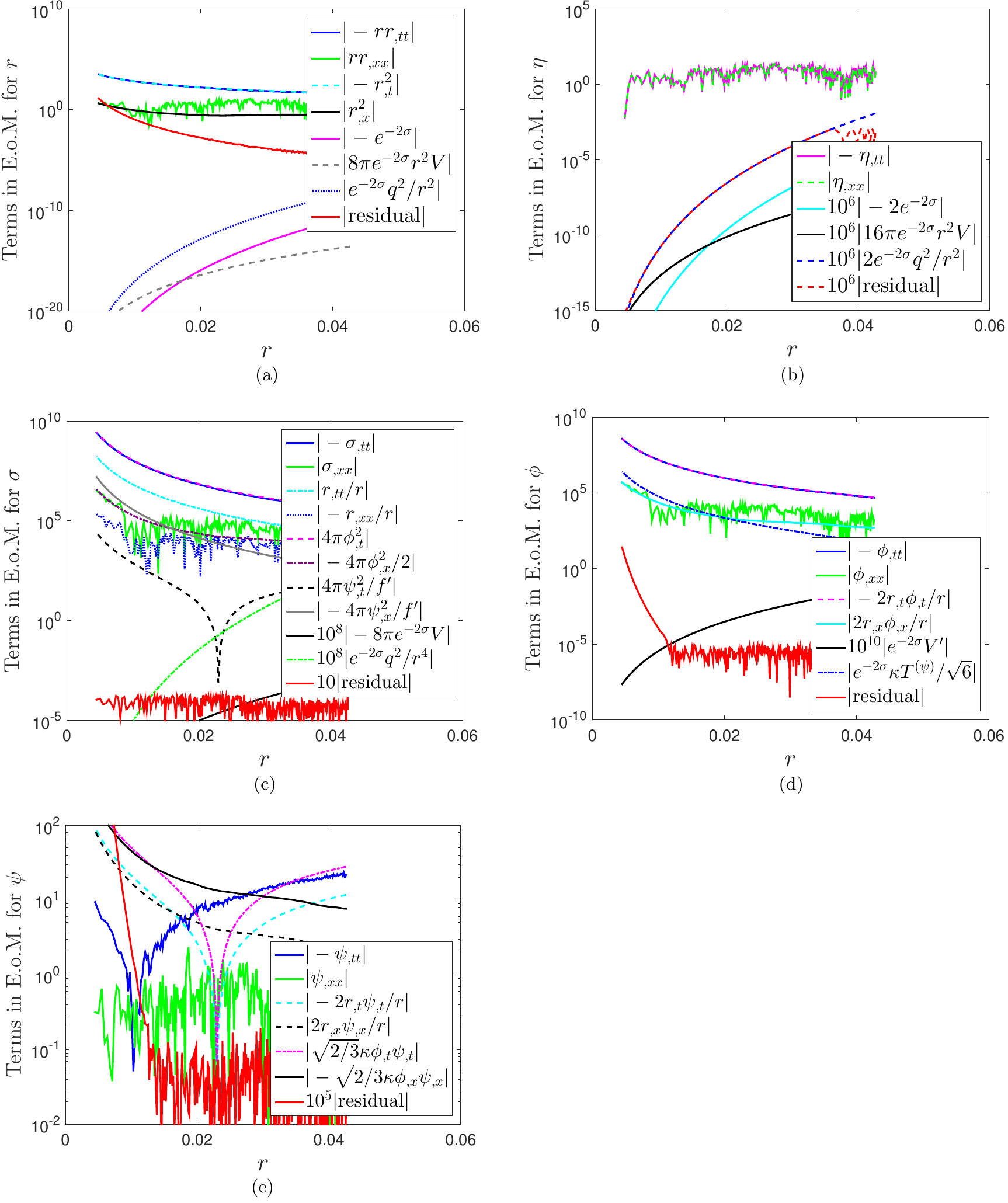, width=0.9\textwidth}
  \caption{(color online). Dynamics along the slice $(x=1.52,t=t)$ near the spacelike, central singularity. Near the singularity, we can approximately rewrite the original equations of motion for $r$, $\eta$, $\sigma$, $\phi$, and $\psi$ as follows.
  (a) $rr_{,tt}\approx-r_{,t}^2$.
  (b) $\eta_{,tt}\approx\eta_{,xx}^2$.
  (c) $\sigma_{,tt}\approx4\pi\phi_{,t}^2$.
  (d) $\phi_{,tt}\approx-2r_{,t}\phi_{,t}/r$.
  (e) $r_{,t}\psi_{,t}{\approx}r_{,x}\psi_{,x}$ and $\phi_{,t}\psi_{,t}\approx\phi_{,x}\psi_{,x}$. Then we have $\psi_{,t}/\psi_{,x}{\approx}r_{,x}/r_{,t}\approx\phi_{,x}/\phi_{,t}$.}
  \label{fig:eom_spacelike}
\end{figure*}

Because of gravity from the black hole and the contribution from the physical scalar field, for some configurations, $f'$ can decrease to zero as the central singularity is approached. [See Eq.~(\ref{equation_phi}) and Figs.~\ref{fig:evolutions}, \ref{fig:eom_spacelike}, and \ref{fig:eom_null}.] Nothing is wrong with this circumstance. However, for certain configurations, $f'$ can be pushed to $1$, as shown in Fig.~\ref{fig:Ricci_singularity}. Correspondingly, the Ricci scalar $R$ becomes singular. We will discuss the former case in this section and the latter case in Sec.~\ref{sec:singularity}.

The evolution of $\psi$ is plotted in Fig.~\ref{fig:evolutions}(d). In the configurations that we consider, as the central singularity is approached, because of the strong suppression from the dark energy scalar $\phi$, $\psi$ asymptotes to constant values.

\subsubsection{Locations of horizons}
We locate the outer and inner horizons using the following equation,
\be g^{\mu\nu}r_{,\mu}r_{,\nu}=e^{2\sigma}(-r_{,t}^2+r_{,x}^2)=1-\frac{2m}{r}+\frac{q^2}{r^2}=0.\ee
The results are plotted in Figs.~\ref{fig:evolutions}(e) and \ref{fig:evolutions}(f). Due to the absorptions of the energies of $\phi$ and $\psi$, the outer horizon increases from the original value of $1.7$ to $3.7$. Note that the results for the inner horizon, especially at regions where $x>2$, are not that accurate. We are aware that the inner horizon is actually at infinity, while $r$ still can be very close to $r_{-}$ even when $x$ and $t$ take moderate values.

\begin{figure*}[t!]
  \epsfig{file=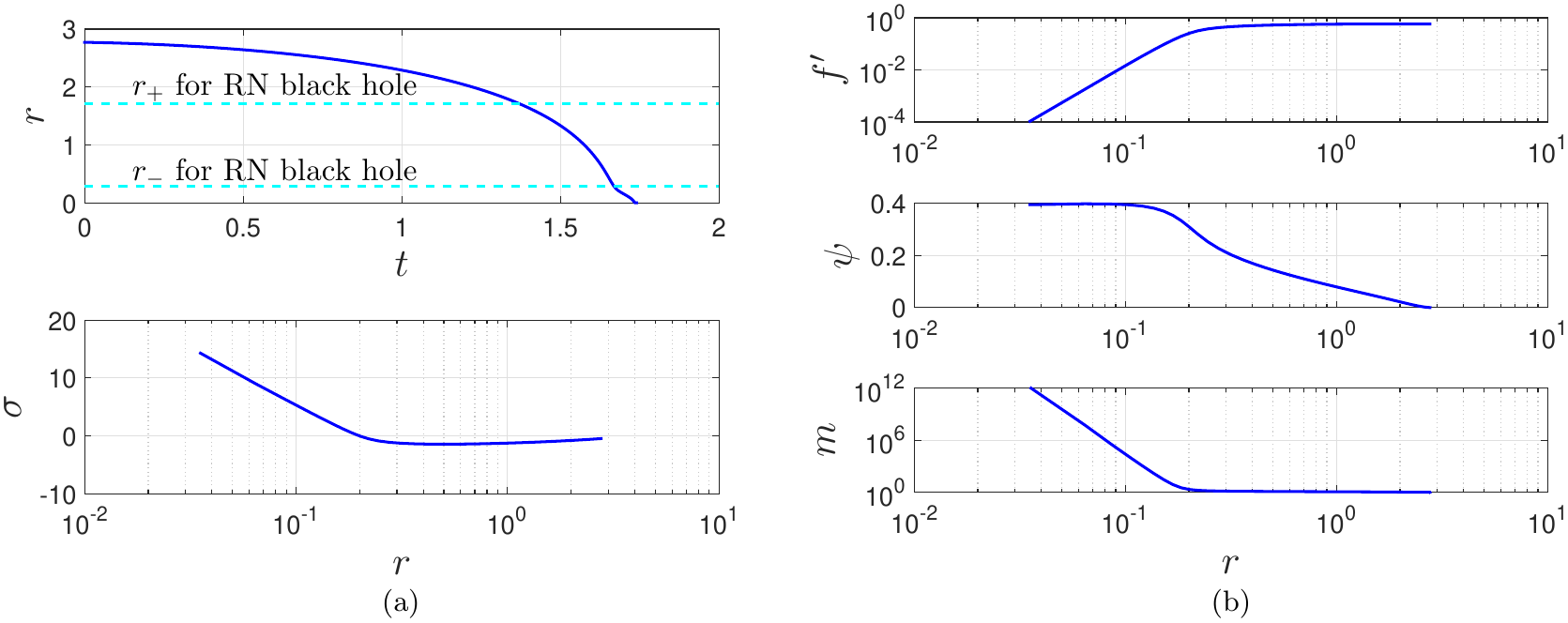, width=0.9\textwidth}
  \caption{Evolutions along the slice $(x=1.52,t=t)$ in one spacelike scattering process.
  (a) evolutions of $r$ and $\sigma$.
  (b) evolutions of $f'$, $\psi$, and $m$.
  In this case, the scalar fields are so strong, such that $r$ can rapidly cross the line of $r=r_{-}$ and approach zero. Other variables ($\sigma$, $f'$, $\psi$, and $m$) also evolve rapidly after $r$ has crossed the line of $r=r_{-}$.}
  \label{fig:evolutions_spacelike}
\end{figure*}

\begin{figure*}[t!]
  \epsfig{file=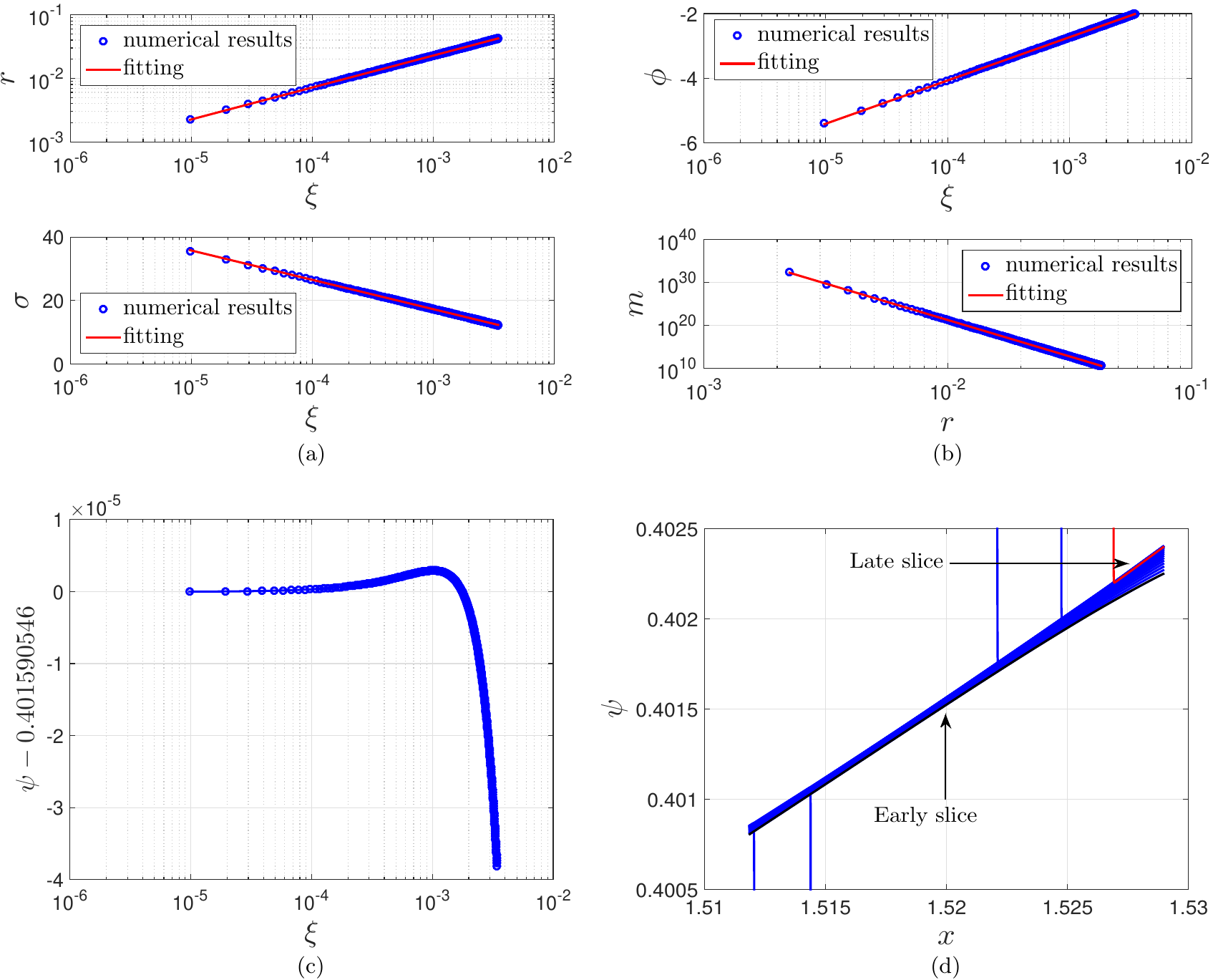, width=0.85\textwidth}
  \caption{Solutions along the slice $(x=1.52,t=t)$ near the spacelike, central singularity.
  (a) ${\ln}r{\approx}a\ln\xi+b$, $a=0.5032\pm0.0001$, $b=-0.3033\pm0.0008$.
      $\sigma{\approx}a\ln\xi+b$, $a=-4.0\pm0.2$, $b=-10\pm2$.
  (b) $\phi{\approx}a\ln\xi+b$, $a=0.5827\pm0.0002$, $b=-1.294\pm0.002$.
      ${\ln}m{\approx}a{\ln}r+b$, $a=-16.937\pm0.003$, $b=-29.04\pm0.01$.
  (c) and (d): $\psi(x,t)$ asymptotes to constant values as the central singularity is approached.}
  \label{fig:asymptotic_solution_spacelike}
\end{figure*}

\subsection{Spacelike scattering\label{sec:spacelike}}
In spacelike scattering, the scalar field is so strong, such that the inner horizon can contract to zero volume rapidly, and the central singularity converts from timelike into spacelike. Taking the slice $(x=1.52,t=t)$ as an example, we plot the evolutions of $r$, $\sigma$, $f'$, $\psi$, and $m$ on this slice in Fig.~\ref{fig:evolutions_spacelike}. The mass function remains equal to the mass of the original Reissner-Nordstr\"{o}m black hole, $m_{0}=1$, until $r$ is very close to $r_{-}$. By then mass inflation takes place. We examine the dynamics near the central singularity via mesh refinement and plot the terms in the dynamical equations in Fig.~\ref{fig:eom_spacelike}.

The strongness of the scalar field causes several consequences as below.
\begin{enumerate}[(i)]
  \item \emph{Motion of $r$.} The quantity $r$ does not decelerate much when it crosses the inner horizon of the given Reissner-Nordstr\"{o}m black hole, and it can approach the center. [See Fig.~\ref{fig:evolutions_spacelike}.]
  \item \emph{Nature of the central singularity.} The central singularity converts from timelike into spacelike.
  \item \emph{The dynamics in spacelike scattering is similar to that in strong, neutral scalar collapse.} The quantity $\sigma$ takes large positive values, such that in the vicinity of the central singularity, compared to other terms, the term $e^{-2\sigma}q^2/r^2$ in the equation for $r$~(\ref{equation_r}) and the term $e^{-2\sigma}q^2/r^4$ in the equation for $\sigma$~(\ref{equation_sigma}) are negligible. As a result, in the vicinity of the central singularity, the dynamics is similar to that in strong, neutral scalar collapse as expressed by Eqs.~(\ref{equation_r_asymptotic})-(\ref{equation_psi_asymptotic}). [See Figs.~\ref{fig:neutral_collapse_x_equal_1} and \ref{fig:eom_spacelike}.] Then the quantities $r$, $\sigma$, $\phi$, and $m$ take similar forms as those in neutral collapse. [See Fig.~\ref{fig:asymptotic_solution_spacelike}.]

      In both strong, neutral scalar collapse and spacelike scattering, the equation of motion for $\eta$ is reduced to $\eta_{,tt}\approx\eta_{,xx}$. [See Figs.~\ref{fig:neutral_collapse_x_equal_1}(b) and \ref{fig:eom_spacelike}(b).]

      Since $\phi_{,t}$ is negative, the term $\sqrt{2/3}\kappa\phi_{,t}\psi_{,t}$ in Eq.~(\ref{equation_psi}) functions as a friction force for $\psi$. Consequently, compared to $\phi$, $\psi$ grows slowly. As the singularity is approached, $\psi$ even approaches constant values.
      [See Figs.~\ref{fig:asymptotic_solution_spacelike}(c) and \ref{fig:asymptotic_solution_spacelike}(d).]
      As shown in Fig.~\ref{fig:eom_spacelike}(e), near the singularity, two major sets of terms can be expressed as
  \be r_{,t}\psi_{,t}{\approx}r_{,x}\psi_{,x}, \hphantom{dd} \phi_{,t}\psi_{,t}{\approx}\phi_{,x}\psi_{,x}.\nonumber\ee
  Alternatively,
  \be \frac{\psi_{,t}}{\psi_{,x}}\approx\frac{r_{,x}}{r_{,t}}\approx\frac{\phi_{,x}}{\phi_{,t}}.
  \label{ratio_mass_inflation}\ee
  \item \emph{Growth of mass function.} In spacelike scattering, the equation of motion for $\phi$ can be simplified as
  \be \sigma_{,tt}\approx4\pi\phi_{,t}^2.\ee
  Consequently, with the results obtained in Sec.~\ref{sec:asymptotics_collapse}, $\sigma$ has the following asymptotic solution:
  \be
  \sigma\approx B\ln\xi+\sigma_0\approx-4{\pi}C^2\ln\xi+\sigma_0,\label{sigma_asymptotic}
  \ee
  with $\phi\approx C\ln\xi$. Then similar to the mass function (\ref{mass_analytic_collapse}) in neutral collapse, with Eq.~(\ref{mass_definition_RN}), the mass function in spacelike scattering can be written as
  \be
  \begin{split}
  m&=\frac{r}{2}\left[1+\frac{q^2}{r^2}+e^{2\sigma}(r_{,t}^2-r_{,x}^2)\right]\\
  &\approx\left[\frac{1}{8}(1-K^2)A^{3}e^{2\sigma_0}\right]\xi^{2B-\frac{1}{2}}\\
  &\approx\left[\frac{1}{8}(1-K^2)A^{4(-B+1)}e^{2\sigma_0}\right]r^{4B-1}\\
  &\approx\left[\frac{1}{8}(1-K^2)A^{4(4{\pi}C^2+1)}e^{2\sigma_0}\right]r^{-16{\pi}C^2-1}.
  \end{split}
  \label{mass_analytic}
  \ee

  Numerical results show that, for the sample slice $(x=1.52,t=t)$, near the central singularity, the slope of the singularity curve $K$ is about $0.04$. As shown in Fig.\ref{fig:asymptotic_solution_spacelike}(d), we linearly fit the numerical results of $m$ via
  \be {\ln}m{\approx}a{\ln}r+b,\label{ln_m_asymptotic}\ee
  obtaining
  \be a=-16.937\pm0.003,\hphantom{dd}b=-29.04\pm0.01.\nonumber\ee
  Fitting numerical results for $\sigma$ according to Eq.~(\ref{sigma_asymptotic}) and combining Eqs.~(\ref{mass_analytic}) and (\ref{ln_m_asymptotic}), we obtain
  \begin{align}
  a_{\scriptsize{\mbox{analytic}}}&=4B-1=17.0\pm0.8,\nonumber\\
  \nonumber\\
  b_{\scriptsize{\mbox{analytic}}}&=\ln\left[\frac{1}{8}(1-K^2)A^{4(-B+1)}e^{2\sigma_0}\right]=-28\pm4.\nonumber
  \end{align}
  Similarly, fitting numerical results for $\phi$ according to $\phi\approx C\ln\xi$, we obtain
  \begin{align}
  a_{\scriptsize{\mbox{analytic}}}&=-16{\pi}C^2-1=18.06\pm0.01,\nonumber\\
  \nonumber\\
  b_{\scriptsize{\mbox{analytic}}}&=\ln\left[\frac{1}{8}(1-K^2)A^{4(4{\pi}C^2+1)}e^{2\sigma_0}\right]=-28\pm4.\nonumber
  \end{align}
  One can see that the above three sets of results match well.
\end{enumerate}

\begin{figure*}[t!]
  \epsfig{file=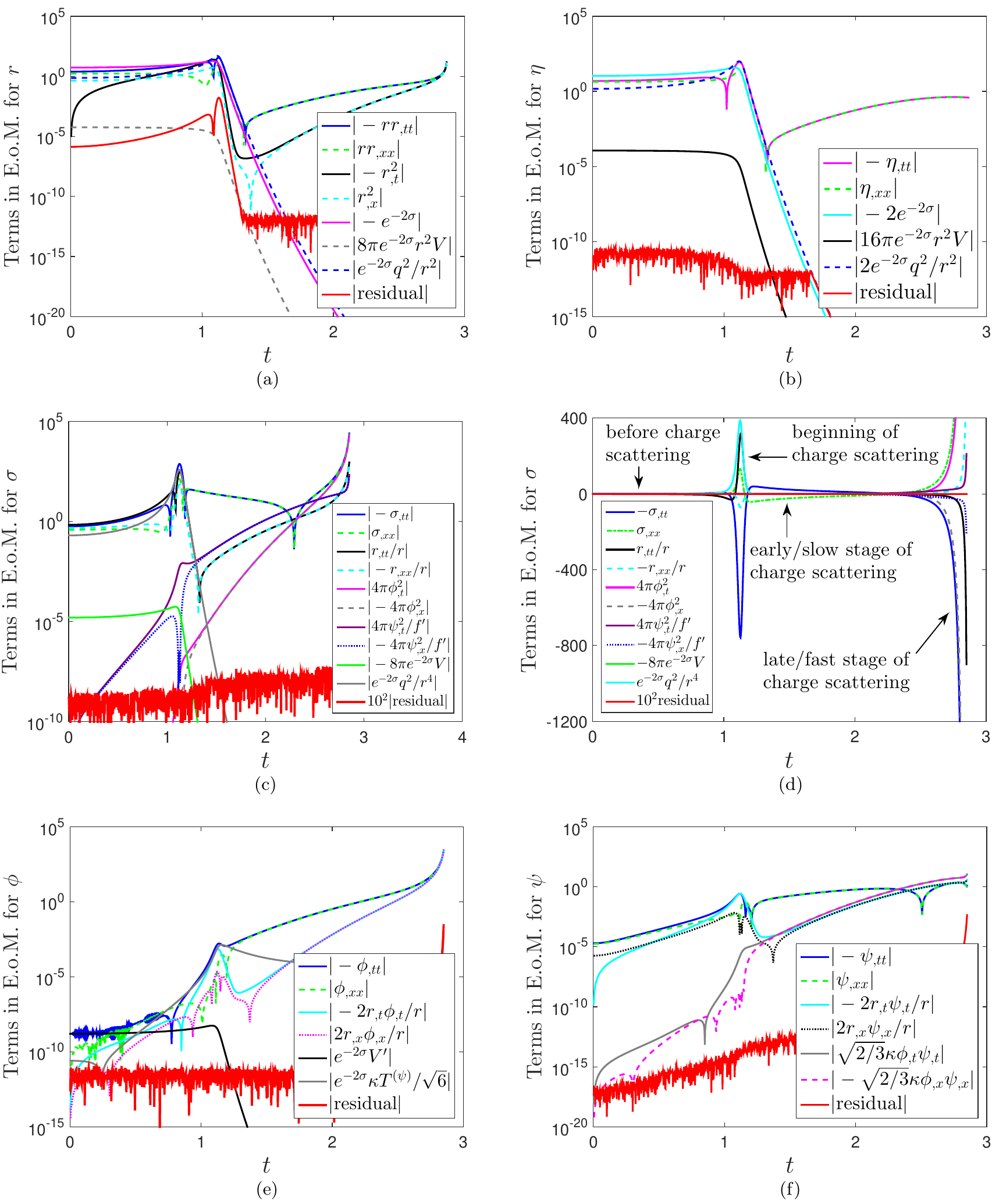, width=0.9\textwidth}
  \caption{(color online). Dynamics along the slice $(x=0.5,t=t)$ in null scattering.
  (a)-(f): dynamical equations for $r$, $\eta$, $\sigma$, $\phi$, and $\psi$.}
  \label{fig:eom_null}
\end{figure*}

\begin{figure*}[t!]
  \epsfig{file=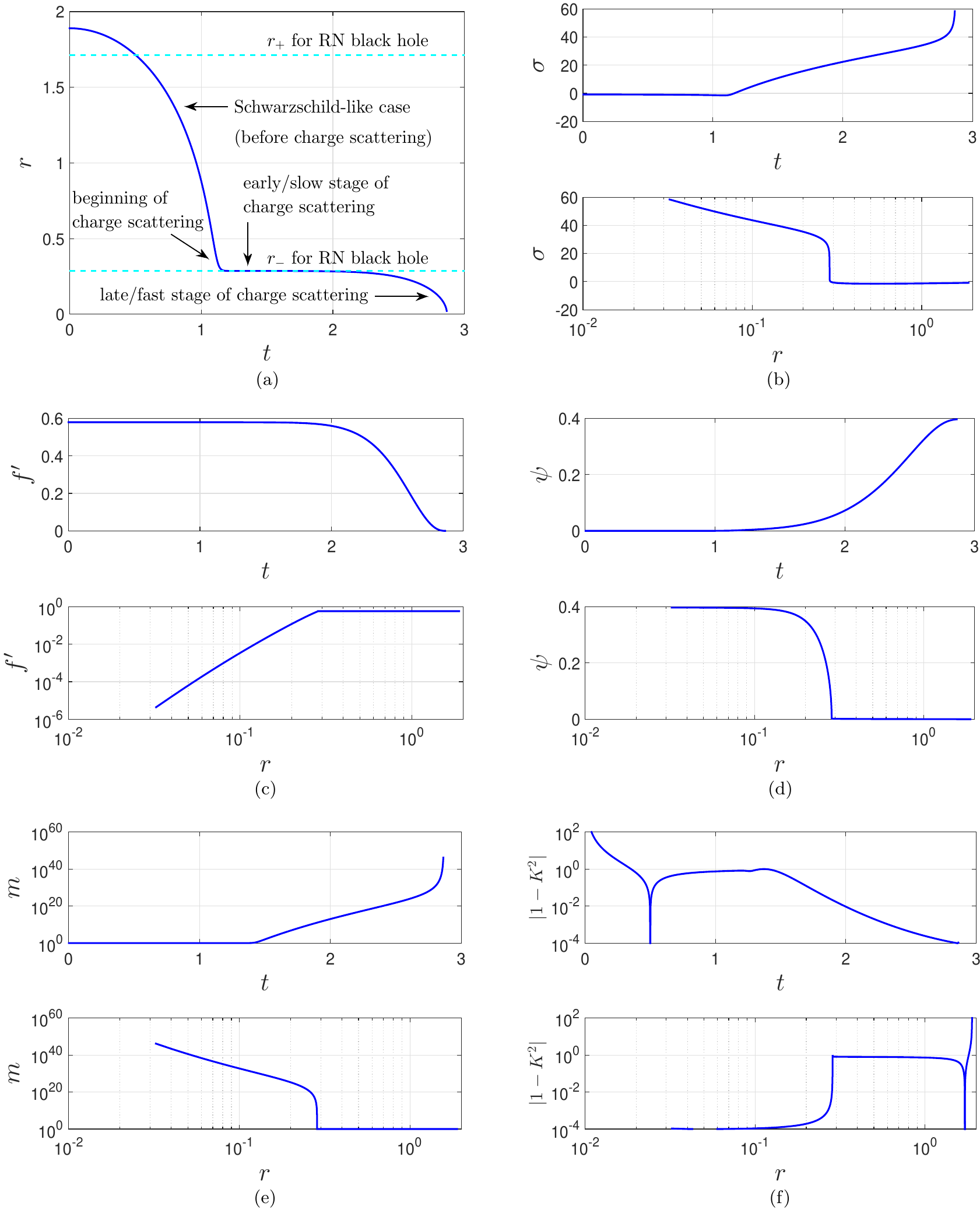, width=0.9\textwidth}
  \caption{Evolutions along the slice $(x=0.5,t=t)$ in null scattering. (a)-(f): evolutions for $r$, $\sigma$, $f$, $\psi$, $m$, and $|1-K^2|$. At the early stage of mass inflation, $1.2<t<2$, $r$ varies slowly; while at the late stage, $t>2$, $r$ varies rapidly toward zero.}
  \label{fig:evolutions_null}
\end{figure*}

\begin{figure*}[t!]
  \epsfig{file=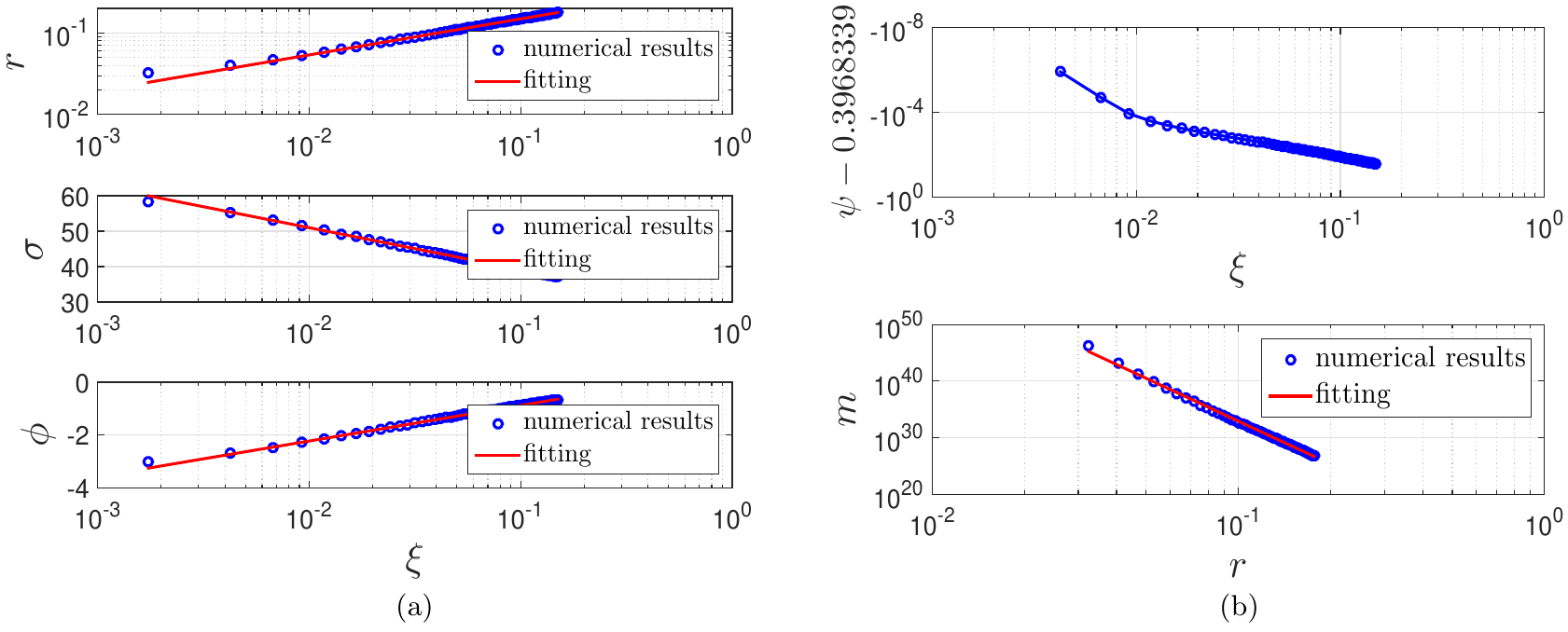, width=0.9\textwidth}
  \caption{Solutions near the null central singularity along the slice $(x=0.5,t=t)$ in null scattering.
  (a) ${\ln}r{\approx}a\ln\xi+b$, $a=0.4421\pm0.0008$, $b=-0.887\pm0.002$.
      $\sigma{\approx}a\ln\xi+b$, $a=-5.15\pm0.01$, $b=27.26\pm0.02$.
      $\phi{\approx}a\ln\xi+b$, $a=0.583\pm0.002$, $b=0.465\pm0.005$.
  (b) $\psi$ approaches a constant value $0.4$.
      ${\ln}m{\approx}a{\ln}r+b$, $a=-25.2\pm0.1$, $b=17.9\pm0.3$.}
  \label{fig:asymptotics_null_fast_stage}
\end{figure*}

\begin{figure*}[t!]
  \epsfig{file=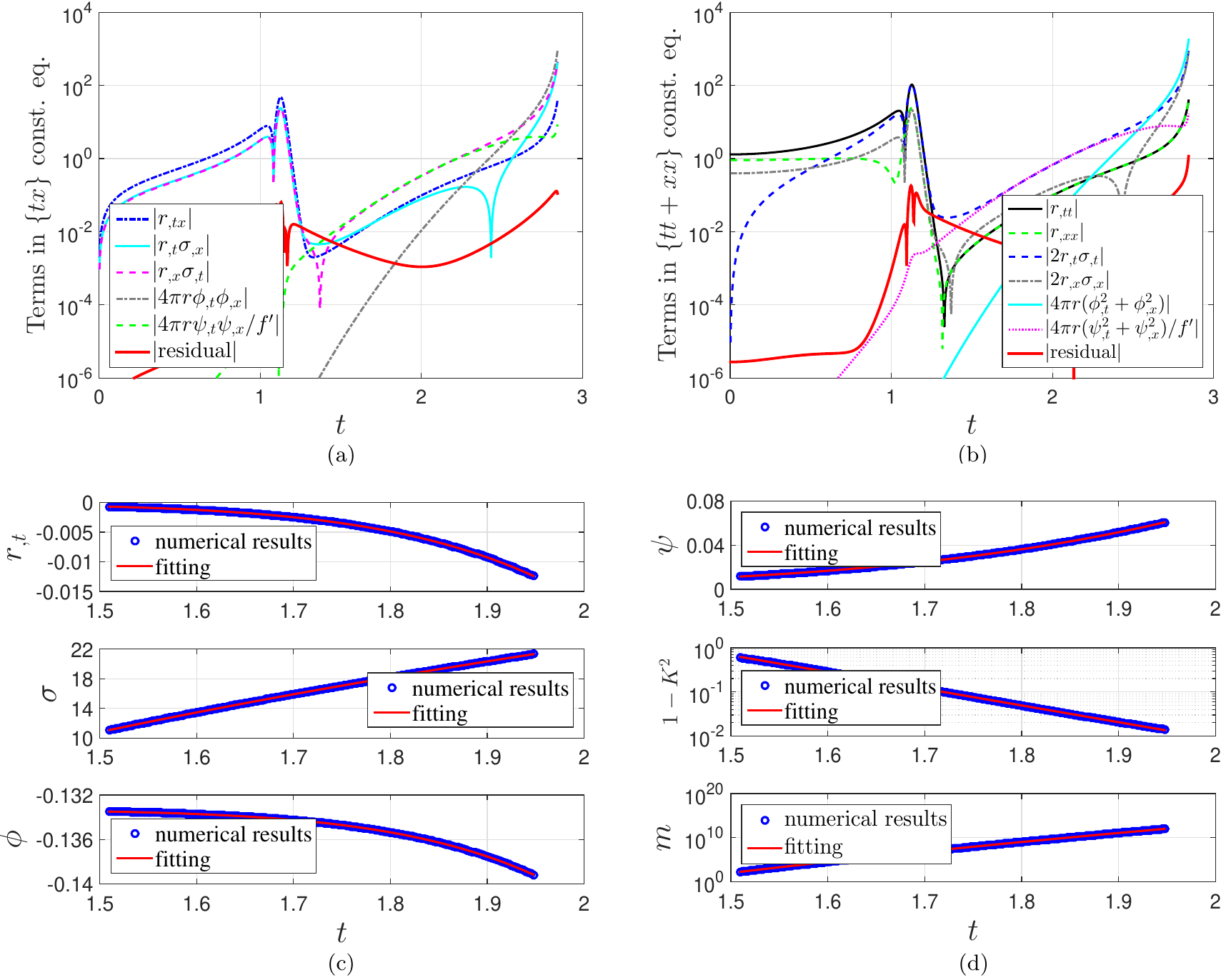, width=0.9\textwidth}
  \caption{(color online). Constraint equations and solutions along the slice $(x=0.5,t=t)$ at the early/slow stage of null scattering.
  (a) and (b): constraint equations (\ref{constraint_eq_xt}) and (\ref{constraint_eq_xx_tt}).
  (c) $r_{,t}{\approx}a(t+b)^c+d$, $a=(-1.67\pm0.01)\times10^{-3}$, $b=-0.6622\pm0.0008$, $c=7.897\pm0.005$, $d=(-2.610\pm0.003)\times10^{-4}$.
      $\sigma{\approx}a\ln(t+b)+c$, $a=34.39\pm0.03$, $b=-0.245\pm0.001$, $c=3.02\pm0.04$.
      $\phi{\approx}a(t+b)^c+d$, $a=(-9.7\pm2.3)\times10^{-4}$, $b=-0.71\pm0.03$, $c=8.3\pm0.2$, $d=-0.133340\pm0.000005$.
  (d) $\psi{\approx}a(t+b)^c+d$, $a=(5.0\pm0.1)\times10^{-3}$, $b=-0.362\pm0.006$, $c=5.40\pm0.02$, $d=(1.19\pm0.02)\times10^{-3}$.
      $\ln(1-K^2){\approx}at+b$, $a\approx-8.73\pm0.01$, $b\approx12.71\pm0.02$.
      ${\ln}m{\approx}a{\ln}(t+b)+c$, $a=77.77\pm0.02$, $b=-0.1994\pm0.0004$, $c=-15.80\pm0.03$.}
\label{fig:asymptotics_null_slow_stage}
\end{figure*}

\subsection{The late/fast stage of null scattering\label{sec:fast_stage}}
When the scalar fields are less strong, the inner horizon may still contract to zero. However, in this case, the central singularity becomes null rather than spacelike. The equations of motion remain null: they have similar forms as free wave equations, e.g., $\phi_{,tt}\approx\phi_{,xx}$.

In the Reissner-Nordstr\"{o}m black hole case, near the center, the repulsive (electric) force dominates gravity, and the central singularity is timelike. In spacelike scattering as discussed in the last subsection, gravity from the scalar field and the background geometry dominates the repulsive force. As a result, the central singularity is spacelike. At the late stage of null scattering that will be studied in this subsection, the scalar field is less strong, and the central singularity is null. Because of this, one may say that null scattering is a critical case of the competition between repulsive and gravitational forces, in which case the two types of forces have a balance.

Take the slice $(x=0.5,t=t)$ as a sample slice, we plot the terms in the field equations for $r$, $\sigma$, $\phi$, and $\psi$ in Fig.~\ref{fig:eom_null}, and the evolutions of $r$, $\sigma$, $\phi$, $\psi$, $m$, and $|1-K^2|$ in Fig.~\ref{fig:evolutions_null}. We investigate the dynamics in the vicinity of the central singularity via mesh refinement and plot the results in Fig.~\ref{fig:asymptotics_null_fast_stage}.

The null scattering has two stages: early/slow and late/fast. As shown in Figs.~\ref{fig:eom_null} and \ref{fig:evolutions_null}, at the beginning of charge scattering, because of the repulsive force from the electric field, $r$, $\sigma$, $\phi$, and $\psi$ evolve slowly. As a result, the mass function $m$ also grows slowly. We call this stage the early/slow stage. Later on, as the center is approached, gravity becomes very strong. Then these quantities evolve faster. We call this stage the late/fast stage.

As shown in Fig.~\ref{fig:eom_null}, when $r$ is very small, the equations of motion for $r$ (\ref{equation_r}), $\sigma$ (\ref{equation_sigma}), and $\phi$ (\ref{equation_phi}) can be rewritten as
\be -rr_{,tt}\approx-rr_{,xx}{\approx}r_{,t}^2{\approx}r_{,x}^2,\label{equation_r_null_late}\ee
\be \sigma_{,tt}\approx\sigma_{,xx}\approx4\pi\phi_{,t}^2\approx4\pi\phi_{,x}^2,\label{equation_sigma_null_late}\ee
\be \phi_{,tt}\approx\phi_{,xx}\approx-\frac{2}{r}r_{,t}\phi_{,t}\approx-\frac{2}{r}r_{,x}\phi_{,x}.\label{equation_phi_null_late}\ee
Since the above three equations have some similarities to the corresponding ones in spacelike scattering, it is natural to guess that the quantities $r$, $\sigma$, $\psi$, and $m$ may have expressions similar to those in spacelike scattering. In fact, this guess is verified by the numerical results plotted in
Fig.~\ref{fig:asymptotics_null_fast_stage}. Then we have
\begin{align}
r&\approx A\xi^{\frac{1}{2}},\\
\nonumber\\
\sigma&\approx B\ln\xi+\sigma_0\approx-4{\pi}C^2\ln\xi+\sigma_0,\\
\nonumber\\
\phi&\approx C\ln\xi,\\
\nonumber\\
m&\approx\left[\frac{1}{8}(1-K^2)A^{3}e^{2\sigma_0}\right]\xi^{2B-\frac{1}{2}}\nonumber\\
&\approx\left[\frac{1}{8}(1-K^2)A^{4(-B+1)}e^{2\sigma_0}\right]r^{4B-1}\nonumber\\
&\approx\left[\frac{1}{8}(1-K^2)A^{4(4{\pi}C^2+1)}e^{2\sigma_0}\right]r^{-16{\pi}C^2-1}.
\end{align}

As shown in Fig.~\ref{fig:evolutions_null}(f), at the late stage, $(1-K^2)$ is around $10^{-4}$. Due to the similarity to spacelike scattering that we discussed in the last subsection, a comparison between numerical and analytical results for the fast stage of null scattering is skipped.

\subsection{The early/slow stage of null scattering\label{sec:slow_stage}}
As shown in Fig.~\ref{fig:eom_null}, at the early/slow stage of null scattering, the equations of motion
for $r$ (\ref{equation_r}), $\sigma$ (\ref{equation_sigma}), $\phi$ (\ref{equation_phi}), and $\psi$ (\ref{equation_psi}) are reduced as follows:
\be r_{,tt}{\approx}r_{,xx}, \hphantom{dd} r_{,t}^2{\approx}r_{,x}^2;\ee
\be
\sigma_{,tt}\approx\sigma_{,xx}, \hphantom{dd} \psi_{,t}^2{\approx}\psi_{,x}^2,
\hphantom{dd} r_{,tt}{\approx}r_{,xx}, \hphantom{dd} \phi_{,t}^2{\approx}\phi_{,x}^2;
\ee
\be \phi_{,tt}\approx\phi_{,xx},\hphantom{dd}r_{,t}\phi_{,t}{\approx}r_{,x}\phi_{,x}; \ee
\be \psi_{,tt}\approx\psi_{,xx},\hphantom{dd}\phi_{,t}\psi_{,t}\approx\phi_{,x}\psi_{,x},\hphantom{dd}r_{,t}\psi_{,t}{\approx}r_{,x}\psi_{,x}. \ee
The above equations are like free scalar wave equations in flat spacetime. The derivatives of one variable ($r$, $\sigma$, $\phi$, and $\psi$) are independent from the derivatives of another. Arbitrary functions of $(t+x)$ or $(t-x)$ can satisfy the above equations, and in principle, the initial conditions right after the collision between the scalar fields and the inner horizon will decide which function each variable can take. On the other hand, we find that, as shown in Fig.~\ref{fig:asymptotics_null_slow_stage}, the constraint equations (\ref{constraint_eq_xt}) and (\ref{constraint_eq_xx_tt}) provide some useful information on the connections between some variables at the early stage of charge scattering:
\be r_{,t}\sigma_{,t}\approx4{\pi}r_{-}\frac{\psi_{,t}^2}{f'}.\ee

As shown in Fig.~\ref{fig:evolutions_null}, the quantities $r$, $\sigma$, $f'$, $\psi$, $|1-K^2|$, and $m$ change dramatically at the beginning of charge scattering where $r{\approx}r_{-}$. Note that, near the central singularity, $r$, $\sigma$, and $f'$ have approximate analytic expressions in terms of $\xi=t_{0}-t$, where $t_{0}$ is the time coordinate of the singularity curve. So it is natural to guess that, at the early stage of charge scattering, the above quantities may also have approximate analytic expressions of $\zeta=t-t_{s}$, where $t_{s}$ is a certain time value related to the early stage of charge scattering. We plot the evolutions of these quantities at the early stage of charge scattering in Fig.~\ref{fig:asymptotics_null_slow_stage}, from which one can see that $r$, $r_{,t}$, and $\sigma$ may have the following approximate analytic expressions:
\begin{align}
r&{\approx}r_{-},\label{r_asymptotic_slow}\\
\nonumber\\
r_{,t}&{\approx}a\zeta^{\lambda},\label{drdt_asymptotic_slow}\\
\nonumber\\
\sigma&{\approx}b\ln\zeta+\sigma_{0}.\label{sigma_asymptotic_slow}
\end{align}

In fact, a logarithmic expression for $\sigma$ is supported by its behavior near the inner horizon in the Reissner-Nordstr\"{o}m black hole case. From Eqs.~(\ref{r_RN_metric}) and (\ref{sigma_RN_metric}), one obtains that, as $r$ approaches the inner horizon $r=r_{-}$, in the case of $t{\gg}x$, $\sigma$ can be approximated by a logarithmic function of $t$. As shown in Fig.~\ref{fig:eom_null}(c), describing the terms in the equation of motion for $\sigma$ in charge scattering, at the very early stage ($t\approx1$) of the collision between the scalar fields and the inner horizon, compared to those from other terms, the contributions from the terms related to $\phi$ and $\psi$ are tiny. Therefore, at this stage, the evolution of $\sigma$ should not be much different from the corresponding one in the Reissner-Nordstr\"{o}m geometry.

As shown in Figs.~\ref{fig:asymptotics_null_slow_stage}(c) and \ref{fig:asymptotics_null_slow_stage}(d), $\phi$ and $\psi$ can be well fitted by power law functions of $\zeta$. Take $\phi$ as an example,
\be \phi\approx c\zeta^d+\phi_{0}.\label{phi_asymptotic_slow}\ee
We plot $(1-K^{2})$ in Fig.~\ref{fig:asymptotics_null_slow_stage}(d) and find that $\ln(1-K^{2})$ can be well fitted linearly with respect to $\zeta$,
\be \ln(1-K^{2}){\approx}f\zeta+h.\ee
Currently we do not have derivations for this linear relation. The good thing is that, in the mass function, $(1-K^2)$ is a minor factor. Therefore, as verified in Fig.~\ref{fig:asymptotics_null_slow_stage}(d), the mass function can be reduced to
\be
\begin{split}
m&=\frac{r}{2}\left[1+\frac{q^2}{r^2}+e^{2\sigma}(r_{,t}^2-r_{,x}^2)\right]\\
&{\sim}\frac{r_{-}}{2}{\cdot}e^{2\sigma}{\cdot}r_{,t}^2\\
&{\sim}\frac{r_{-}}{2}e^{2\sigma_{0}}\cdot\zeta^{2b}{\cdot}a^{2}\zeta^{2\lambda},
\end{split}
\ee
where $a$, $b$, $\lambda$, and $\sigma_{0}$ are defined in Eqs.~(\ref{drdt_asymptotic_slow}) and (\ref{sigma_asymptotic_slow}).

We list the fitting results below:
\begin{enumerate}[(i)]
  \item $r_{,t}{\approx}a(t+b)^c+d$, $a=(-1.71\pm0.01)\times10^{-3}$, $b=(-5.877\pm0.008)\times10^{-1}$, $c=7.880\pm0.005$,
        $d=(-2.639\pm0.003)\times10^{-4}$.
  \item $\sigma{\approx}a\ln(t+b)+c$, $a=34.37\pm0.03$, $b=(-1.72\pm0.01)\times10^{-1}$, $c=3.06\pm0.04$.
  \item $\phi{\approx}a(t+b)^c+d$, $a=(-2.7\pm0.6)\times10^{-3}$, $b=(-7.6\pm0.3)\times10^{-1}$, $c=7.4\pm0.2$, $d=(-1.3336\pm0.0001)\times10^{-1}$.
  \item $\psi{\approx}a(t+b)^c+d$, $a=(5.1\pm0.1)\times10^{-3}$, $b=(-2.91\pm0.06)\times10^{-1}$, $c=5.38\pm0.02$, $d=(1.21\pm0.02)\times10^{-3}$.
  \item $\ln(1-K^2){\approx}at+b$, $a\approx-8.73\pm0.01$, $b\approx12.05\pm0.02$.
  \item ${\ln}m{\approx}a{\ln}(t+b)+c$, $a=77.69\pm0.02$, $b=(-1.268\pm0.004)\times10^{-1}$, $c=-15.65\pm0.03$.
\end{enumerate}

\begin{figure*}[t!]
  \epsfig{file=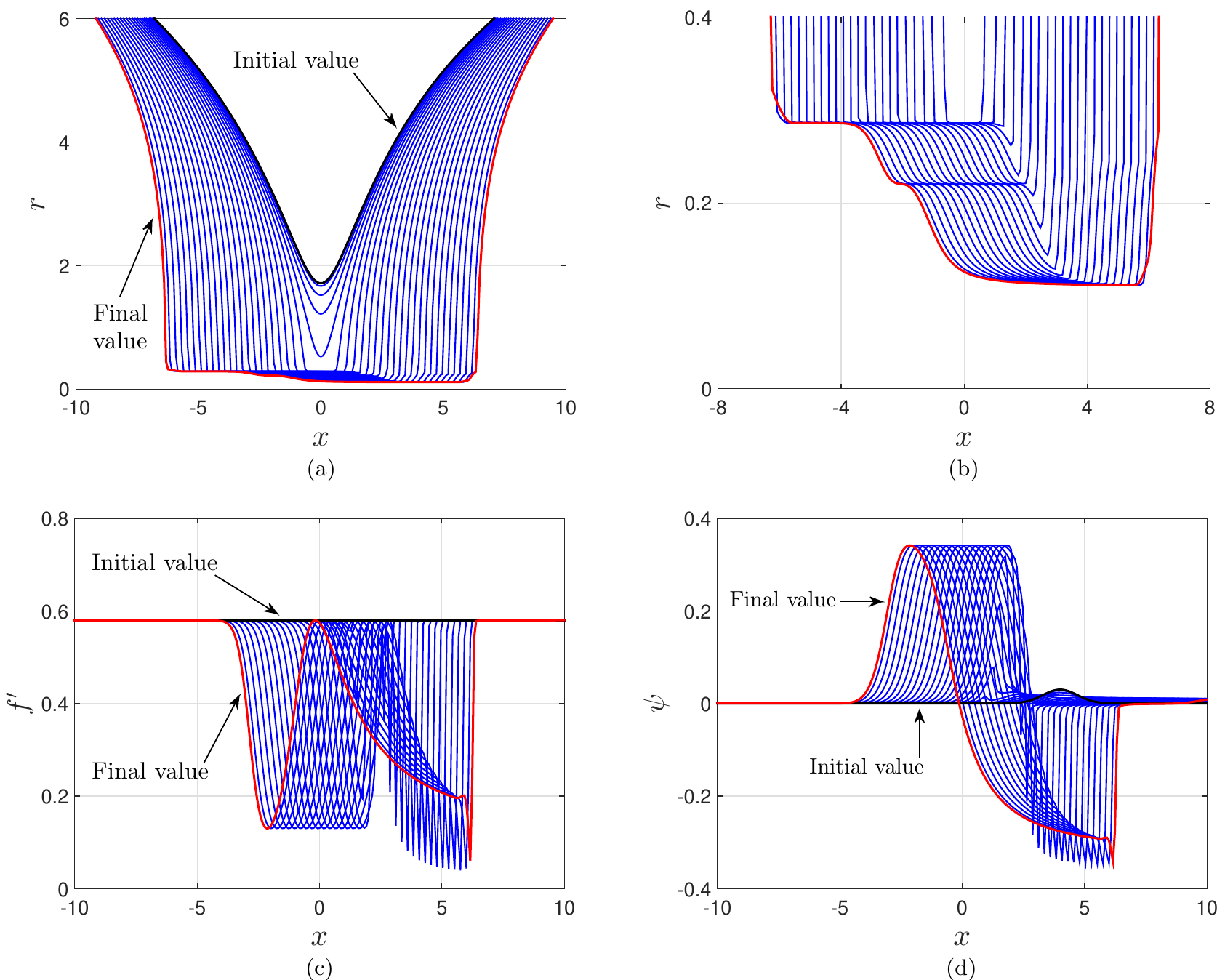, width=0.9\textwidth}
  \caption{Evolutions for charge scattering with a weak scalar field.
  (a)-(d): evolutions for $r$, $f'$, and $\psi$. The time interval between two consecutive slices is $120{\Delta}t=0.24$.
  The central singularity is not approached.}
  \label{fig:evolutions_weak}
\end{figure*}

\begin{figure*}[t!]
  \epsfig{file=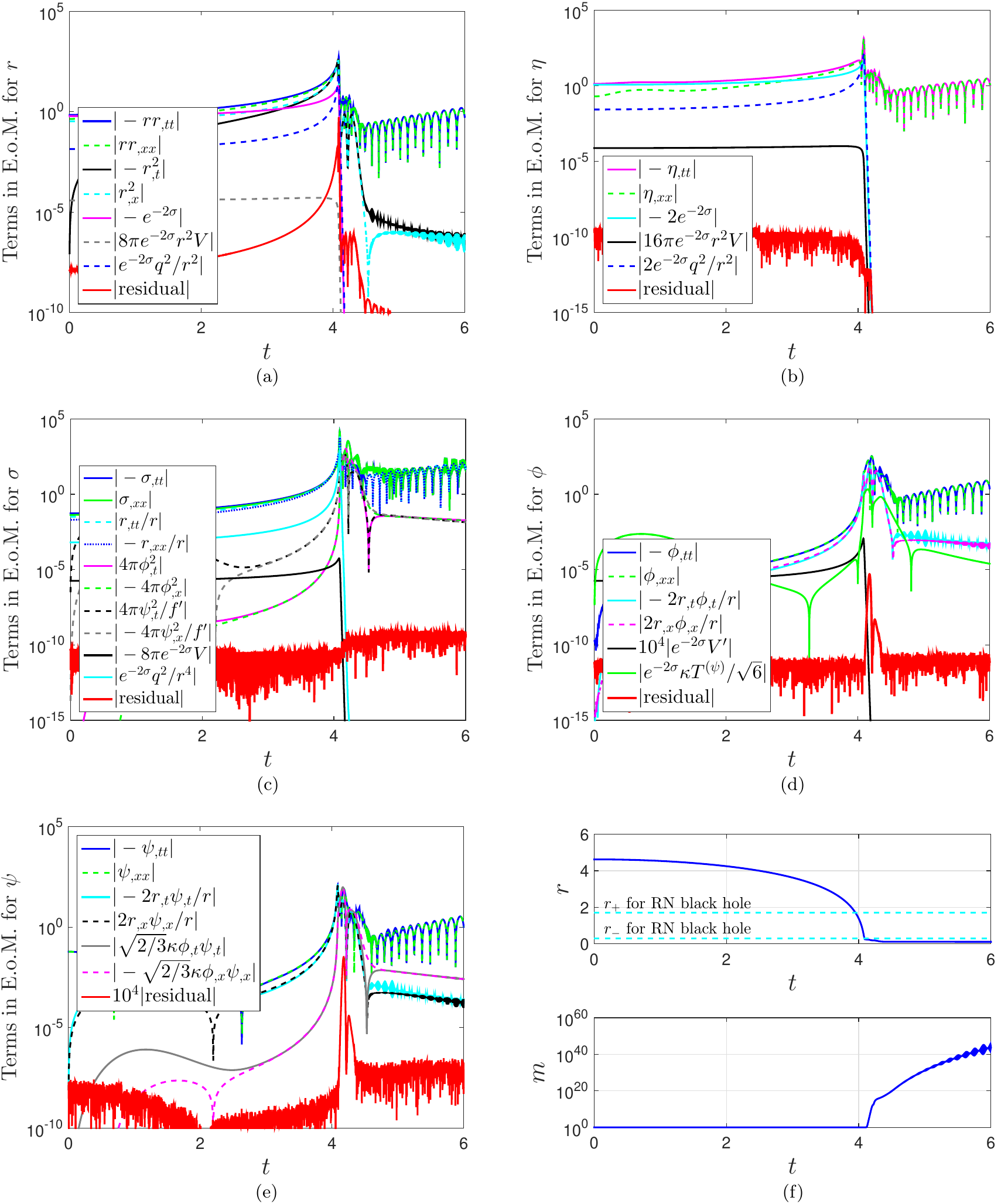, width=0.9\textwidth}
  \caption{(color online). Dynamics and evolutions on the slice $(x=4,t=t)$ in weak scattering.
  (a)-(e): dynamical equations for $r$, $\eta$, $\sigma$, $\phi$, and $\psi$.
  (f) evolutions of $r$ and $m$.}
  \label{fig:weak_scattering_eom}
\end{figure*}

\begin{figure*}[t!]
  \epsfig{file=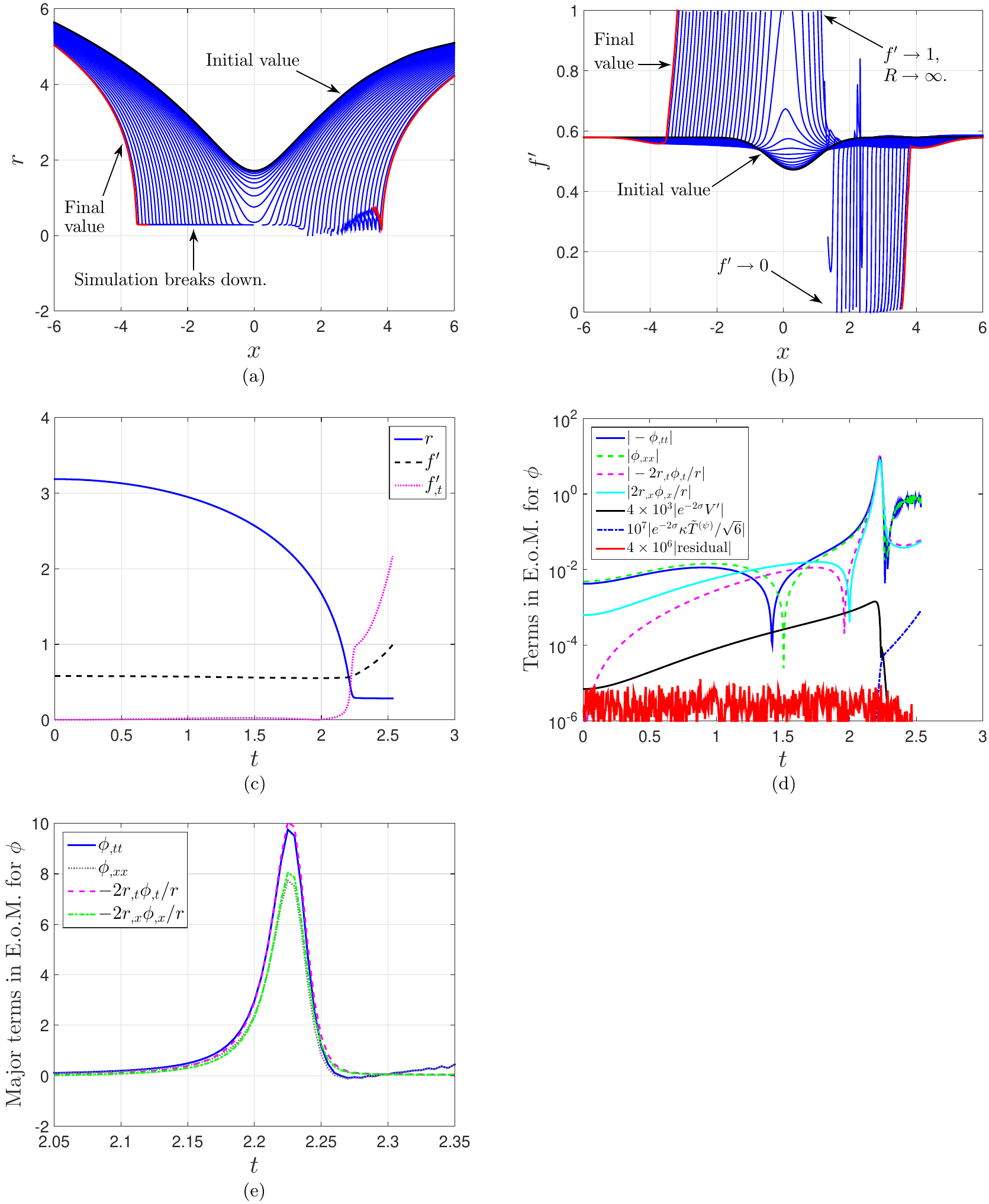, width=0.9\textwidth}
  \caption{(color online). A singularity problem in charge scattering for the Hu-Sawicki model~(\ref{f_R_Hu_Sawicki}), $f(R)=R-DR_{0}R/(R+R_{0})$.
   (a) and (b): evolutions of $r$ and $f'$. The time interval between two consecutive slices is $20{\Delta}t=0.01$.
   (c) evolutions of $r$, $f'$, and $f'_{,t}$ on the the slice $(x=-2,t=t)$.
   (d) and (e): dynamical equation for $\phi$ on the slice $(x=-2,t=t)$.
   In (e), as the inner horizon is approached, $\phi_{,tt}\approx-2r_{,t}\phi_{,t}$. This equation describes a positive feedback, since $-2r_{,t}/r$ is positive. As a result, when $\phi_{,t}$ is positive, $\phi$ can be accelerated to zero rapidly. Correspondingly, $f'$ goes to $1$ as plotted in (b) and (c), and the Ricci scalar $R$ becomes singular. Then the simulation breaks down as shown in (a).}
  \label{fig:Ricci_singularity}
\end{figure*}

\begin{figure*}[t!]
  \epsfig{file=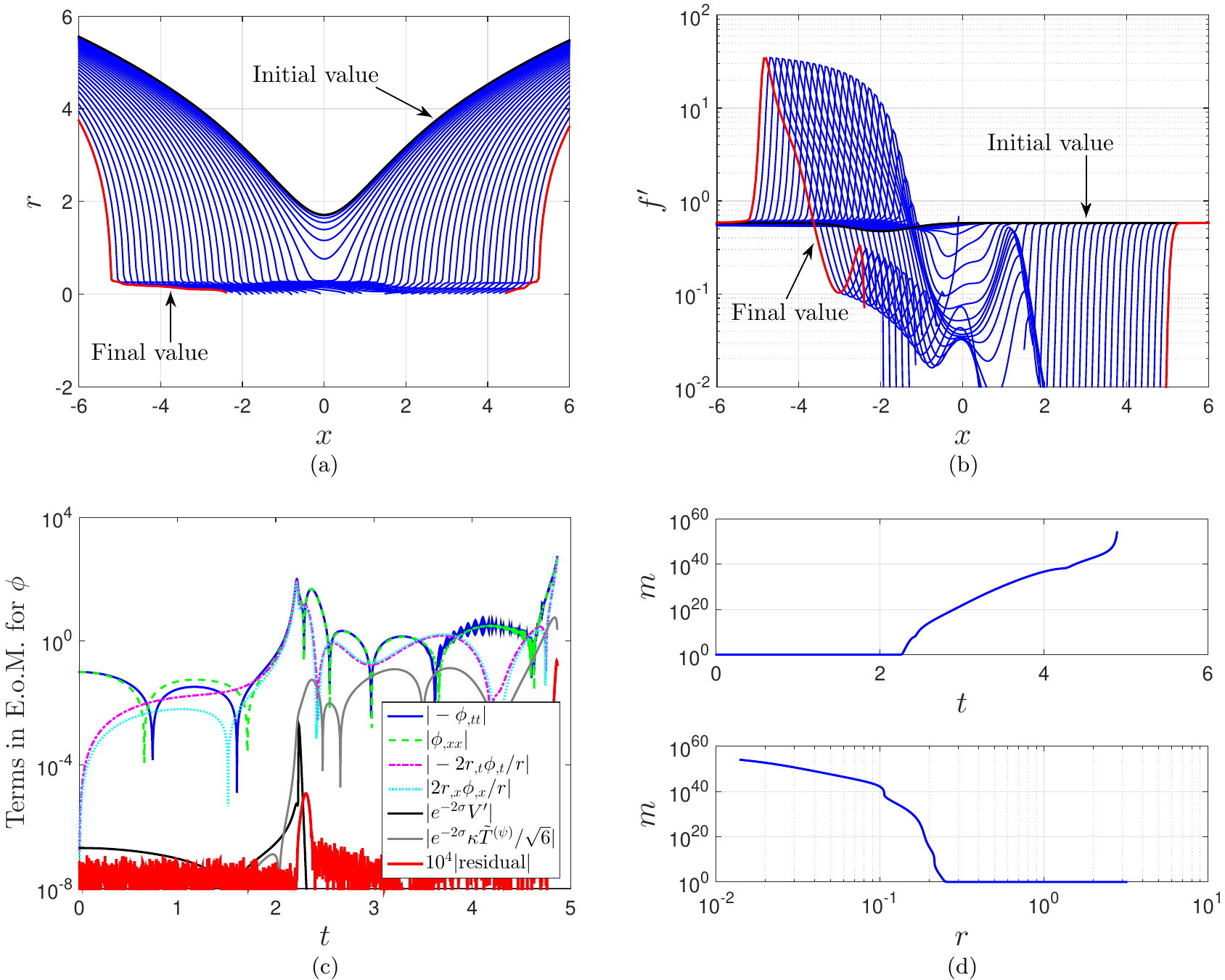, width=0.9\textwidth}
  \caption{(color online). Avoidance of the singularity problem in charge scattering for the combined model~(\ref{f_R_combined_charge}), $f(R)=R-DR_{0}R/(R+R_{0})+{\alpha}R^2$.
  (a) and (b): evolutions of $r$ and $f'$. The time interval between two consecutive slices is $60{\Delta}t=0.15$.
  (c) dynamical equation for $\phi$ on the slice $(x=-2,t=t)$.
  (d) evolution of $m$ on the slice $(x=-2,t=t)$.}
  \label{fig:Ricci_singularity_avoidance}
\end{figure*}

\section{Weak scalar charge scattering\label{sec:weak_scattering}}
In this section, we consider charge scattering with a weak scalar field. Parameter settings in this section are almost the same as those in the last section with the following exceptions:
\begin{enumerate}[(i)]
  \item Physical scalar field:
  $\psi(x,t)|_{t=0}=a\cdot\exp\left[-(x-x_{0})^2/b\right]$, $a=0.03$, $b=1$, and $x_{0}=4$.
  \item Grid. Spatial range: $x\in[-12~12]$. Grid spacings: ${\Delta}x={\Delta}t=0.002$.
\end{enumerate}

As discussed in the above section, in some spacetime regions where the scalar field is strong, the inner horizon can contract to zero volume, and the central singularity becomes spacelike. However, this does not always necessarily happen. After all, it takes energy for the inner horizon to contract. When the scalar field carries less energy, the inner horizon may only contract to a nonzero value. This is confirmed by our numerical results plotted in Fig.~\ref{fig:evolutions_weak}. The evolution of $\sigma$ is similar to that in strong scalar field case and is skipped. These results are in agreement with the mathematical proof in Ref.~\cite{Dafermos_2014} and the numerical work in Ref.~\cite{Hansen_2005}. Since in this case the inner horizon is not totally destructed, one needs to reconsider whether the strong cosmic censorship conjecture is valid here. In addition, in Ref.~\cite{Hong_0808}, it was argued that, when Hawking radiation is taken into account, this censorship may also be violated. Note that in Ref.~\cite{Kawai_1409} the interior of a Schwarzschild black hole was also discussed with the backreaction from the Hawking radiation being taken into account.

The dynamics for the quantities $r$, $\eta$, $\sigma$, $\phi$, and $\psi$ are plotted in Figs.~\ref{fig:weak_scattering_eom}(a)-\ref{fig:weak_scattering_eom}(e), respectively. The numerical results show that, at the late stage, the field equations for such quantities become null, in the sense that the temporal and spatial derivatives are almost equal, i.e., $\sigma_{,tt}\approx\sigma_{,xx}$. Moreover, the derivatives have oscillations. As shown in Fig.~\ref{fig:weak_scattering_eom}(f), the mass function keeps growing even as $r$ approaches a constant value. Further details are skipped.

\section{Dark energy $f(R)$ singularity problem in charge scattering\label{sec:singularity}}
Up to this point, in the numerical simulations of charge scattering that we have implemented in this paper, the scalar degree of freedom $f'$ asymptotes to zero as the center is approached. Note that, as discussed in Sec.~\ref{sec:neutral_scattering}, in neutral scattering in dark energy $f(R)$ gravity, when the initial velocity or acceleration of $f'$ is large enough, $f'$ can go to $1$ before the central singularity is approached. Consequently, the Ricci scalar $R$ goes to infinity, and the simulation breaks down. Next we will show that such a problem also happens in charge scattering.

In the simulation, the values of the parameters are the same as those described at the beginning of Sec.~\ref{sec:results}, including grid spacings ${\Delta}x={\Delta}t=0.005$. However, the initial value for $\phi$ takes the following format: $\phi(x,t)|_{t=0}=a\exp[{-(x-x_{0})^2}]+\phi_{0}$,
with $V'(\phi_{0})=0$, $a=-0.05$, and $x_{0}=0.3$. We plot the numerical results in Fig.~\ref{fig:Ricci_singularity}. As shown in Fig.~\ref{fig:Ricci_singularity}(b), as the inner horizon is approached, $f'$ goes to $1$, and the Ricci scalar $R$ becomes singular. The simulation breaks down.

Now we explore the causes of this singularity problem. As an example, in Figs.~\ref{fig:Ricci_singularity}(d) and \ref{fig:Ricci_singularity}(e), we plot the terms in the dynamical equation for $\phi$ on the slice $(x=-2,t=t)$, and find that initially because of the contribution from $\phi_{,xx}$, $\phi_{,t}$ changes from $\phi_{,t}|_{t=0}=0$ to
\\
\\
\\
$\phi_{,t}>0$ at late time. Near the inner horizon, there is
\be \left(\phi_{,tt}\approx-\frac{2}{r}r_{,t}\phi_{,t}\right)\approx\left(\phi_{,xx}\approx-\frac{2}{r}r_{,x}\phi_{,x}\right).
\label{positive_feedback_singularity}\ee
Then $\phi$ is accelerated by gravity. After the inner horizon is met, there is $\phi_{,tt}\approx\phi_{,xx}>0$. Eventually, $\phi$, $f'$, and $R$ go to $0$, $1$, and $+\infty$, respectively. The simulation breaks down.

Similar to neutral scattering, the singularity problem can be avoided by adding an $R^2$ term to the Hu-Sawicki model,
\be f(R) = R-\frac{DR_{0} R}{R+R_{0}}+{\alpha}R^2. \label{f_R_combined_charge}\ee
The parameters take the same values as in the last subsection with the following exceptions:
\begin{enumerate}[(i)]
  \item Grid spacings: ${\Delta}x={\Delta}t=0.0025$.
  \item Physical scalar field $\psi$:
  $\psi(x,t)|_{\scriptsize{t=0}}=a\cdot\exp\left[-(x-x_{0})^2/b\right]$, $a=0.05$, $b=1$, and $x_{0}=2$.
  \item Scalar degree of freedom $\phi$:
  $\phi(x,t)|_{t=0}=a\cdot\exp[-(x-x_{0})^2]+\phi_{0}$, with $V'(\phi_{0})=0$, $a=-0.05$, and $x_{0}=-2$.
  \item $f(R)$ model (\ref{f_R_combined_charge}): $D=1.2$, $R_{0}=10^{-5}$, and $\alpha=1$.
\end{enumerate}

The numerical results for charge scattering for this modified model are plotted in Fig.~\ref{fig:Ricci_singularity_avoidance}. As shown in Fig.~\ref{fig:Ricci_singularity_avoidance}(b), for this model, $f'$ can cross $1$ without difficulty. The simulation can run smoothly.

\section{Summary\label{sec:summary}}
In this paper, we studied scalar collapses in flat, Schwarzschild, and Reissner-Nordstr\"{o}m geometries in $f(R)$ gravity numerically. Approximate analytic solutions for different types of collapses were partially obtained. One dark energy $f(R)$ singularity problem was discussed. We summarize our work on computational and physical issues separately below.

\subsection{Computational issues}
\begin{enumerate}[(i)]
  \item \emph{The Jordan frame vs the Einstein frame.} The field equations for $f(R)$ gravity in the Jordan frame are more complex than those in general relativity. Therefore, for ease of computation, we transform $f(R)$ gravity from the Jordan frame into the Einstein frame, in which the formalism can be formally treated as Einstein gravity coupled to a scalar field.

  \item \emph{$dudv$ vs $(-dt^2+dx^2)$ in double-null coordinates.} In the studies of mass inflation, the $dudv$ format of the Kruskal-like coordinates, $ds^{2} = 4e^{-2\sigma}dudv+r^2d\Omega^2$, is usually used. In the field equations, many terms are mixed derivatives of $u$ and $v$, e.g., $r_{,uv}$.
      In this paper, we used the $(-dt^2+dx^2)$ format instead, $ds^{2}=e^{-2\sigma}(-dt^2+dx^2)+r^2d\Omega^2$, with $u=(t-x)/2=\text{const}$ and $v=(t+x)/2=\text{const}$. In the $(t,x)$ line element, one coordinate is timelike, and the rest are spacelike. We are used to this setup. It is more convenient and more intuitive to use this set of coordinates. Moreover, for the $(t,x)$ choice, spatial and temporal derivatives are usually separated, e.g., $(r_{,tt}-r_{,xx})$.

      We set the initial conditions close to those in a Reissner-Nordstr\"{o}m geometry. With this setup, it is convenient to test the code. Removing the terms related to the scalar fields, we can test our code by comparing the numerical results to the analytic ones in a Reissner-Nordstr\"{o}m geometry. Moreover, by comparing numerical results for charge scattering to the dynamics in the Reissner-Nordstr\"{o}m geometry, we can obtain intuitions as to how the scalar fields affect the geometry.

  \item \emph{Cauchy horizon: infinite or local regions?} As implied by Eq.~(\ref{r_RN_metric}), the exact inner horizon $r=r_{-}$ is at the regions where $uv$ and $(t^2-x^2)$ are infinite. However, $r$ still can be very close to the inner horizon even when $uv$ and $(t^2-x^2)$ take moderate values. Consequently, at regions where $uv$ and $(t^2-x^2)$ take some moderate values, the scalar fields and the inner horizon still can have strong interactions, resulting in mass inflation.
\end{enumerate}

\subsection{Physical issues}
\begin{enumerate}[(i)]
  \item \emph{Scalar collapse in $f(R)$ gravity vs scalar collapse in general relativity.} In scalar collapse, the scalar degree of freedom $\phi(\equiv\sqrt{3/2}\ln{f'}/\sqrt{8{\pi}G})$ plays a similar role as a physical scalar field in general relativity. Regarding the physical scalar field in $f(R)$ case, when $\phi_{,t}$ is negative (positive), the physical scalar field is suppressed (magnified) by $\phi$.

  \item \emph{The inner horizon in a Reissner-Nordstr\"{o}m black hole vs the central singularity in a Schwarzschild black hole.} These two share some similarities.

     For Reissner-Nordstr\"{o}m and Schwarzschild black holes, throughout the whole spacetime, the Misner-Sharp mass function is constant. When a scalar field impacts the inner horizon of a Reissner-Nordstr\"{o}m black hole, the scalar field can modify the geometry in the vicinity of the inner horizon significantly, especially on $r_{,t}$. The inner horizon contracts and mass inflation takes place. In neutral scalar collapse toward a Schwarzschild black hole formation, the scalar field can also modify the geometry in the vicinity of the central singularity dramatically, especially on the metric component $\sigma$~\cite{Guo_1312}. Then mass inflation also happens.

     The Belinskii, Khalatnikov, and Lifshitz (BKL) conjecture is an important result on dynamics in the vicinity of a spacelike singularity~\cite{Belinskii_1970,Belinskii_1973,Belinski_1404,einstein_online}. The first statement of this conjecture is that as the singularity is approached, the dynamical terms dominate the spatial terms in the field equations. In other words, the way gravity changes over time is more important than the variation of the gravitational field from one location to the next~\cite{einstein_online}. We would like to say that, to a large extent, later evolutions in a strong gravitational field largely erase away the initial information on the connections between neighboring points. As discussed in Ref.~\cite{Guo_1312} and also in this paper, in double-null coordinates, using the above argument, one can interpret the following behaviors displayed in numerical simulations: near the central singularity of a Schwarzschild black hole and also near the inner horizon of a Reissner-Nordstr\"{o}m black hole, there are
     \be \frac{\psi_{,x}}{\psi_{,t}}{\approx}\frac{r_{,x}}{r_{,t}}<1,\label{ratio_derivatives}\ee
     \be \phi_{,tt}{\approx}-\frac{2r_{,t}}{r}\phi_{,t}.\label{positive_feedback}\ee
  In this paper, it was shown that Eq.~(\ref{ratio_derivatives}) can explain the causes of mass inflation, while Eq.~(\ref{positive_feedback}) can explain the dark energy $f(R)$ singularity problem in collapse.

  The second and third statements of the BKL conjecture are that i) the metric terms will dominate the matter field terms, while the matter field may not be negligible if it is a scalar field; ii) the dynamics of the metric components and the matter fields is described by the Kasner solution. These two statements were confirmed in simulations of neutral scalar collapse in $f(R)$ gravity in Ref.~\cite{Guo_1312} and in general relativity in Ref.~\cite{Guo_1507}. The second statement was also verified in charge scattering in this paper. However, the third statement on Kasner solution may not apply to the dynamics near the inner horizon in charge scattering.

  \item \emph{Compact stars vs black holes in $f(R)$ gravity.} The internal structure of compact stars is usually in an equilibrium state and is static. In $f(R)$ gravity, inside compact stars, the scalar degree of freedom, $f'$, can be coupled to the energy density of the stars and then is not very free to move. However, due to strong gravity, the internal structure of black holes is dynamical. $f'$ and the matter fields are decoupled. As a result, $f'$ is more free to move than in the compact stars case. It can keep increasing or decreasing until singularities are met.

  \item \emph{Dark energy $f(R)$ singularity problem: cosmology (or static compact objects) vs black hole physics.} In dark energy $f(R)$ gravity, the Ricci scalar $R$ can be singular in both cosmology and black hole physics. We consider a homogeneous cosmological model. Using the flat Friedmann-Robertson-Walker metric,
      \be ds^2=-dt^2+a^2(t)d\vec{x}^2,\ee
      the equation of motion for $f'$ is
      \be \ddot{f''}+3H\dot{f'}+U(f')+\frac{8\pi}{3}T=0,\ee
      where $H$ is the Hubble parameter. Due to the finiteness of the potential barrier, the force from a perturbation of $8{\pi}T/3$ may push $f'$ to $1$. Correspondingly, the Ricci scalar goes to singularity~\cite{Frolov_2008}. In the black hole case that we discussed in this paper, the equation of motion for $\phi({\equiv}\sqrt{3/2}{\ln}f'/\sqrt{8{\pi}G})$ has more complex structure~(\ref{equation_phi}),
     \be
     \begin{split}
     &-\phi_{,tt}+\phi_{,xx}+\frac{2}{r}(-r_{,t}\phi_{,t}+r_{,x}\phi_{,x})\\
     &=e^{-2\sigma}\left[V'(\phi)+\frac{1}{\sqrt{6}}\kappa T^{(\psi)}\right].
     \end{split}
     \ee
     As the central singularity of a Schwarzschild black hole or the inner horizon of a Reissner-Nordstr\"{o}m black hole is approached, the above equation can be simplified as~(\ref{positive_feedback}), and gravity from the black hole, $-2r_{,t}\phi_{,t}/r$, can cause a similar singularity problem as in cosmology or static compact stars.

  \item \emph{The combined $f(R)$ model and the $R^2$ model: singular or non-singular?} At the center of a Schwarzschild black hole and in the very early Universe, the tidal forces are singular, and general relativity fails. Taking into account quantum-gravitational effects, Starobinsky obtained an $R^2$ model, $f(R)=R+{\alpha}R^2$. This model has a non-singular de Sitter solution, which is unstable both to the past and to the future~\cite{Starobinsky_1980,Vilenkin_1985}. In Sec.~\ref{sec:neutral_scattering}, scalar collapse in a Schwarzschild geometry for the combined model (a combination of dark energy model and $R^2$ model) was explored. A new Schwarzschild black hole, including a new central singularity, can be formed. Moreover, under certain initial conditions, $f'$ and $R$ can be pushed to infinity as the central singularity is approached. In Ref.~\cite{Guo_2015}, scalar collapse in flat geometry for the $R^2$ model was simulated. Similar results were obtained. Namely, the classical singularity problem, which is present in general relativity, remains in collapse in these models.

  \item \emph{Inside vs outside black holes: local vs global.} Throughout the whole spacetime of stationary Schwarzschild and Reissner-Nordstr\"{o}m black holes, the Misner-Sharp mass function is equal to the black hole mass. For a gravitational collapsing system, at asymptotic flat regions, the mass function describes the total mass of the dynamical system. However, in this system, near the central singularity of a Schwarzschild black hole or near the inner horizon of a Reissner-Nordstr\"{o}m black hole, the dynamics is local. Then the mass function does not provide global information on the mass of the collapsing system.
\end{enumerate}

In summary, in this paper, we studied scalar collapses in flat, Schwarzschild, and Reissner-Nordstr\"{o}m geometries in $f(R)$ gravity. For convenience and intuitiveness, in simulating scalar collapses in Schwarzschild and Reissner-Nordstr\"{o}m geometries, Kruskal and Kruskal-like coordinates were used, respectively. Approximate analytic solutions for different types of collapses were partially obtained. Causes and avoidance of a dark energy $f(R)$ singularity problem in collapse were discussed.

\section*{Acknowledgments}
The authors are grateful to Andrei V. Frolov, Jos\'{e} T. G\'{a}lvez Ghersi, Ken-ichi Nakao, Dong-han Yeom, and the referees for the helpful discussions and comments. JQG would like to thank Simon Fraser University where part of this work was done.

\appendix*
\section{Equations of motion for a physical scalar field and $f'$ in the Einstein frame\label{sec:appendix:EoM_EM}}
In this appendix, based on the equations of motion for a massless physical scalar field $\psi$ and the scalar degree of freedom $f'$ in the Jordan frame for $f(R)$ gravity, we derive the corresponding equations in the Einstein frame. In the transformation from the Jordan frame into the Einstein frame,
we use $\tilde g_{\mu\nu}=\chi\cdot g_{\mu\nu}$ and $\kappa\phi\equiv\sqrt{3/2}\ln\chi$, where a tilde denotes that the quantity is in the Einstein frame.

For a scalar field $\psi$, the first covariant derivatives of $\psi$ are equal in two frames, since they are both equal to the partial derivative~\cite{Carroll_2003}:
\be \tilde{\nabla}_{\mu}\psi=\nabla_{\mu}\psi=\partial_{\mu}\psi.\ee
Then for the first contravariant derivative of $\psi$, there is
\be \nabla^{\mu}\psi=g^{\mu\nu}\nabla_{\nu}\psi=\chi\cdot\tilde{\nabla}^{\mu}\psi.\ee
For $\Box\psi$, we have~\cite{Landau_1971}
\be
\begin{split}
\Box\psi &= \frac{1}{\sqrt{|g|}}\partial_\mu \left(\sqrt{|g|}g^{\mu\nu}\partial_\nu \psi\right) \\
  &= \frac{1}{\sqrt{|\tilde g|\cdot \chi^{-4}}}\partial_\mu \left(\sqrt{|\tilde g|\cdot \chi^{-4}}\cdot\chi\cdot\tilde g^{\mu\nu}\partial_\nu \psi\right)\\
  &= \chi\big[\tilde \Box \psi - \tilde g^{\mu\nu} \partial_\mu \psi\partial_\nu (\ln\chi)\big]\\
  &= \chi\left(\tilde \Box\psi -\sqrt{\frac{2}{3}} ~\kappa \tilde g^{\mu\nu} \partial_\mu \phi\partial_\nu \psi \right).
\end{split}
\label{Box_psi_JF_EF}
\ee
In the Jordan frame, for a massless scalar field $\psi$, there is
\be \Box\psi=0,\label{eom_psi_appendix}\ee
Combining Eqs.~(\ref{Box_psi_JF_EF}) and (\ref{eom_psi_appendix}) yields the equation of motion for $\psi$ in the Einstein frame (\ref{fR:Box_psi}).

In the special case of $\psi=\chi\equiv\exp(\sqrt{2/3}\kappa\phi)$, there is
\be \Box\chi=\sqrt{\frac{2}{3}}\kappa\chi^{2}\tilde{\Box}\phi.\label{Box_chi_JF_EF}\ee
Combining Eqs.~(\ref{v_prime}), (\ref{trace_eq2}), (\ref{T_JF_EF}), (\ref{V_prime_EF}), and (\ref{Box_chi_JF_EF}) gives the equation of motion for $\phi$ in the Einstein frame (\ref{fR:Box_phi}).



\begin{thebibliography}{99}
\bibitem{Burko_1997_book}
{\it Internal Structure of Black Holes and Spacetime Singularities,}
edited by L. M. Burko and A. Ori.
(Institute of Physics Publishing, Bristol, UK; and The Israel Physical Society, Jerusalem, Israel. 1998).

\bibitem{Brady_1999}
P. R. Brady,
{\it ``The Internal Structure of Black holes,''}
Prog. Theor. Phys. Suppl. {\bf 136}, 29 (1999).

\bibitem{Berger_2002}
B. K. Berger,
{\it ``Numerical Approaches to Spacetime Singularities,''}
Living Rev. Relativity {\bf 5}, 1 (2002).
[\arXiv{gr-qc/0201056}]

\bibitem{Joshi_2007}
P. S. Joshi,
{\it Gravitational Collapse and Spacetime Singularities}
(Cambridge University Press, Cambridge, UK, 2007).

\bibitem{Henneaux_2008}
M. Henneaux, D. Persson, and P. Spindel,
{\it ``Spacelike Singularities and Hidden Symmetries of Gravity,''}
Living Rev. Relativity {\bf 11}, 1 (2008).
[\arXiv[hep-th]{0710.1818}]

\bibitem{Price}
R. H. Price,
{\it ``Nonspherical Perturbations of Relativistic Gravitational Collapse. I. Scalar and Gravitational Perturbations,''}
Phys. Rev. D {\bf 5}, 2419 (1972).

\bibitem{Clifton_1106}
T. Clifton, P. G. Ferreira, A. Padilla, and C. Skordis,
{\it ``Modified gravity and cosmology,''}
Physics Reports {\bf 513} (2012) 1.
[\arXiv[astro-ph]{1106.2476}]

\bibitem{Sotiriou_0805}
T. P. Sotiriou and V. Faraoni,
{\it ``f(R) Theories Of Gravity,''}
Rev. Mod. Phys. {\bf 82}, 451 (2010).
[\arXiv[gr-qc]{0805.1726}]

\bibitem{Felice_1002}
A. D. Felice and S. Tsujikawa,
{\it ``f(R) Theories,''}
Living Rev. Relativity {\bf 13}, 3 (2010).
[\arXiv[gr-qc]{1002.4928}]

\bibitem{Nojiri_1002}
S. Nojiri and S. D. Odintsov,
{\it ``Unified cosmic history in modified gravity: from F(R) theory to Lorentz non-invariant models,''}
Phys. Rept. {\bf 505}, 59 (2011).
[\arXiv[gr-qc]{1011.0544}]

\bibitem{Capozziello_1108}
S. Capozziello and M. De Laurentis,
{\it ``Extended Theories of Gravity,''}
Phys. Rept. {\bf 509}, 167 (2011).
[\arXiv[gr-qc]{1108.6266}]

\bibitem{Cruz_0907}
A. de la Cruz-Dombriz, A. Dobado, and A. L. Maroto,
{\it ``Black Holes in f(R) theories,''}
Phys. Rev. D {\bf 80}, 124011 (2009).
[\arXiv[gr-qc]{0907.3872}]

\bibitem{Olmo_1110}
G. J. Olmo and D. Rubiera-Garcia,
{\it ``Palatini f(R) Black Holes in Nonlinear Electrodynamics,''}
Phys. Rev. D {\bf 84}, 124059 (2011).
[\arXiv[gr-qc]{1110.0850}]

\bibitem{Nojiri_1301}
S. Nojiri and S. D. Odintsov,
{\it ``Anti-Evaporation of Schwarzschild-de Sitter Black Holes in F(R) gravity,''}
Classical Quantum Gravity {\bf 30}, 125003 (2013).
[\arXiv[hep-th]{1301.2775}]

\bibitem{Sebastiani_1305}
L. Sebastiani, D. Momeni, R. Myrzakulov, and S.D. Odintsov
{\it ``The instabilities and (anti)-evaporation of Schwarzschild-de Sitter black holes in modified gravity,''}
Phys. Rev. D {\bf 88}, 104022 (2013).
[\arXiv[gr-qc]{1305.4231}]

\bibitem{Nojiri_1405}
S. Nojiri and S. D. Odintsov,
{\it ``Instabilities and Anti-Evaporation of Reissner-Nordstr\"{o}m Black Holes in modified F(R) gravity,''}
Phys. Lett. B {\bf 735}, 376 (2014).
[\arXiv[gr-qc]{1405.2439}]

\bibitem{Avelino_2009}
P. P. Avelino, A. J. S. Hamilton, and C. A. R. Herdeiro,
{\it ``Mass Inflation in Brans-Dicke gravity,''}
Phys. Rev. D {\bf 79}, 124045 (2009).
[\arXiv[gr-qc]{0904.2669}]

\bibitem{Borkowska_2011}
A. Borkowska, M. Rogatko, and R. Moderski,
{\it ``Collapse of Charged Scalar Field in Dilaton Gravity,''}
Phys. Rev. D {\bf 83}, 084007 (2011).
[\arXiv[gr-qc]{1103.4808}]

\bibitem{Hwang_1110}
D.-i. Hwang, B.-H. Lee, and D.-h. Yeom,
{\it ``Mass inflation in f(R) gravity: A conjecture on the resolution of the mass inflation singularity,''}
J. Cosmol. Astropart. Phys. {\bf 12} (2011) 006.
[\arXiv[astro-ph]{1110.0928}]

\bibitem{Guo_1312}
J.-Q. Guo, D. Wang, and A. V. Frolov,
{\it ``Spherical collapse in $f(R)$ gravity and the Belinskii-Khalatnikov-Lifshitz conjecture,''}
Phys. Rev. D {\bf 90}, 024017 (2014).
[\arXiv[gr-qc]{1312.4625}]

\bibitem{Simpson_1973}
M. Simpson and R. Penrose,
{\it ``Internal instability in a Reissner–Nordstr\"{o}m black hole,''}
Int. J. Theor. Phys. {\bf 7}, 183 (1973).

\bibitem{Dafermos_2003}
M. Dafermos,
{\it ``Stability and instability of the Cauchy horizon for the spherically symmetric Einstein-Maxwell-scalar field equations,''}
Ann. Math. (N. Y.) {\bf 158}, 875 (2003).

\bibitem{Dafermos_2014}
M. Dafermos,
{\it ``Black holes without spacelike singularities,''}
Commun. Math. Phys. {\bf 332}, 729 (2014).
[\arXiv[gr-qc]{1201.1797}]

\bibitem{Ringstrom_2015}
H. Ringstr\"{o}m,
{\it ``Origins and development of the Cauchy problem in general relativity,''}
Classical Quantum Gravity {\bf 32}, 124003 (2015).

\bibitem{Isenberg_2015}
J. Isenberg,
{\it ``On Strong Cosmic Censorship,''}
\arXiv[gr-qc]{1505.06390}

\bibitem{Poisson_1989}
E. Poisson and W. Israel,
{\it ``Inner-horizon instability and mass inflation in black holes,''}
Phys. Rev. Lett. {\bf 63}, 1663 (1989).

\bibitem{Poisson_1990}
E. Poisson and W. Israel,
{\it ``Internal structure of black holes,''}
Phys. Rev. D {\bf 41}, 1796 (1990).

\bibitem{Barrabes_1990}
C. Barrabes, W. Israel, and E. Poisson
{\it ``Collision of light-like shells and mass inflation in rotating black holes,''}
Classical Quantum Gravity {\bf 7}, L273 (1990).

\bibitem{Gnedin_1991}
N. Yu. Gnedin and M. L. Gnedina,
{\it ``Instability of the internal structure of a Reissner-Nordstr\"{o}m black hole,''}
Sov. Astron. {\bf 36}, 296 (1992) [Astron. Zh. {\bf 69}, 584 (1992)].

\bibitem{Gnedin_1993}
M. L. Gnedin and N. Y. Gnedin,
{\it ``Destruction of the Cauchy horizon in the Reissner-Nordstrom black hole,''}
Classical Quantum Gravity {\bf 10}, 1083 (1993).

\bibitem{Brady_1995}
P. R. Brady and J. D. Smith,
{\it ``Black hole singularities: a numerical approach,''}
Phys. Rev. Lett. {\bf 75}, 1256 (1995).
[\arXiv{gr-qc/9506067}]

\bibitem{Burko_1997}
L. M. Burko,
{\it ``Structure of the black hole's Cauchy horizon singularity,''}
Phys. Rev. Lett. {\bf 79}, 4958 (1997).
[\arXiv{gr-qc/9710112}]

\bibitem{Burko_1997b}
L. M. Burko and A. Ori,
{\it ``Late-time evolution of nonlinear gravitational collapse,''}
Phys. Rev. D {\bf 56}, 7820 (1997).
[\arXiv{gr-qc/9703067}]

\bibitem{Hansen_2005}
J. Hansen, A. Khokhlov, and I. Novikov,
{\it ``Physics of the interior of a spherical, charged black hole with a scalar field,''}
Phys. Rev. D {\bf 71}, 064013 (2005).
[\arXiv{gr-qc/0501015}]

\bibitem{Hod_1997}
S. Hod and T. Piran,
{\it ``Mass Inflation in Dynamical Gravitational Collapse of a Charged Scalar Field,''}
Phys. Rev. Lett. {\bf 81}, 1554 (1998).
[\arXiv{gr-qc/9803004}]

\bibitem{Oren_2003}
Y. Oren and T. Piran,
{\it ``Collapse of charged scalar fields,''}
Phys. Rev. D {\bf 68}, 044013 (2003).
[\arXiv{gr-qc/0306078}]

\bibitem{Burko_1998}
L. M. Burko and A. Ori,
{\it ``Analytic study of the null singularity inside spherical charged black holes,''}
Phys. Rev. D {\bf 57}, R7084 (1998).
[\arXiv{gr-qc/9711032}]

\bibitem{Burko_1999}
L. M. Burko,
{\it ``Strength of the null singularity inside black holes,''}
Phys. Rev. D {\bf 60}, 104033 (1999).
[\arXiv{gr-qc/9907061}]

\bibitem{Hu_0705}
W. Hu and I. Sawicki,
{\it ``Models of f(R) Cosmic Acceleration that Evade Solar-System Tests,''}
Phys. Rev. D {\bf 76}, 064004 (2007).
[\arXiv[astro-ph]{0705.1158}]

\bibitem{Cai_0910}
R.-G. Cai, L.-M. Cao, Y.-P. Hu, and N. Ohta,
{\it ``Generalized Misner-Sharp Energy in f(R) Gravity,''}
Phys. Rev. D {\bf 80}, 104016 (2009).
[\arXiv[hep-th]{0910.2387}]

\bibitem{Guo_1507}
J.-Q. Guo and P. S. Joshi,
{\it ``Interior dynamics of neutral and charged black holes,''}
Phys. Rev. D {\bf 92}, 064013 (2015).
[\arXiv[gr-qc]{1507.01806}]

\bibitem{Starobinsky_1980}
A. A. Starobinsky,
{\it ``A new type of isotropic cosmological models without singularity,''}
Phys. Lett. {\bf 91B}, 99 (1980).

\bibitem{Vilenkin_1985}
A. Vilenkin,
{\it ``Classical and quantum cosmology of the Starobinsky inflationary model,''}
Phys. Rev. D {\bf 32}, 2511 (1985).

\bibitem{Nojiri_0804}
S. Nojiri and S. D. Odintsov,
{\it ``The future evolution and finite-time singularities in F(R)-gravity unifying the inflation and cosmic acceleration,''}
Phys. Rev. D {\bf 78}, 046006 (2008).
[\arXiv[hep-th]{0804.3519}]

\bibitem{Bamba_0807}
K. Bamba, S. Nojiri, and S. D. Odintsov,
{\it ``Future of the universe in modified gravitational theories: Approaching to the finite-time future singularity,''}
J. Cosmol. Astropart. Phys. {\bf 10} (2008) 045.
[\arXiv[hep-th]{0807.2575}]

\bibitem{Capozziello_0903}
S. Capozziello, M. De Laurentis, S. Nojiri, and S. D. Odintsov,
{\it ``Classifying and avoiding singularities in the alternative gravity dark energy models,''}
Phys. Rev. D {\bf 79}, 124007 (2009).
[\arXiv[hep-th]{0903.2753}]

\bibitem{Appleby_0909}
S. A. Appleby, R. A. Battye, and A. A. Starobinsky,
{\it ``Curing singularities in cosmological evolution of F(R) gravity,''}
J. Cosmol. Astropart. Phys. {\bf 06} (2010) 005.
[\arXiv[astro-ph]{0909.1737}]

\bibitem{Bamba_1012}
E. Elizalde, S. Nojiri, S. D. Odintsov, L. Sebastiani, and S. Zerbini,
{\it ``Non-singular exponential gravity: a simple theory for early- and late-time accelerated expansion,''}
Phys. Rev. D {\bf 83}, 086006 (2011).
[\arXiv[hep-th]{1012.2280}]

\bibitem{Bamba_1101}
K. Bamba, S. Nojiri, and S. D. Odintsov,
{\it ``Time-dependent matter instability and star singularity in F(R) gravity,''}
Phys. Lett. B {\bf 698}, 451 (2011).
[\arXiv[gr-qc]{1101.2820}]

\bibitem{Poisson_2004}
E. Poisson,
{\it A Relativist's Toolkit: The Mathematics of Black-Hole Mechanics}
(Cambridge University Press, Cambridge, UK, 2004).

\bibitem{Frolov_2004}
A. V. Frolov,
{\it ``Is It Really Naked? On Cosmic Censorship in String Theory,''}
Phys. Rev. D {\bf 70}, 104023 (2004).
[\arXiv{hep-th/0409117}]

\bibitem{Graves_1960}
J. C. Graves and D. R. Brill
{\it ``Oscillatory Character of Reissner-Nordstr\"{o}m Metric for an Ideal Charged Wormhole,''}
Phys. Rev. {\bf 120}, 1507 (1960).

\bibitem{Reall_2015}
H. Reall,
{\it Lecture Notes on Black Holes,}
\url{http://www.damtp.cam.ac.uk/user/hsr1000/black_holes_lectures_2015.pdf} (accessed on August 29, 2015).

\bibitem{Nunez}
A. Nunez and S. Solganik,
{\it ``The content of f(R) gravity,''}
\arXiv{hep-th/0403159}

\bibitem{Dolgov}
A. D. Dolgov and M. Kawasaki,
{\it ``Can modified gravity explain accelerated cosmic expansion?''}
Phys. Lett. B {\bf 573}, 1 (2003).
[\arXiv{astro-ph/0307285}]

\bibitem {Amendola}
L. Amendola, R. Gannouji, D. Polarski, and S. Tsujikawa,
{\it ``Conditions for the cosmological viability of f(R) dark energy models,''}
Phys. Rev. D {\bf 75}, 083504 (2007).
[\arXiv{gr-qc/0612180}]

\bibitem{Guo_1305}
J.-Q. Guo and A. V. Frolov,
{\it ``Cosmological dynamics in f(R) gravity,''}
Phys. Rev. D {\bf 88}, 124036 (2013).
[\arXiv[astro-ph.CO]{1305.7290}]

\bibitem{Justin1}
J. Khoury and A. Weltman,
{\it ``Chameleon Fields: Awaiting Surprises for Tests of Gravity in Space,''}
Phys. Rev. Lett. {\bf 93}, 171104 (2004).
[\arXiv{astro-ph/0309300}]

\bibitem{Justin2}
J. Khoury and A. Weltman,
{\it ``Chameleon Cosmology,''}
Phys. Rev. D {\bf 69}, 044026 (2004).
[\arXiv{astro-ph/0309411}]

\bibitem{Chiba}
T. Chiba, T. L. Smith, and A. L. Erickcek,
{\it ``Solar System constraints to general f(R) gravity,''}
Phys. Rev. D {\bf 75}, 124014 (2007).
[\arXiv{astro-ph/0611867}]

\bibitem{Tamaki_0808}
T. Tamaki and S. Tsujikawa,
{\it ``Revisiting chameleon gravity - thin-shells and no-shells with appropriate boundary conditions,''}
Phys. Rev. D {\bf 78}, 084028 (2008).
[\arXiv[gr-qc]{0808.2284}]

\bibitem{Tsujikawa_0901}
S. Tsujikawa, T. Tamaki, and R. Tavakol,
{\it ``Chameleon scalar fields in relativistic gravitational backgrounds,''}
J. Cosmol. Astropart. Phys. {\bf 05} (2009) 020.
[\arXiv[gr-qc]{0901.3226}]

\bibitem{Guo_1306}
J.-Q. Guo,
{\it ``Solar system tests of f(R) gravity,''}
Int. J. Mod. Phys. D {\bf 23}, 1450036 (2014).
[\arXiv[astro-ph.CO]{1306.1853}]

\bibitem{Pretorius}
F. Pretorius,
{\it ``Numerical Relativity Using a Generalized Harmonic Decomposition,''}
Classical Quantum Gravity {\bf 22}, 425 (2005).
[\arXiv{gr-qc/0407110}]

\bibitem{Sorkin}
E. Sorkin and T. Piran,
{\it ``Effects of Pair Creation on Charged Gravitational Collapse,''}
Phys. Rev. D {\bf 63}, 084006 (2001).
[\arXiv{gr-qc/0009095}]

\bibitem{Golod}
S. Golod and T. Piran,
{\it ``Choptuik's Critical Phenomenon in Einstein-Gauss-Bonnet Gravity,''}
Phys. Rev. D {\bf 85}, 104015 (2012).
[\arXiv[gr-qc]{1201.6384}]

\bibitem{Misner}
C. W. Misner and D. H. Sharp,
{\it ``Relativistic Equations for Adiabatic, Spherically Symmetric Gravitational Collapse,''}
Phys. Rev. {\bf 136}, B571 (1964).

\bibitem{Hayward}
S. A. Hayward,
{\it ``Gravitational energy in spherical symmetry,''}
Phys. Rev. D {\bf 53}, 1938 (1996).
[\arXiv{gr-qc/9408002}]

\bibitem{Baumgarte}
T. W. Baumgarte and S. L. Shapiro,
{\it Numerical Relativity: Solving Einstein's Equations on the Computer}
(Cambridge University Press, Cambridge, UK, 2010).

\bibitem{Csizmadia}
P. Csizmadia and I. Racz,
{\it ``Gravitational collapse and topology change in spherically symmetric dynamical systems,''}
Classical Quantum Gravity {\bf 27}, 015001  (2010).
[\arXiv[gr-qc]{0911.2373}]

\bibitem{Garfinkel}
D. Garfinkle,
{\it ``Choptuik scaling in null coordinates,''}
Phys. Rev. D {\bf 51}, 5558 (1995).
[\arXiv{gr-qc/9412008}]

\bibitem{Frolov_2008}
A. V. Frolov,
{\it ``A Singularity Problem with f(R) Dark Energy,''}
Phys. Rev. Lett. {\bf 101}, 061103 (2008).
[\arXiv[astro-ph]{0803.2500}]

\bibitem{Guo_2015}
J.-Q. Guo and P. S. Joshi, in preparation.

\bibitem{Hong_0808}
S. E. Hong, D.-i. Hwang, E. D. Stewart, and D.-h. Yeom,
{\it ``The causal structure of dynamical charged black holes,''}
Classical Quantum Gravity {\bf 27}, 045014 (2010).
[\arXiv[gr-qc]{0808.1709}]

\bibitem{Kawai_1409}
H. Kawai and Y. Yokokura,
{\it ``Phenomenological Description of the Interior of the Schwarzschild Black Hole,''}
Int. J. Mod. Phys. A {\bf 30}, 1550091 (2015).
[\arXiv[hep-th]{1409.5784}]

\bibitem{Belinskii_1970}
V. A. Belinskii, I. M. Kalathnikov, and E. M. Lifshitz,
{\it ``Oscillatory Approach to a Singular Point in the Relativistic Cosmology,''}
Adv. Phys. {\bf 19}, 525 (1970) [Sov. Phys. Usp. {\bf 13}, 745 (1971)].

\bibitem{Belinskii_1973}
V. A. Belinskii and I. M. Khalatnikov,
{\it ``Effect of scalar and vector fields on the nature of the cosmological singularity,''}
Zh. Eksp. Teor. Fiz. {\bf 63}, 1121 (1972) [Sov. Phys. JETP {\bf 36}, 591 (1973)].

\bibitem{Belinski_1404}
V. A. Belinskii,
{\it ``On the cosmological singularity,''}
Int. J. Mod. Phys. D {\bf 23}, 1430016 (2014).
[\arXiv[gr-qc]{1404.3864}]

\bibitem{einstein_online}
BKL conjecture.
Available online: \url{http://www.einstein-online.info/dictionary/bkl-conjecture} (accessed on August 29, 2015).

\bibitem{Carroll_2003}
S. Carroll,
{\it Spacetime and Geometry: An Introduction to General Relativity}
(Addison-Wesley, San Francisco, U.S.A. 2003).

\bibitem{Landau_1971}
L. D. Landau and E. M. Lifshitz,
{\it The Classical Theory of Fields}, Course of Theoretical Physics Series Vol.2
(Pergamon Press, Oxford, UK 1971), 4th ed.
\end{thebibliography}
\end{document}